# Coronographic Methods for the Detection of Terrestrial Planets

Conclusions from a workshop held February 02-06, 2004 at Leiden University

Edited by A. Quirrenbach

Based on contributions from the workshop participants (see p. 78)

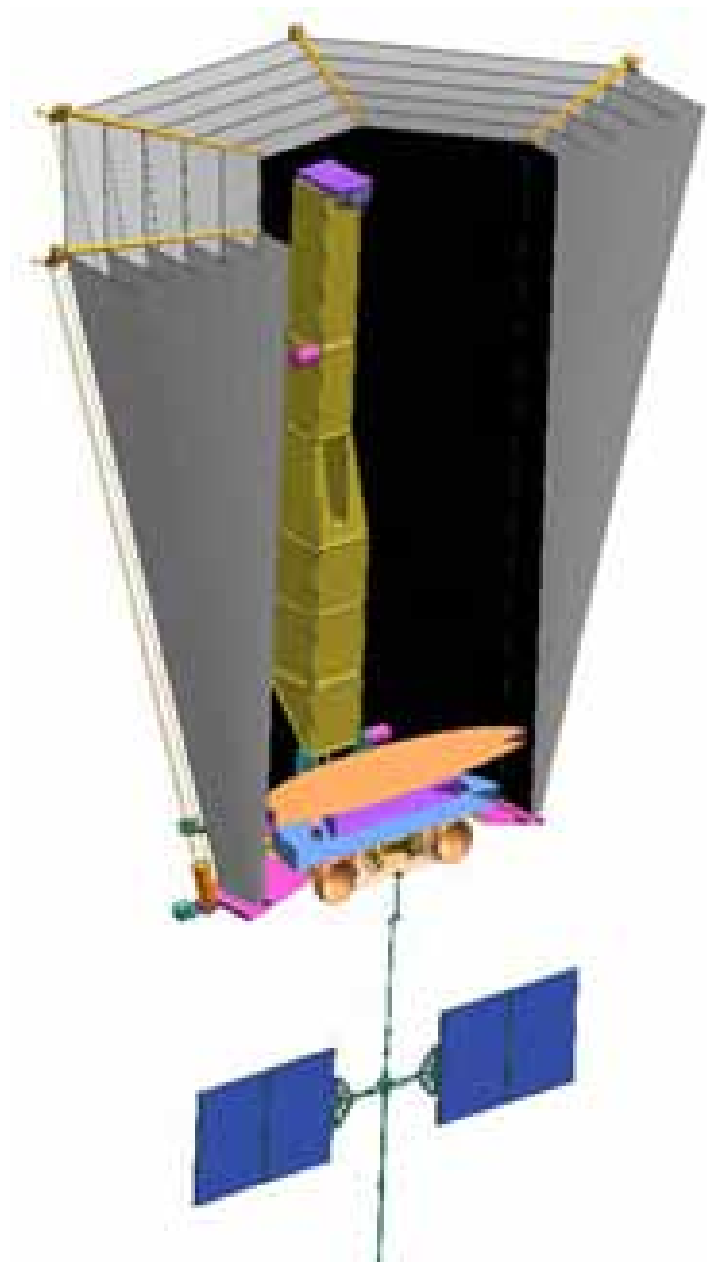

# Table of Contents

# 1. Introduction

We are privileged to live in an age when we can realistically expect to search for Earth-like planets around nearby stars, and to look for evidence of life on those planets. For thousands of years people have wondered about these questions. We need no longer wonder. For the first time, we now possess the technical ability to carry out the dream of really finding out.

A dozen of methods have been proposed to search for planets, but to date only four indirect methods have contributed positive results. These are the Doppler shift of the parent star induced by the planet in its orbit, occultation of the star by the planet, pulsar timing, and gravitational microlensing. An additional promising method is currently under development: looking for the astrometric motion of the star, induced by its planet.

These methods are exceedingly valuable in their ability to detect and measure the masses and orbits of giant planets, but they are very poor at detecting Earth-sized planets. Furthermore these methods cannot tell us anything about the surface and atmosphere of a planet (except for the occultation technique which can detect some species in the extreme upper reaches of a planet's atmosphere).

To detect and characterize Earth-sized planets, and to search for evidence of life, we must directly detect the information-carrying photons from the planet itself. Future extremely large ground-based telescopes will play a role in detecting planets, but it is at present unclear whether ground-based techniques can be pushed into the realm of Earth-like planets. It is therefore generally assumed that space missions will provide the most direct path towards these goals. Two main types of instruments have been proposed to do this: an interferometer operating at thermal infrared wavelengths, and a coronograph operating at visible wavelengths. The scientific aspects of both techniques were discussed at length during the joint US-European meeting held in Heidelberg in April 2003.

After the Heidelberg meeting, many of us felt that it would be productive to hold a second joint US-European meeting, this time as a workshop devoted specifically to the coronograph technique. We realized that researchers in the US and Europe both had already made very strong contributions to the optical theory of coronographs. We also recognized that the technical aspects of coronographs had been pursued more strongly in the US, owing to funding considerations. Therefore we decided to focus the second joint meeting on taking an inventory of current efforts in this area, and on fostering a stronger European level of contribution to the technical side of coronographs, in the universities, in industry, and at national levels. That is the goal of the present report. The recent announcement by NASA of an intended launch of a coronographic option for TPF (TPF-C) in 2014 reinforces the pertinence of our effort.

This report was generated during a week-long workshop held at the University of Leiden in the Netherlands, 2-6 February 2004. Attendance was limited in number to match the facilities available. Approximately 50 people attended, in about equal numbers from the US and Europe. We were very fortunate to be offered the full use of the wonderful facilities and generous hospitality of the Lorentz Center at the University of Leiden.

The workshop itself was opened by a few introductory talks, followed by a round of self-introductions by the participants. Most of the week was spent in focus groups, developing the material in this report. By the end of the last day of the workshop, every participant had contributed to the report, and the first drafts were all in hand. Polishing and editing required another few months.

Our goal was to quickly produce a comprehensive snapshot of the state of the art in coronography, complete with clear suggestions as to areas where more research and technical development are required. It is our hope that many of these areas will grow and flourish in Europe and in the US.

We should also mention at the outset some of the limitations of this report. It is evident that it is incomplete in many respects, and biased towards the views of the workshop participants. The lists of technologies, institutions with specific experience etc. that we provide are very likely not exhaustive. We may have missed important developments in areas where we lacked collective expertise. In some areas the workshop participants felt knowledgeable, but could not reach consensus on all issues. Specifically, we did not attempt to conduct a fair assessment of the merits of visible-light coronography vis-à-vis mid-infrared interferometry for planet detection, and we have completely left out a discussion of other scientific programs that could be carried out with such missions.

# 2. Overview of the Scientific and Technical Merits of the Coronographic Approach

## 2.1. Introduction

If asked to design the ideal mission to search for and characterize extrasolar terrestrial planets, most scientists would say that visible as well as infrared wavelengths should be required. The reason is that we understand that the visible and infrared can measure quite different properties of an object, as is dramatically illustrated in a comparison of visible Hubble Space Telescope and thermal infrared Spitzer Space Telescope images.

In the Leiden workshop reported here, we approach the question of visible vs. infrared by considering the scientific and technical merits presented by a coronograph operating in the visible. (By "visible" we mean a wavelength band lying well within the range from the ultraviolet to the thermal infrared.) We do not formally compare the visible and thermal infrared approaches, because adequate preparation of such a direct comparison would require a substantial amount of work, well beyond the scope of the present document. However from what we know today, we see certain inherent advantages to a visible coronograph approach. Although both types of instruments are undoubtedly required in the long run, these inherent advantages suggest to us that it is worth exploring the option that the first mission could be a coronograph, for the following scientific and technical reasons.

## 2.2. Detection Advantages

The first main advantage of visible wavelengths ($\lambda$) is that they are short compared to the thermal infrared, and this means that the size (D) of the collecting optics can be scaled correspondingly, for a given angular resolution ($\lambda/D$). The resulting collecting mirror size is still large (~ 6–12 m diameter) by conventional standards, but it is relatively small compared to the array baselines needed for the thermal infrared (a factor of about 3 larger). The second advantage of visible wavelengths is that the telescope itself will not emit such photons into the detector. The result is that the telescope can be operated near room temperature, giving a strong engineering advantage over a thermal infrared instrument that typically would be cooled to 40 Kelvin. The third advantage is that there are numerous coronograph designs that can be used with a monolithic or segmented primary mirror. This means that we may be able to build several types of coronographs into a given telescope, and select the best design for a given application, depending on the target at hand. A fourth advantage is the simple aperture shape and excellent point spread function, allowing direct application of the same large telescope to conventional astronomical studies. In summary, the visible coronograph can be built as a single telescope, operating at room temperature, with flexible options available so that an optimal coronograph mask can be selected to match the target.

## 2.3. Characterization Advantages

An advantage of visible wavelengths is that they provide a richer palette of spectral signatures than the infrared – at least when observations over a large wavelength range, and with high spectral resolution and signal-to-noise are available. Therefore we can obtain information about a planet in the visible that is not accessible in the infrared. For a planet like the present Earth, we can measure spectral features of water ($H_2O$), ozone ($O_3$), oxy-gen ($O_2$), Rayleigh scattering (column abundance of all gases above the surface and clouds), "red edge" (indicating plant leaves on dry land), color (blue, green, red, infrared bands indicating whether the planet is similar to Venus, Earth, Mars, or Jupiter), brightness (indicating whether terrestrial or Jovian in size), time variations in brightness (giving the rotation rate, and indicating weather patterns, and the presence of large ocean and land masses), and polarization (characteristic of a molecular atmosphere and of Venus-like cloud droplets). Furthermore these measured values can be used to infer properties including the planet's temperature, diameter, mass, surface gravity, and atmospheric pressure. If the planet is like the primitive Earth, then we have a chance to measure methane ($CH_4$) and carbon dioxide ($CO_2$) as well. By comparison, in the thermal infrared the list of measurable values is shorter. In summary, the visible coronograph can measure a long list of planet properties, including several indicators of life, as well as the environmental conditions conducive to generating and maintaining life.

## 2.4. Planetary System Advantages

A visible coronograph has advantages in characterizing the planetary system. It can measure the level of zodiacal light (indicating the presence of an asteroid belt or comet cloud), and giant planets at Jupiter-like distances (indicating possible protective gravitational deflectors of harmful asteroids). In the event that Earth-like planets are infrequent or absent, by measuring the planetary system we may learn about a possible connection between terrestrial planets and other system properties. These properties can be measured in the thermal infrared as well, but with the caveat that as the individual infrared telescope diameters are increased to collect more light, their field of view shrinks. This makes it ironically problematic whether the outer planets can be seen. A visible coronograph does not suffer from this effect.

# 3. Scientific Drivers and Requirements

## 3.1. Fundamental Scientific Goals

The TPF/Darwin projects would not have come into existence without the goal of searching for life outside the solar system. The prospect of a realistic search for life on terrestrial exoplanets was indeed the inspiration for the Darwin and TPF proposals. If we search for life based on organic chemistry, the most promising location is small (0.5 $M_\oplus$ < M < 10 $M_\oplus$) exoplanets in the "water zone" (Habitable Zone, HZ) of K, G, and F stars. (M stars are probably excluded due to their high X-ray/UV flux.) Mars and Venus-like planets are extreme cases in this respect. It is even possible to detect spectral "biosignatures", revealing the possible presence of life on these planets.

This goal leads to 4 specific objectives:

- Planet detection,
- Physical characterization,
- Chemical characterization and biosignatures,
- Planetary system architecture.

## 3.2. Scientific Background

### 3.2.1. The Frequency $\eta_\oplus$ of Terrestrial Planets in the Habitable Zone

There is currently (i.e., before the launch of space-borne transit missions) no objective empirical method to estimate the fraction of stars with terrestrial planets. The study of circumstellar disks provides only weak observational constraints on the average number of terrestrial planets in the habitable zone per star, a parameter known as $\eta_\oplus$. Evidence from sub-millimeter surveys shows that about half of young stars possess disks with sufficient mass to form a planetary system like our own. It is unclear if stars without disks never had them, or if their disks simply dissipated very rapidly. It is also unclear if each disk that we see will form a planetary system, or if they might dissipate without forming any planets. Still, the disk observations probably allow setting an upper limit of 0.5 for $\eta_\oplus$. The newly launched Spitzer Space Telescope has the sensitivity to trace disk evolution in Solar-type stars; its measurements of the dust inventory as a function of time will constrain models for the formation of terrestrial planets, and through those models, the theoretical expectation value for $\eta_\oplus$. In 2007 we will have from the CoRoT mission the distribution of planets in the (a,R) plane up to a = 0.3 AU and down to R = 1.5 … 2 Earth radii. The extrapolation of the distribution $f$ (a,R) of detected planets to a = 1 AU and R = 1 … 1.5 Earth radii will give an order of magnitude estimate for the fraction of terrestrial planets in the HZ. The transit missions (CoRoT and Kepler) can only detect planets at distances of order 500 pc, but this should provide a good statistical estimate of the local frequency of old planets with distances up 25 pc, because the Galaxy is well-mixed on these scales.[1]

---

[1] Note though that the CoRoT mission will also explore two different parts of the Galaxy, which could give some insight into the dependence on planet frequency as a function of location in the Galaxy.

### 3.2.2. Types of Planets

We believe that life requires liquid water, and therefore that life – or at least life forms similar to our own – can only exist in a "habitable zone", in which liquid water can exist on the planet's surface. We also believe that plate tectonic activity is necessary to sequester and recycle $CO_2$. We do not know, however, the precise boundary of the habitable zone, in terms of semi-major axis, eccentricity, albedo, obliquity, and greenhouse effect; therefore the range of semi-major axis space to be searched is uncertain. Nevertheless, a primary search zone (to be scaled by $L_*^{1/2}$) is from 0.9 AU to 1.1 AU, and a desirable search zone extends from 0.7 AU to 2 AU. We do not know the minimum mass of a planet required to maintain plate tectonics, therefore we do not know the minimum diameter of a terrestrial planet, and thus its relative reflecting area. But a minimum radius can be set at R = 0.4 Earth radii, beyond the envisaged minimum detectable radius of 0.7 to 1 Earth radii. Planets with masses beyond 10 Earth masses (i.e., 2 Earth radii for solid planets and 3 Earth radii for water planets) would retain their hydrogen and likely turn into giant planets.

**Suggestions for European work:** We need better definitions of the range of orbital semi-major axis, planet radius, and other characteristics for "habitable" extrasolar terrestrial planets.

## *3.3. Specific Aspects of the Coronographic Approach*

The goal of detecting and characterizing Earth-like planets can be reached by observing either the thermal emission of the planet (heated by its parent star) or the stellar light, in the visible, reflected by the planet. The relative merits of the two approaches will probably be debated for many years to come; we will not attempt resolve this debate here. The instrumentation for the reflected light regime is generally considered to be a coronograph associated with a single telescope (although it could be associated with an interferometer). This is why we here speak about the "coronographic approach", and hereafter call the corresponding mission concept "Terrestrial Exoplanet Coronograph" (TEC).

The coronographic approach has the specific capability to measure the planet flux and its characteristics (color, spectrum, time variation, polarization). From these observables, one can in principle derive the following:

- The amount of Rayleigh scattering,
- The atmospheric gas content – including oxygen as a biosignature,
- Cloud coverage and its possible time variation,
- The planet temperature,
- The planet size (radius and mass),
- Surface properties: ocean, rocks, possibly "vegetation" signature,
- The period of rotation (by analyzing the time series of the signal),
- The presence of associated objects (moons and rings).

Some of these characteristics are more difficult to obtain than the others, depending on the SNR for different spectral resolutions and exposure times, and some require the help of models of the planets (surface, clouds, and atmosphere).

In addition, coronographic imaging can easily accommodate auxiliary science such as imaging the environments of stars, the brightest optically identified gamma ray bursts, and quasars.

### 3.3.1. Detection

Secure detection of planets requires at least three exposures at different epochs, from which one can derive the orbital elements. The coronographic approach has the specificity that the reflected light flux is correlated with the orbital phase, making the identification of the detected object as a planet easier.

### 3.3.2. Physical Characterization

**Status:** An important task for a TEC is to physically characterize terrestrial planets. The TEC should be able to use color information to recognize if a planet is spectrally similar to a Solar System planet, or if it has distinctively different colors. To do this, it should be able to measure the colors of planets, at a spectral resolution of $R \sim 5$. TEC measurements should permit the planet's mass, radius, and temperature to be estimated, on the basis of analogy with Solar System objects. Observations of Rayleigh scattering will constrain the atmospheric density. Time variations of brightness can potentially tell us the length of day, the presence of oceans and land masses, and the degree of cloud variability. Also, we should be able to measure unanticipated color signatures (similar to the unexplained features in the Venus and Jupiter atmospheres or peculiar reflectance spectral features), because we do not know what extrasolar planets will look like.

**Open issues:** We do not have a complete spectral library of potential constituents. We do not have a good instrument concept for measuring 3 or more colors in parallel.

**Suggestions for European work:** We need studies of the accuracy needed for spectral band measurements. We need a library of reflectances of potential planetary surfaces and atmospheres. We need a design for a focal-plane instrument to measure colors, and also photometric variability.

### 3.3.3. Chemical Characterization and Biomarkers

**Status:** The driving goal of a TEC is to chemically characterize the planet and search for signs of life (biomarkers). To accomplish this, the TEC must have spectroscopic capability, with a spectral resolution of $R \sim 70$, to measure the equivalent widths of features of $H_2O$, $O_2$, and $O_3$ in Earth-like abundances, features of $CH_4$ and $CO_2$ under early-Earth conditions, and potentially the "Vegetation Red Edge". We should be able to measure unanticipated spectral signatures.

A particularly interesting – but challenging – observation would be the detection of the "Vegetation Red Edge" (VRE), which occurs at 725 nm for terrestrial vegetation, but could be at a different wavelength for extrasolar vegetation. Trying to identify such small features at unknown wavelengths in an extrasolar planet spectrum may be very difficult. The exact wavelength and strength of the spectroscopic "red edge" depends on the plant species and environment. The feature is very strong for an individual plant leaf but averaged over a spatially unresolved hemisphere of Earth, the spectral feature is diluted from this high reflectivity down to a few percent. The main factors affecting the strength of the VRE are forest canopy architecture, soil characteristics, non-continuous coverage of vegetation across Earth's surface, and the presence of clouds which obscure view of the surface. In addition the reflectance of vegetation is anisotropic and so the illumination conditions and viewing angle are important. Although the chances are very small that another planet has developed the exact same vegetation as Earth, finding the same signatures would be thrilling, but might be unlikely. But 1) it would still be very interesting to find a type of "vegetation" different from terrestrial vegetation, and 2) if a "VRE" is detected at a position incompatible with any shoulder in the libraries of spectra of minerals, it would be a promising signature of a non-terrestrial biology.

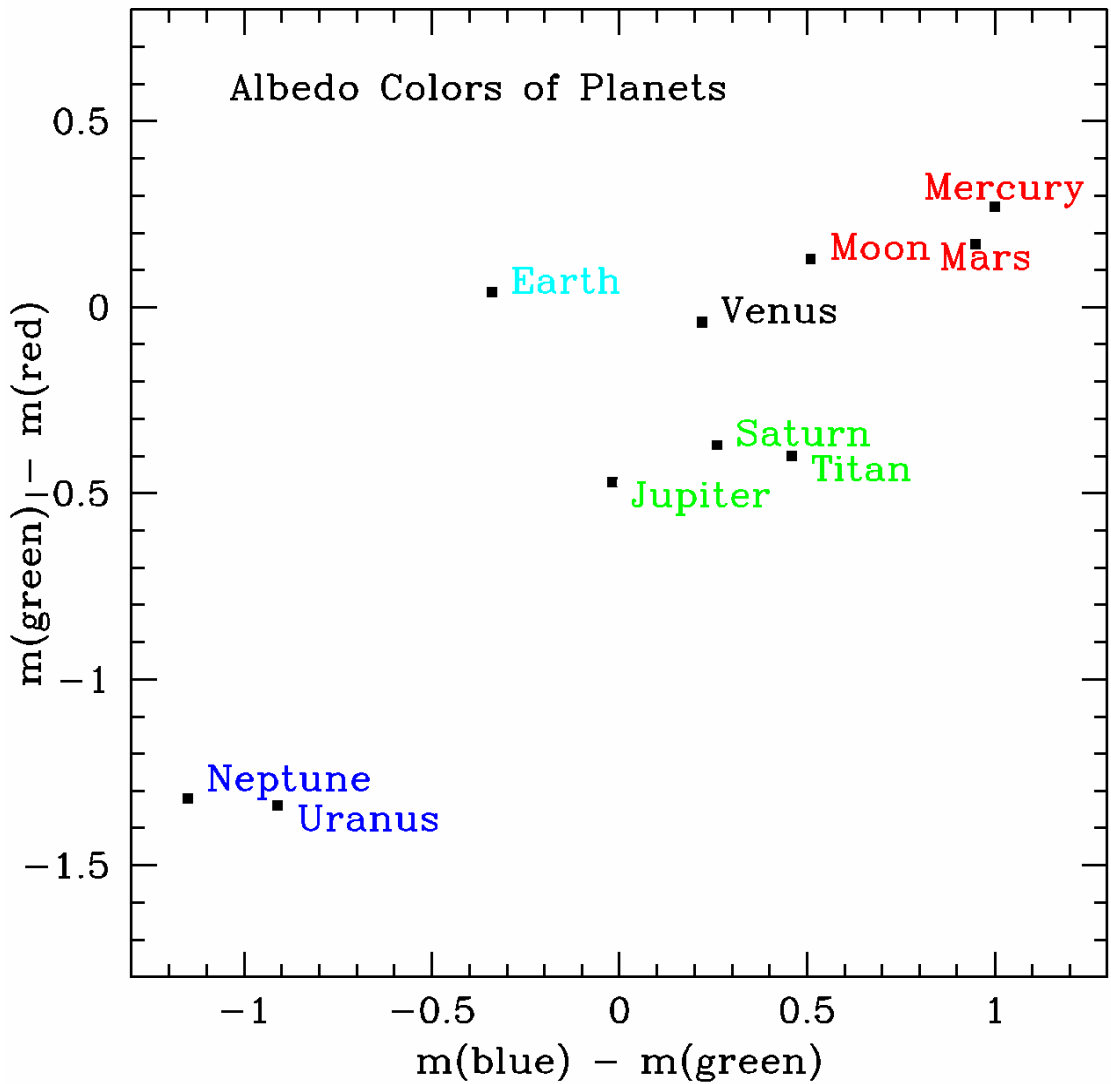

Figure 1: This color-color diagram illustrates how filter photometry could be used to distinguish among various types of planets, using the Solar System planets as a model. The colors here are the intrinsic planet colors, i.e., after the star spectrum has been divided out. Notice that the following groups or types can be separated: rocky planets, Earth-like planets, cloud-covered planets, Jupiter-like giants, and Uranus-like giants.

**Open issues:** We do not know how accurately these features should be measured in order to interpret them. We do not know if the red edge is a unique signature of plant life. Planetary diversity (i.e., planets different from Solar System planets, such as "methane planets") has not been sufficiently explored. Also atmospheric models under unusual conditions should be investigated. Figure 3 presents a simplified first attempt for a planet characterization flow chart.

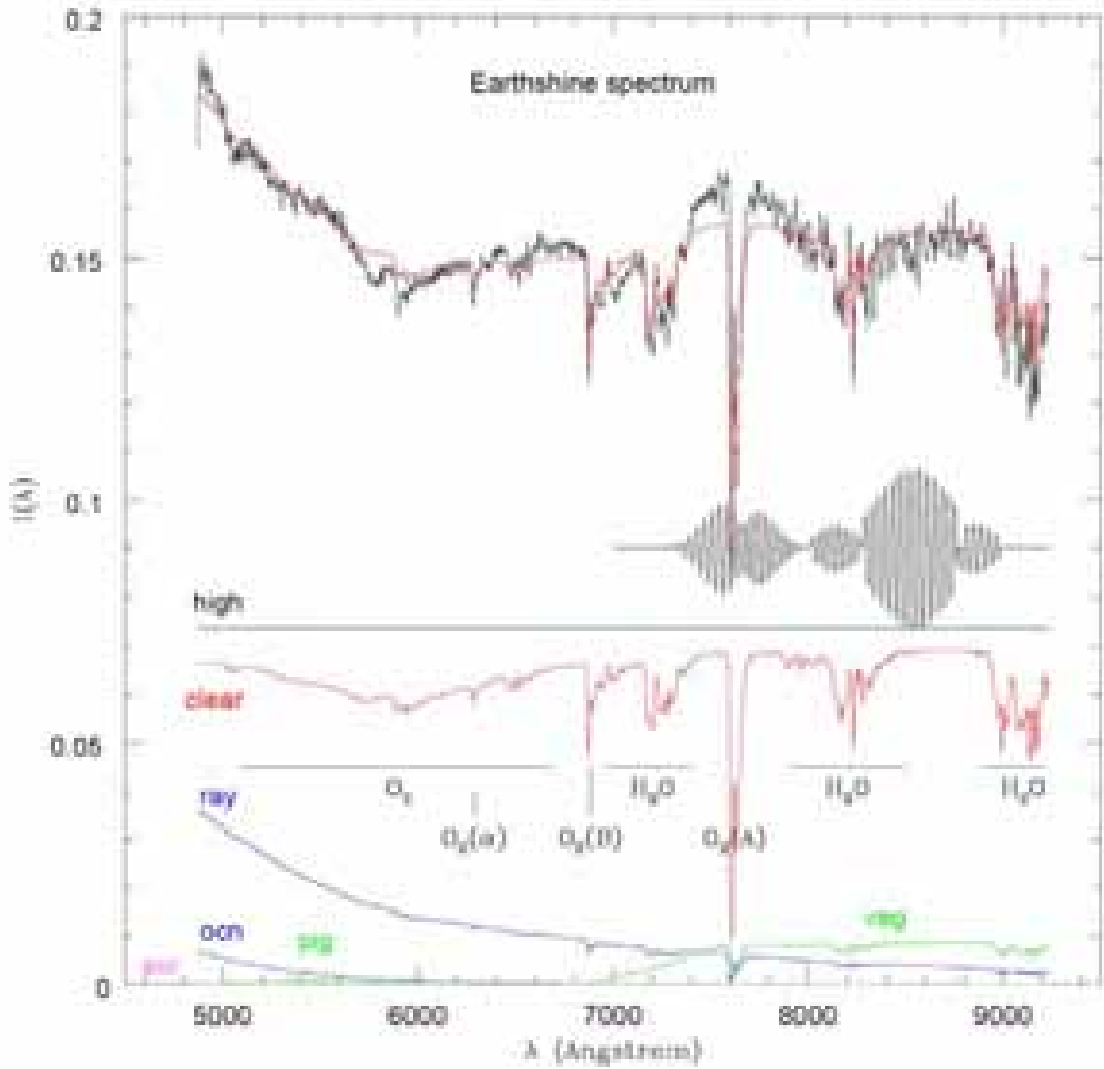

**Figure 2:** This spectrum of the reflected light from the Earth shows what kind of data we might anticipate in a favorable signal-to-noise case, and what components we might expect to be able to extract from the spectrum. The dominant features are water, ozone, oxygen, Rayleigh scattering, and a modest vegetative red edge.

**Suggestions for European work:** We need an instrument design for planet spectroscopy, such as a focal-plane integral-field spectrometer and detector system, or alternatively a wavelength-resolving detector. We need theoretical studies on planet modeling.

### 3.3.4. Polarization Issues

The detection of the planet in polarized light provides a precious complement to the planet spectrum. The planet flux is polarized by the reflection of the stellar light off different components of the planet: atmosphere (Rayleigh scattering), clouds, surface. The polarization of Rayleigh scattering is $Pol_{Ray} = \sin^2 \varepsilon_S/(1+\cos^2 \varepsilon_S)$, where $\varepsilon_S$ is the scattering angle. It depends only on orbital phase, not on planetary properties or wavelength. The polarization due to clouds and surface depends on their physical composition (ice content for clouds, nature of soils for the surface). They introduce an extra phase factor $g(\varepsilon_S, \varepsilon)$ in the polarization.

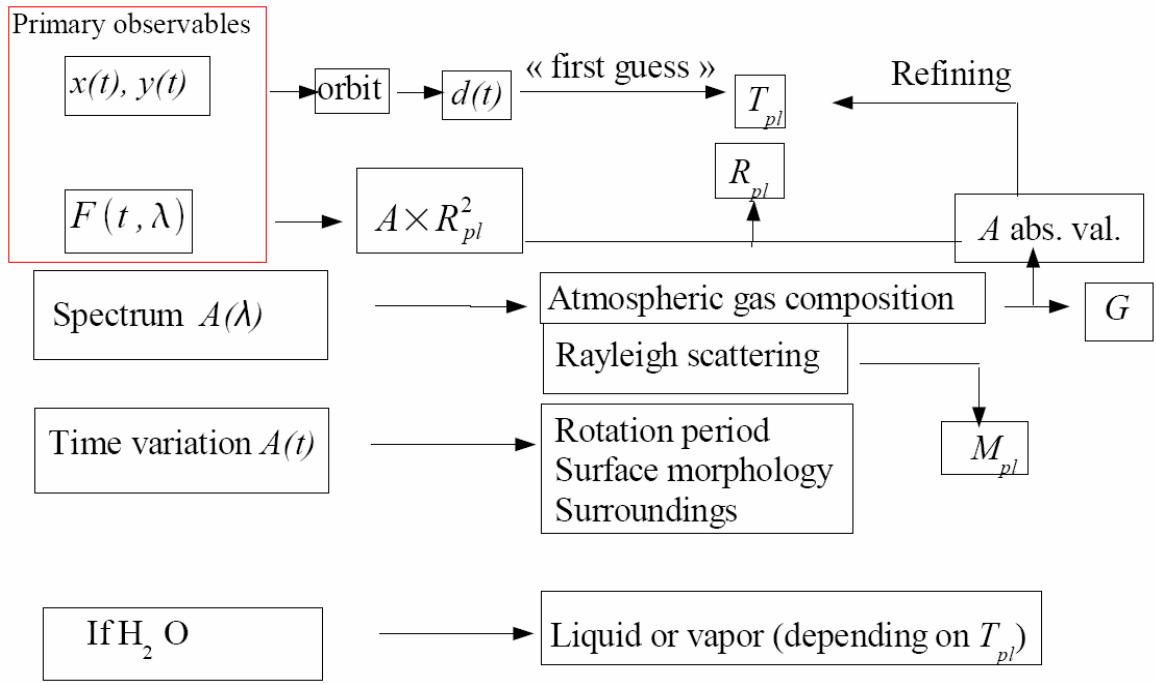

**Figure 3: Flow chart for deriving physical and chemical information about a planet from primary observables.**

**Suggestions for European work:** Investigate further the science issues of polarization, in particular the possibility to disentangle the components of the measured global polarization. Europe is developing polarimetric instruments to detect exoplanets. Push this effort further.

### 3.3.5. Planetary Systems

**Status:** The fourth goal of TEC is to characterize the planetary systems of its target stars, by searching for zodiacal dust and giant planets.

**Open issues:** We do not know if the target stars have small or large levels of zodiacal dust compared to the Solar System, or if they have giant planets. We do not know if giant planets at Jupiter-like distances are needed to shield a terrestrial planet from bombardment by debris, or if this is a necessity for the formation of life. We do not know if there are systems where giant planets are common but terrestrial ones are not, or the inverse, or if these types frequently occur together. If no terrestrial planets are found in a given system, we want to know why; measurements of the zodiacal light and giant planets may provide clues.

**Suggestions for European work:** We need precursor measurements of zodiacal brightness, and the presence of giant planets in Jupiter-like orbits. CoRoT will provide the semi-major axis distribution of big rocky planets (R > 2 Earth radii) up to a = 0.3 AU. The radial-velocity measurements will have sufficient sensitivity to detect giant planets up to 1-2 AU for the same systems. We need the results of these surveys.

### *3.4. Artifacts and Ambiguities, and some Possible Solutions*

Determining the physical and chemical properties of the planet, its surface, the atmosphere, and the presence of a biomarker will be complicated by possible ambiguities in interpreting the spectra. The severity of these ambiguities depends on the SNR. In particular, biosignatures have such an important philosophical impact that they must be extremely robust before one can claim conclusively that indications of extraterrestrial life have been found. This means that for biomarkers, more than for any other characteristics, all possible artifacts must be explored (this caveat holds also for the thermal infrared approach). Here we discuss some of them, and indicate possible solutions.

### 3.4.1. Planet Total Albedo

The surface albedo $A$ can vary from 0.05 (for oceans or a Moon-like surface) to 0.4. But the planets a TEC is capable of detecting will be sufficiently massive to retain an atmosphere; then the total albedo is dominated by the atmosphere (clouds with $A = 0.7$ or Rayleigh scattering) and the uncertainty from the surface albedo is negligible. There is nevertheless the possibility that a low mass ($< 1\ M_\oplus$), low density frozen "water planet" (with $R > 1\ R_\oplus$), with an albedo of 0.9 can be confused with an Earth-like planet with high altitude clouds.

### 3.4.2. Planet Radius

If the planet has a giant moon, its radius is overestimated by a factor $(1+r^2)^{1/2}$, where $r$ is the moon-to-planet radius ratio. If it has a ring with a radius $R_{Ring}$ and an inclination $i$ its radius is overestimated by a factor $(R_{Ring}/R_{pl})\sqrt{\cos i}$. Both can be detected by the time variation of the planet flux.

### 3.4.3. Biosignatures

The triangular shape of the ozone Chappuis bands (520-580 nm) can be mimicked by the crossing of the Rayleigh scattering (decreasing with increasing wavelength) and a Mars-like surface reflectance (increasing with the wavelength). But then there should be no oxygen line. An additional concern is the possibility of abiotic generation of a large amount of oxygen (and hence in most cases also ozone). This possibility has to be addressed by detailed modeling of the planetary atmosphere, which requires gathering as much ancillary information about physical properties of the planet and its atmospheric composition as possible.

**Suggestions for European Work:** All the ambiguities discussed here would benefit from extensive theoretical modeling to establish the extent to which they could be resolved or discounted with spectra of the quality expected. It would be especially valuable if such modeling could include a range of non-Solar System-like, but still plausible, planet surfaces and atmospheres.

### *3.5. Specifications*

An Earth in the HZ of a G star with magnitude $m = 5$ has $m = 30$ and gives, in the 400-800 nm total range $N = 100\varepsilon A$ photons/hour, where $\varepsilon$ is the end-to-end efficiency and A the collecting area in square meters. For $\varepsilon = 5\%$ and A = 3.5m × 7m, $N = 100$ photons/hour. Assuming that the photon noise dominates speckle noise and detector noise, for an SNR = 7, the detection times for different features (Rayleigh scattering with $R = 3$, $CH_4$, vegetation red edge, water) range from 2 to 80 hours.

### 3.5.1. Photometric Precision

To detect albedo variations of 30% with SNR = 7, the required exposure time is 400 h. The exposure can be fragmented into elementary short exposures (for instance 30 min) in which the planet is not detected individually. We infer that a photometric precision of at least 4% is required.

### 3.5.2. Polarization

Solar System planets have linear polarizations ranging from 10% to 30%. To detect a linear polarization of 30% with SNR = 5, a precision of 6% for the polarization is required. Note that good precision on the absolute polarization is not required. Since stars are not polarized, measuring the polarization relative to the star is sufficient.

### 3.5.3. Inner Working Angle (IWA)

The specifications on IWA follow from the preliminary science requirements for observing a core group of 35 stars and an extended sample of 165 stars. The IWA is 80 mas for observing the core group and 50 mas for the extended sample.

### 3.5.4. Wavelength Range

The baseline is 500 … 800 nm. An extension down to 300 nm (Rayleigh scattering, Huggins ozone bands at 330 nm, although the latter are blended with $SO_2$) is desirable. Access to the near-infrared (to 1.1 μm or 1.3 μm) is equally important because of the $H_2O$ and $CO_2$ bands in this spectral region.

### 3.5.5. Flexibility

To characterize the most promising planets, it may be necessary to revisit them frequently (say once a week or once a month) or at specific given epochs. It is thus necessary to minimize the Sun exclusion angle, and to allow for frequent re-orientations of the spacecraft.

## 3.6. Mission Planning and Observation Strategy

A TEC should be able to carry out a program of auxiliary astrophysical science, in addition to its planet-finding program. However, the design of a TEC should be optimized for the terrestrial planet program outlined above. The fraction of the mission time devoted to auxiliary science will depend of the status of science at the time of launch. The TEC mission should be designed for a minimum 5-year operational life, with a goal of 10 years. We recognize that by the time TEC flies, nominally about 2015, several other relevant missions will have obtained data relevant to the TEC mission. These missions include SIM, which may find about 5 Earth analogs in the Solar neighborhood, and Kepler, which may find at least 20 Earth-like planets at 1 AU (if all stars have one Earth at that distance). Ground-based radial-velocity searches may find dozens of Jupiter-like planets. Ground-based interferometers (the Keck Interferometer, the Large Binocular Telescope, and the VLTI) and possible 30m class telescopes will measure dust down to about 10 zodis. A TEC mission should be planned to take advantage of the results of these missions and projects.

By 2012-15, SIM and GAIA should be able to search for planets down to a few Earth masses (if they exist) at less than 25 pc. Although this information will arrive too late to drive the design of a TEC, it will provide important targets.

### 3.6.1. Target Selection: Binary Stars

Binary stars are interesting because they can provide important information about the conditions under which planets can form in circumstellar disks. It would also be instructive for planetary system dynamics to compare the binary orbital plane with the planet orbit. Some coronographic masks allow the rejection of both components of a binary star. Linear occulters in Lyot coronographs may allow planet detection in binaries. The exact capability depends on the brightness ratio and binary separation. Another possibility to enable the detection of planets in binaries is the use of specially shaped pupils (Jacquinot-Spergel or square apodized).

### 3.6.2. Observing Strategy

Expanding the sample of known "Earth-like" planets from three (Venus, Earth, Mars) to a significantly larger number for comparative planetology will be an important goal for a TEC. We feel that as soon as one or a few planets looking like an Earth (distance to the star, total flux) are detected, subsequent observations should focus, for months and years if necessary, on searching for and validating possible biosignatures.

## *3.7. Recommendations*

- Perform an independent European analysis of the reflected light approach. This should include the expected range of properties of the surface as well as the atmosphere of Earth-like planets. Investigate in particular the potential ambiguities and artifacts.

- Update the report "Biomarkers for TPF" (DesMarais et al. 2001).

- Investigate the polarization issues (science, instrumentation design and performance).

- Include the coronographic approach in the deliberations of the regular ESA/NASA contacts; in particular include an American coronography-oriented representative in the Darwin Science Team (TE-SAT) to keep the Darwin Science Team apprised of progress in the US relevant to coronograph technology. Conversely, keep the TPF Science Working Group informed about relevant developments in Europe.

# 4. Overview of "Coronograph Space"

## *4.1. Coronograph Concept*

As initially conceived by Lyot (1939), a coronograph is a device to suppress instrumental light diffraction by the use of a sequence of stops for the specific purpose of observing the Solar corona. Since the Lyot coronograph addresses light diffraction, a "coronograph" has become a generic term for a system to suppress diffraction and scattered light in a telescope for astronomical purposes. In common usage, any system to achieve high contrast with a single aperture telescope is referred to as a coronograph, even in cases where there is no physical or historical basis for the connection to the original problem of observing the solar corona. A practical definition that encompasses the current usage of the term is that a coronograph is a device to suppress the noise associated with stellar light by rejecting it from an area of interest in the focal plane of a telescope. This light must be rejected because of the associated noise processes: wavefront distortions that produce artifacts in the image that resemble planets (speckle noise), and photon shot noise. Although in principle it is possible to treat these problems simultaneously in a completely general physical model, mostly the conceptual design of coronographs treat diffraction with Fourier optical theory, assuming that a corrected wavefront can somehow be obtained with a separate wavefront control subsystem. The sensitivity of essentially all coronographic concepts is exquisite given a sufficiently large telescope with a perfect wavefront. A primary issue is then the development of a wavefront sufficiently close to perfect, which is treated in detail in Sections 6 and 8. There are a large number of possible devices to tackle this problem. Recent years have seen extensive development of new concepts, all of which strive to search for the ideal coronograph, a device that confines the stellar light as tightly as possible, with the maximum possible efficiency to reveal for the first time images of extrasolar planets.

## *4.2. Physical Limitations of Coronography*

For a conventional circular aperture telescope, the envelope of diffracted light falls off with the angular distance from the optical axis as $\theta^{-3}$, which completely prohibits achieving the necessary contrast for planet detection for any reasonably sized telescope. The ultimate coronograph would be completely unaffected by light from a region of exclusion (e.g., the stellar disk), yet exquisitely sensitive over a region of interest outside this area. In general, with a finite telescope pupil of diameter D, it would appear impossible to confine light with any sequence of manipulations to a scale less than $\lambda/D$, nor possible to produce a null in the PSF closer than $\lambda/D$ from a transmission peak. Due to the limited availability of large telescopes, the coronograph design problem is then to somehow minimize the focal plane region of high stellar flux, without sacrificing throughput and sensitivity in the region of interest. The problem is fundamentally one of how to avoid "throwing out the baby with the bathwater".

## *4.3. Inner Working Angle*

The Inner Working Angle (IWA), commonly scaled to $\lambda/D$, is one way to describe quantitatively how close a coronograph design is to meeting this theoretical goal. The benefit of the smallest possible IWA is to minimize the required telescope size. There is no universal definition of IWA for a coronograph, but a common usage is the minimum angular offset from the star at which the star flux suppression matches the Earth-Sun contrast.

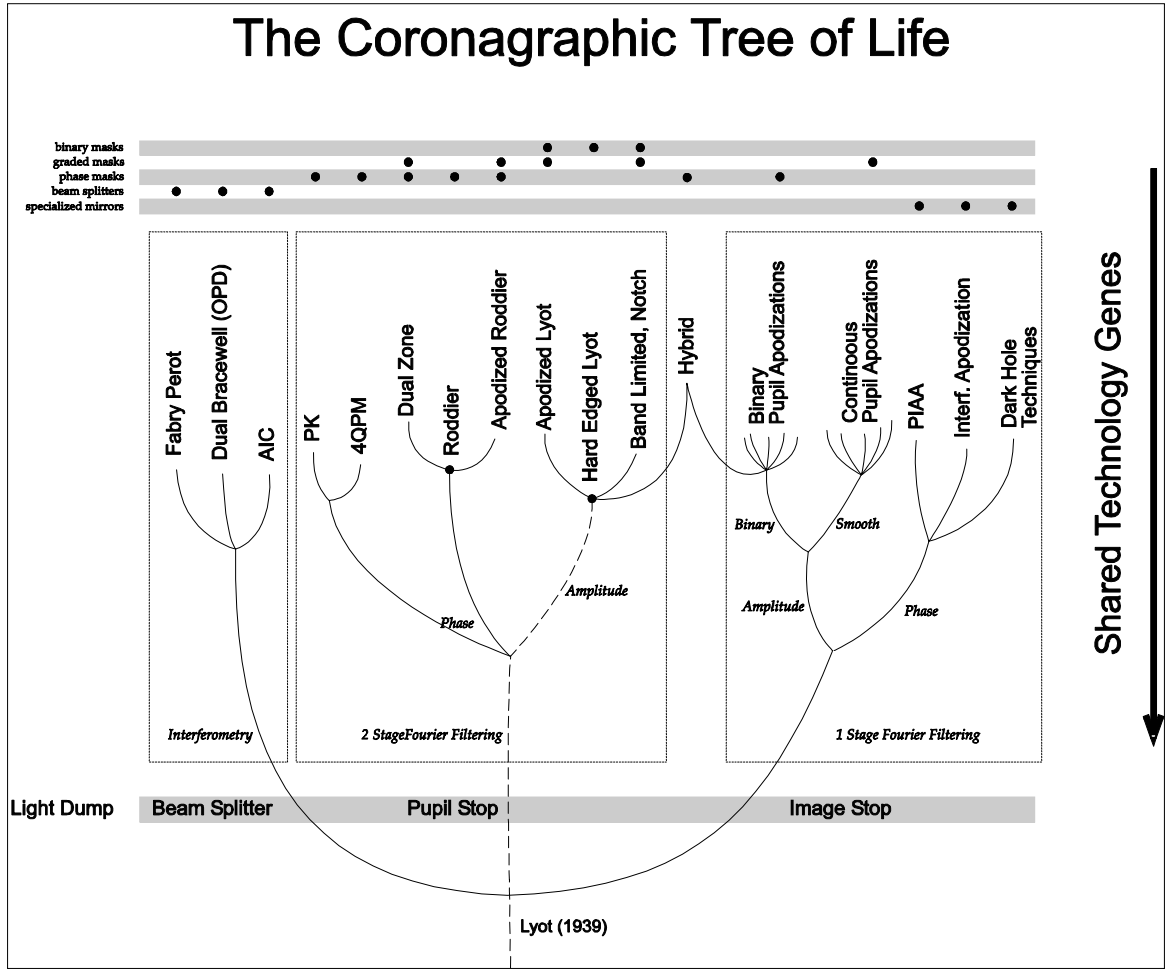

Figure 4: Classification of coronographic concepts.

## *4.4. Systematic Classification of Approaches*

A literature search reveals many concepts in coronography. A short description and nomenclature is shown in Table 1, with references to introductory articles and an attempt to document the available discussion of the concept in the literature. A preliminary classification of the families of coronographs that have developed is presented in Figure 4. This tree-based classification captures several salient differences between coronograph designs. A useful analogy is to the evolutionary tree of life. Certain clades can be identified as clearly common traits between designs, which share properties and technological inheritance. A primary division is between the mechanism the coronographs use as a "light dump", the method by which the stellar light is segregated: dumping the light in a sequence of stops in the image and exit pupil plane, as with a classical Lyot coronograph; dumping the light in a confined region of the image plane or image stop; and one rejecting light by interference beam combination at a beam splitter, with a close parallel in nulling interferometry. The image stop solutions are generally one-stage Fourier spatial filters: application of a method for controlling spatial frequencies in the pupil plane is used to confine the light in the image plane. The pupil stop family of coronographs shares the property of having two (or in some cases more) stages of Fourier spatial filtering, one in the first image plane, and one in the exit pupil plane. Hybrid approaches can be generalized from these to an expanded number of filtering steps, and there is currently fertile development in mutually optimizing these designs. External occultation has been used for Solar coronographs. This approach, while very efficient at rejecting light before diffraction has become the problem, is not widely considered to be realistic, because planet detection requires that the free-flying occulter or generalized diffraction aperture must be positioned thousands of kilometers from the telescope to achieve a sufficient IWA.

Table 1: A selection of key coronograph designs

| Name | Alternate Name | Short Description | Selected References, by 1st author & year | Comprehensive References |
|---|---|---|---|---|
| Lyot | Hard Edged Lyot, Classical Lyot | Image plane stop, undersized pupil stop | Lyot (1939), Sivaramakrishan et al. (2001) | |
| Visible Nulling Coronograph | | Pupil shearing interferometry with fibers | | Mennesson et al. (2003) |
| AIC | Achromatic Interfero-Coronograph | Imaging nulling (cat's-eye) Michelson interferometer. | Baudoz et al. (2000) | |
| PK | Phase Knife | Image plane $\pi$ phase shift, half plane geometry | Abe et al. (2001) | Abe et al. (2003) |
| 4QPM | Four Quadrant Phase Mask | Image plane $\pi$ phase shift, quadrant geometry | Rouan et al. (2000) | Riaud et al. (2000, 2001, 2003), Lloyd et al. (2003) |
| Roddier | | Image plane $\pi$ phase shift, circular geometry | Roddier & Roddier (1997) | Guyon et al. (1999) |
| Dual Zone | Dual Zone Phase Mask | Image plane variable phase shift, circular geometry, multiple zones, pupil apodization | Soummer et al. (2003) | |
| Apodized Roddier | Apodized entrance pupil Roddier | Smooth apodized pupil, image plane $\pi$ phase shift, circular geometry | Aime et al. (2002) | |
| Apodized Lyot | Apodized focal stop Lyot | Hard edged entrance pupil, smooth image plane stop | Watson et al. (1991) | |
| Band Limited | Band Limited Mask | Smooth image plane apodization with band limited function. | Kuchner et al. (2002) | |
| Notch | Notch filter | Binary image plane mask band limited over a region in the exit pupil | Kuchner et al. (2003) | |
| Apodized Pupil Lyot | | Apodized entrance pupil, hard edged image stop | Aime et al. (2002) | Soummer et al. (2003), Soummer (2005) |
| Binary Pupil Apodization | | Binary mask, ranging from simple to highly complex structure | Jacquinot et al. (1964), Kasdin et al. (2003) | |

| Name | Alternate Name | Short Description | Selected References, by 1st author & year | Comprehensive References |
|---|---|---|---|---|
| Continuous Pupil Apodizations | | | Jacquinot et al. (1964), Kasdin et al. (2003) | Nisenson et al. (2001), Soummer et al. (2003) |
| PIAA | Phase Induced Amplitude Apodization | Pupil apodization by light redistribution | Guyon (2003), Traub (2003) | |
| Interferometric Apodization | | | Aime et al. (2001) | |
| Phase Apodization | Dark Hole | Complex (phase) apodization of pupil plane | Malbet et al. (1995) | |
| External Apertures | External Occulter, Lunar Eclipse, Shaped pinhole camera | Blocking of light before entering any telescope diffraction apertures | | |

## *4.5. System Architecture, Mission Concept*

The architecture of planet finding coronographic mission concepts consists of a telescope, optimized for precise wavefront control, and with a pupil shape chosen to match the output coronograph. A wavefront control subsystem and diffraction control subsystem are used in concert to achieve the requisite suppression of speckle and photon noise and thus to meet the mission requirements for the contrast.

# 5. Critical Technological Areas for Coronography and Prospects for Advances

## 5.1. Introduction

Identification of the critical technology challenges for the Terrestrial Exoplanet Coronograph (TEC) is based on the need to measure a planet signal, and therefore to suppress the residual (after masks) scattered starlight speckle pattern at the focal plane, a result of wavefront error (WFE). Adequate suppression requires correcting distortions in the stellar wavefront phase, as well as limiting amplitude and polarization variations across the final pupil plane. The required correction and uniformity levels must be *maintained* during the several hours of integration time required for each detection of a terrestrial planet.

Due to the critical importance of maintaining high optical performance over long times, the main technological challenges for the TEC are optical quality, and mechanical and thermal stability, all at levels that in our opinion are within the grasp of a reasonable technology development effort. We identify a list of technological improvements to be made and demonstrated for the U.S. and European partners to ensure that the visible-light TEC will be manufacturable, deployable, and, above all, stable enough in space to perform successfully.

## 5.2. Optical Aspects

Starlight reaches the planet pixel not only by diffraction from the aperture rim, but also by optical aberration and scatter from wavefront defects (in phase and amplitude). Maintaining a nearly aberration-free wavefront is thus an essential condition in any coronographic setup: optics design, manufacturing, control, and assembly are of paramount importance. Also important is the control of spurious light, such as sunlight, stellar ghost images etc.

To detect planets at the expected angular separations, we must suppress stray starlight, concentrating on the effects of wavefront ripples in phase and amplitude, especially at critical spatial frequencies (~ 3 to 100 cycles per aperture, higher frequencies than those characterizing the basic aberrations – focus, coma, spherical, trefoil, etc.). Adequate suppression requires that distortions in the stellar wavefront phase be corrected to better than 0.1 nm, and that stellar wavefront amplitude and polarization variations across the final pupil plane be less than 0.03%. These correction and uniformity levels must be maintained during the hours of integration time required for the detection of a terrestrial planet. The following fields are all relevant to this family of problems.

### 5.2.1. Mirrors

Production and control of an optical system with a large primary mirror meeting the stated specifications presents a real challenge that requires development of new material, manufacturing, control, etc., technologies.

SiC space-compatible mirrors are an important European technology thrust (Herschel). Key properties of ULE, Zerodur, SiC, and other potential mirror materials need further exploration. These properties include figurability, coatability, strength, CTE, thermal conductivity, surface density, and long-term stability.

Fabrication of ultra-smooth aspheric optical surfaces has recently seen great advances due to enormous commercial and technology efforts in the semiconductor equipment industry of Europe and the U.S. The main business driver is the advent of "Extreme Ultraviolet Lithography" for 13.5 nm soft X-rays, demanding high aberration and straylight control for this very short wavelength. The results for low-spatial frequency and for mid-spatial frequency roughness are up to 0.2 nm surface RMS. Experience and results have been gained in Europe using several low-expansion materials. Equipment for computer-driven polishing, ion-beam figuring, and appropriate optical metrology is readily available in Europe, as is expertise in manufacturing and active control of large monolithic mirrors (such as ESO's VLT).

### 5.2.2. Masks

In the past few years, on both sides of the Atlantic, a wealth of new coronographic concepts using pupil or focal plane masks have been developed, followed by technical efforts toward producing real devices for lab and/or space experiments. The high levels of performance required have resulted in the exploration of the technology capabilities of European industries and research labs. In Europe, manufacturing phase masks using techniques of the semiconductor industry (ion and electron etching, lift-off) or multi-layer deposition is ongoing and has already produced operational devices (France, Germany). Programs partly funded by some national space agencies are also aiming at solving the question of achieving achromatic phase shifts through different approaches (half-wave plates, zero-order gratings [ZOGs], multi-layers). Also relevant to the coronographic masks issue is the strong R&D effort to obtain a quasi-achromatic phase shift for the Darwin program.

For shaped-pupil masks, there is also the possibility adding metrology and control systems to adjust for amplitude non-uniformities and low-order aberrations.

### 5.2.3. Wavefront Sensing and Control

Reshaping the wavefront by means of active optics is required to achieve very low phase errors, to locally produce dark holes at the focal plane. Europe has been working on adaptive optics since the 1980's. Several groups have developed extensive expertise in systems, subsystems, and optimum control algorithms, while industry has been able to provide deformable mirrors (DMs) of reasonably high density and stability. Ongoing programs on large adaptive secondary mirrors or, at the other extreme, on MEMS (electrostatic, magnetic), are currently being conducted in view of the development of extremely large telescopes (ELTs) on the ground. On the same topic, Europe has good experience in spatial filtering using mono-mode fiber optics through development of several interferometric instruments (FLUOR, VLTI, OHANA, Darwin).

Wavefront sensing options for the TEC are primarily those using science focal plane data acquired before science observations, but other concepts using measurements of pick-off beams elsewhere in the optical system should be evaluated, especially for real-time control loops using *in situ* measurement of low-order aberrations during science observations.

### 5.2.4. Polarization

Polarization of the light introduced by optical elements may become a strong limitation and has to be studied thoroughly. One possible concept is to make use of the polarization properties, rather than trying to minimize them, in order to reach the nulling level: this may have special advantages in terms of achromaticity. It is one of the paths followed in Europe, for instance through the development of ZOG or achromatic half-wave plates. It is worthwhile noting that expertise in mono-mode fiber optics with polarization preservation has been gained in astronomical interferometry.

Modeling efforts can lead to the design of compensation schemes applicable to mirror coatings, filters, optical forms, and mask materials, in order to minimize the impact of polarization effects.

### 5.2.5. Spectroscopy and Detectors

Low-spectral resolution characterization of Earth-like planets is one of the most challenging tasks of TPF/Darwin, and several possible paths for the TEC should be explored. Europe, in particular through VLT and JWST instrumentation, has developed extensive expertise in spectroscopy of faint objects, especially in systems using Integral Field Unit (IFU) spectrometers (with an image slicer or fiber optics). Other technologies to be considered for TEC spectroscopy include filters and wavelength-sensitive detectors. For detectors, the important parameters are to have frame rejection, noise-free performance, sensitivity to wavelengths from 0.4 to 1.0 microns, and radiation tolerance.

### 5.2.6. Stray Light Control

Limiting the stray light from the Sun, other stars, or internal reflections, is essential to achieve the desired level of contrast sensitivity. Baffles and light traps, anti-reflection coatings, contamination control, and careful modeling are the key elements needed to minimize stray light. Europe has some expertise on those aspects, thanks to the development of space experiments such as CoRoT, and infrared/sub-mm missions (Planck, Herschel, ISO) where similar problems occur, although though not at a same level nor the same wavelengths.

## 5.3. Mechanical / Thermal Aspects

A critical performance requirement for the coronograph optical system is that for speckle rejection the wave fronts must be stable to sub-nanometer level RMS throughout the hours required for on-target integration. Technologies to achieve this stability involve controlling thermal, mechanical, and pointing disturbances, either passively, or actively, or both.

### 5.3.1. Thermal Control

Milli-Kelvin thermal stability will be required to maintain wavefront stability. The concept currently favored for the TEC is to operate the large primary mirror and smaller secondary mirror close to room temperature so that, during observation, thermal effects on the mirrors are similar to the effects occurring during fabrication. Also, smaller temperature changes result from a given small heat disturbance to the optical system.

A thermal system conceptual design has been developed to optimize passive thermal control and includes a large, light-weight sunshield involving deployable specular v-groove vanes (similar to the thermal shield developed for JWST), actively heated, insulated, warm enclosures around the mirrors and coronograph instrument components, thermal isolation between structures, and ULE glass mirrors. Active control will be added as needed, either open or closed loop, and involving distributed heating elements and thermal sensors. It is challenging to test the performance of the deployed integrated thermal system on the ground.

One of the great challenges of the Planck mission is the cryogenic control of the detectors and their environment at the milli-Kelvin level. A strong effort has been made in this field that could be of interest for TEC.

### 5.3.2. Vibration Control

During observation, vibration from the reaction wheel assemblies (RWAs) affects the observatory in two ways – disturbing the pointing line-of-sight (attitude control), and causing optical component rigid-body movement and element deformation. Both these effects impair coronographic wavefront control. Small fast reaction wheels such as used in the Hubble Space Telescope minimize the input jitter. Passive isolation systems may be optimized to damp vibrations adequately or, failing that, active isolation systems being developed for other programs appear able to damp the RWA inputs enough to meet the stability requirements. Line-of-sight pointing stability requirements will demand a fine steering mirror as well as isolation and spacecraft attitude control systems, although pointing jitter appears nearly adequate without an active fine steering mirror.

### 5.3.3. Structural Material Properties

Micro-snap (spontaneous release of strain) is difficult to isolate because it can originate throughout the structure. Micro-snap effects can be minimized by optimal system design using integrated modeling to understand design sensitivities and interdependent requirements. Another phenomenon is micro-creep or gradual dimensional change of structures over time. This is best controlled through material selection, knowledge of precise material properties, and by applying stress relieving processes to the fabricated structures. The most significant technology needed to achieve the required nano- and pico-meter structural stability is a modeling capability precise enough to predict the impact of each concern so that designs can be optimized to minimize those impacts. To model the system properly, we must measure structural material properties, such as strength, Young's Modulus, coefficient of thermal expansion (CTE), conductivity, damping, uniformity, and temporal stability. These must be understood relative to all axes through the material, precisely enough for sufficient accuracy of the modeling results. Since many light-weight materials might be considered for TEC structures, newer light-weight fabrication techniques using non-standard methods are worth evaluating, including multi-axis properties of the resulting structure.

The European VIRGO gravitational wave detector utilizes vibration control, extremely precise metrology, and is concerned with controlling micro-snap effects. Developments from this effort may be applicable to TEC.

### 5.3.4. Mechanisms and Joints

The TEC Observatory will be large enough that deployment after launch will be required. This will involve joints and mechanisms that must move into precise locations and remain stable throughout the remainder of the mission. Robust, reliable deployment mechanisms to achieve these results are challenging, since precision location in three dimensions often conflicts with space-quality, low-friction, low-stress, highly reliable mechanism operation.

There will also be a need for adjustment and calibration of the secondary mirror after deployment. This will require very smooth, precise motion. Some technology development can be borrowed from the development of mirror segment control mechanisms for JWST.

Mechanism motion requirements can be traded off based on calibration capability, but stability after motion is critical. Design and development of concepts and hardware that can validate performance will be required before the TEC system error budgets will be able to accurately predict the residual errors that will result from joints and mechanisms.

The European space industry has developed frictionless mechanisms for deployment and clamping of subsystems in space. In addition, shape memory alloys have been used and may be applicable in some circumstances.

## *5.4. Operational Aspects*

### 5.4.1. Launch

Volume and mass available to launch the TEC are constrained by the launch vehicle and fairing size. Increased mass and volume capabilities have an impact on the design and capabilities of the TEC observatory. The current conceptual design is based on the Delta IV Heavy spacecraft with a 5 m diameter × 19 m long fairing, capable of lifting a 9245 kg load to Earth drift-away orbit (third coefficient of energy equation = 0.4). Larger diameter fairings would enable more circular primary mirror designs that have distinct advantages. The current TPF Coronograph team has chosen to constrain the design to fit into existing fairings. It is risky to allow the TEC mission feasibility to depend on the development of extremely expensive large fairings.

Larger fairings and more capable spacecraft that exist or are planned would be of great interest. The number of stars that can be searched by TEC depends on aperture size, which directly relates to launch vehicle and shroud capability.

### 5.4.2. Operational Metrology

The most important metrology task is to sense the position of the secondary mirror with respect to the primary mirror. The secondary alignment has critically tight tolerances – around 1 nanometer in z, 8 nanometer in x and y, and as low as 5 nanoradians around $\theta_x$ and $\theta_y$. These tolerances depend on coronograph form and system requirement details, including type of mask, contrast approach, and inner working angle. The current TPF coronograph concept employs laser metrology to measure the position of the secondary mirror relative to the primary mirror, smoothly and precisely adjusting the secondary mirror to the correct position within acceptable tolerances. Any alternative method of accurate position sensing across a distance of up to 10 to 20 meters would also be of interest.

Sensing of shapes of optics may be possible by monitoring thermal changes and interpreting those measurements via validated models for open-loop control using micro-heaters. There may be ways to use laser metrology to monitor optical surface changes, too.

## *5.5. Mission Design and Assurance Aspects*

### 5.5.1. Modeling

Development and verification of high-fidelity models is critical to the complex TEC, and many aspects of the models may be needed to supplement the ground testing elements that will be possible to achieve.

### 5.5.2. Test Facilities

Facilities will be needed allowing as much as possible of a full end-to-end performance test on the ground. They main challenging factors in such facilities will be the thermal and vibrational elements in the very large vacuum chamber needed for a full-aperture ground test.

## *5.6. Prospects for Advances*

In the following we provide an assessment of the current state of development of the critical technologies. It is obvious that there is a very large range of maturities from technologies that are almost at hand now, to aspects that have hardly been developed beyond the conceptual level. It is clear that the latter will require intense attention in the near future. Most importantly, potential showstoppers have to be identified, and the highest risks have to be retired as soon as possible.

**Table 2: Definition of maturities in terms of Technology Readiness Levels (TRL)**

| | |
|---|---|
| 1 | Basic principles observed and reported |
| 2 | Technology concept and/or application formulated |
| 3 | Analytical and experimental proof-of-concept |
| 4 | Component or breadboard validation in laboratory environment |
| 5 | Component or breadboard validation in relevant environment |
| 6 | System/subsystem prototype demo in a relevant environment |
| 7 | System prototype demo in a space environment |
| 8 | Actual system "flight qualified" |
| 9 | Actual system "flight proven" |

**Table 3: Status of Critical Technologies**

| | U.S. Effort & TRL | European Effort & TRL |
|---|---|---|
| **Optical Areas** | | |
| **Mirrors** | | |
|   Developmental material options | | |
|     Compatible with space environment | 9 | 8 |
|     Properties (strength, CTE, conductivity, surf. density) | - | - |
|       ULE, Zerodur properties | 9 | 9 |
|       SiC properties | 5 | 7 |
|       Other materials | 3 | 3 |
| **Fabrication** | | |
|   Developmental methods | 4 | |
|   Development of plan for metrology of mirror surface | 4 | |
|   Development of fixturing concept | 2 | |
|   Facilities | 2 | |
| **Coating** | | |
|   Achievable uniformity | 3 | |
|   Waveband performance | 9 | |
|   Polarization and straylight impact | 1 | |
|   Durability | 9 | |
|   Metrology | 3 | |
| **Surface control** | | |
|   Mechanical | 5 | |
|   Thermal | 2 | |
| **Masks** | | |
| *Analysis* | | |
|   Focal plane masks with Lyot stop | 4 | 5 |
|   Pupil plane mask systems | 3 | 5 |
|   Tolerances and Sensitivity of masks | 3 | 4 |
|   Thickness and material effects | 3 | 4 |
|   Fabrication of masks | 4 | 6 |
|   Measurement of masks - phase and amplitude | 4 | 6 |
|   Achromaticity | 3 | 4 |

|  | U.S. Effort & TRL | European Effort & TRL |
|---|---|---|
| **Wavefront Control** | | |
| *High density deformable mirror technologies* | 5 | 3 |
|     Stability of DM | 5 | 4 |
|     Wavefront sensing options | 5 | 4 |
|     Optimum control algorithms | 5 | 4 |
| **Polarization** | | |
|     *Coatings* | 3 | 3 |
|     Optical forms to minimize impact | 2 | 3 |
|     Mask material impact | 2 | 3 |
|     Modeling | 2 | 3 |
| **Spectroscopy and Colors** | | |
| *Low-resolution spectrograph* | 9 | 9 |
|     Filters | 9 | 9 |
|     Wavelength-sensitive detectors | 2 | 4 |
| **Scattered Light Control** | | |
| *Baffles* | 9 | 9 |
|     Contamination | 4 | 4 |
|     Modeling | 3 | 3 |
| **Mechanical/Thermal Areas** | | |
| **Mechanical** | | |
| *Mechanisms and joints* | | |
|     Stability after deployment | 4 | |
| *Materials for structures* | | |
|     High temporal stability | 4 | |
|     Composites or metallics? | 4 | |
|     Light weight fabrication techniques | 3 | |
|     Predictable - | | |
|         Thermal response | 6 | |
|         Stress relief | 4 | |
|         Damping | 4 | |
| **Thermal** | | |
| *Sun shade development* | 4 | 3 |
|     Milli-Kelvin thermal control | 2 | 1 |
|     Thermal sensor development | 3 | 2 |

|  | U.S. Effort & TRL | European Effort & TRL |
|---|---|---|
| **Operational Areas** | | |
| **Launch** | | |
| *Fairing size* | 4 | 7 |
| **Operational Metrology** | | |
| *Sensing of positions of optics* | 7 | 6 |
| Sensing of shapes of optics | 4 | 4 |
| **Pointing and Propulsion** | | |
| *FEEPs* | | 7 |
| Other Electric Propuls. | 9 | |
| Reaction Wheels | 9 | 9 |
| Damping | 6 | 7 |
| **Mission Design and Assurance Areas** | | |
| **Modeling** | | |
| **Test Facilities** | | |
| *Star, planet, and background sources for testbed use* | 3 | 4 |

# 6. Elements and Operation of a Coronograph

A modern diffraction-limited coronograph combines a Diffracted Light Suppression System (DLSS) and a mechanism for wavefront sensing and control (WFSC), possibly including amplitude control. "Diffracted light" is starlight left in the planet's detector pixel, even with perfect wavefronts, due to spreading of the stellar image by aperture and image masks in the telescope. The DLSS narrows or removes this diffracted light, while the WFSC suppresses the "scattered light", starlight appearing in the planet pixel due to wavefront imperfections. Light scattering can occur anywhere in the optical path, including inside the detector.

Recent years have seen the rapid development of DLSS concepts. An infinite variety of DLSS designs exist, and we can always derive more through numerical optimization (e.g Vanderbei et al. 2003). The variety of these concepts corresponds indirectly to the variety of different possible configurations of a nulling interferometer. We attempt to provide an overview of the rapid progress in the field of DLSS development and ferret out the key open questions and stumbling blocks.

A number of specific coronograph designs are discussed in Appendix 11.2.

## *6.1. DLSS Characteristics*

One configuration of an interferometer may be well suited for imaging a diffuse interstellar cloud, while another may be better suited for nulling the light from a nearby star. Our palette of DLSS designs offers analogous tradeoffs, and other tradeoffs beyond those available to interferometers. As we learn more about the rest of planet-finding trade space, we may find ourselves pushed to some corner of trade space. But for now we have a cornucopia of possibilities.

Here we describe some key DLSS characteristics, and attempt to highlight which different designs offer high performance in each area—generally at the cost of performance in some other area. Table 13 in the appendix summarizes the characteristics of some leading coronograph designs.

### 6.1.1. Theoretical Limitations

**Theoretical Maximum Extinction:** The extinction of a monochromatic on-axis point source provided by the DLSS given a perfect wavefront. This figure-of-merit has limited use, since real stars are neither monochromatic nor point-sources, nor can they be aligned perfectly with the optical axis. We list it in Table 13 to emphasize how many design options we have that meet the minimum requirement of Theoretical Maximum Extinction better than $10^{-10}$.

**Inner Working Angle (IWA):** The closest a planet can be to a star and still be conveniently detectable using a particular Diffracted Light Suppression System. Rather than trying to define "convenience", Table 13 quotes the planetary throughput of each device at the quoted IWA.

**Throughput at Inner Working Angle:** The detectable fraction of the flux from a planet located at the inner working angle. An ordinary telescope with no coronograph is considered to have unity throughput.

**Throughput Outside Inner Working Angle:** The detectable fraction of the flux from a planet located in the search area far from the inner working angle. DLSS designs that use pupil-plane masks generally have lower-than-average throughput (and lower-than-average sensitivity to low-order aberrations, to compensate)

**Outer Working Angle:** The farthest a planet can be from a star and still be conveniently detectable. The outer working angle of most DLSSs is limited by classical optical design constraints. In this case, the outer working angle is "large". In a few cases, the DLSS concept itself imposes a smaller limit to the outer working angle.

**Achromaticity:** DLSSs need to be able to work across a broad band pass. Some designs aim to achieve broadband diffracted light control using special materials with intrinsic chromatic properties (e.g. coronographs using phase masks). Such systems may have advantages in throughput and IWA, but potentially face manufacturing, tolerancing, and integration time problems.

**Planet PSF FWHM:** The Full Width at Half Maximum (FWHM) of the Point Spread Function (PSF) of the image of the planet. Concentrating the planet flux in a small number of pixels improves SNR by reducing detector noise contributions and the acceptance angle for stellar and exozodiacal backgrounds. In some designs, the PSF shape varies with the location of the planet. But usually the core of the planet's PSF has a reasonably uniform width for planets outside the IWA.

Although to first order the shape of the image is not crucial, any extension of the image decreases the contrast between planet images and residual host flux, making both detection and spectral analysis harder. The effective contrast ratio achieved is reduced by the number of times the image size exceeds that of the diffraction-limited image. This parameter also indicates the sensitivity of the coronograph to exozodiacal light; when the image quality is poor, the coronograph detects relatively more flux in each pixel from this extended source.

**Search Space / Useable FOV:** The focal plane region in which diffracted and scattered starlight are well-suppressed may be restricted in azimuth angle (roll orientation around the central star). The numbers in this row indicate roughly how much of the area in the image plane between the IWA and OWA is useable for planet hunting.

**Number of Telescope Roll Positions Needed:** When the Search Space is not 100%, rolling the telescope (or perhaps just the mask) can provide access to the blocked part of the image plane. This row suggests how many roll angles need to be sampled in order to search the whole annulus between the IWA and the OWA.

**Double / Field Star Compatibility:** Some DLSSs can easily be adapted to block the light from two stars at once – usually at the cost of some search space.

**Sensitivity to Pointing / Stellar Size:** The ability of the DLSS to suppress starlight given pointing errors and stars with large angular diameter. When possible, we have indicated the order of the null (see Section 6.2.1). Second order nulls are probably inadequate for terrestrial planet finding, because of extremely tight pointing requirements. Fourth order nulls still present a pointing challenge. Devices that create nulls of $8^{th}$ order and higher are highly robust. Relatively little is usually known about sensitivity to other low order aberrations: see discussion in Section 6.2.2.

**Photometric Efficiency:** The product of the numbers in the *Throughput Outside Inner Working Angle* row and the Search Space / useable FOV row. For the planet search mode, this quantity summarizes the overall search efficiency.

**Sensitivity to Telescope Reflectivity / Transmission:** How robust is the DLSS to moderate (~ 0.1%) amplitude errors across the pupil?

**Sensitivity to Red Leak:** When a DLSS is designed for some reference wavelength, often its suppression degrades sharply for longer wavelengths. Starlight at these longer wavelengths must be removed by a filter. This row flags DLSSs with this potential drawback.

**Telescope Pupil Shape:** Some DLSSs are only compatible with special telescope primary mirror shapes. This row flags those DLSSs.

**Compatibility with Segmented / Diluted Pupil:** Can the DLSS, or a modified version of it, work with a segmented primary mirror? This compatibility issue requires more research, as we discuss below.

**Compatibility with On-Axis Telescope Design:** For an on-axis telescope, the secondary mirror and its support structure block some of the primary mirror in an on-axis telescope. Can the DLSS, or a modified version of it, work effectively with such obscurations?

Besides these DLSS features, Table 13 compares some characteristics of DLSS technology pertinent to each design. These technology issues do not follow tradeoffs set by the principles of optics. On the other hand, we can probably trade time and money for progress in these areas!

### 6.1.2. Fabrication Issues

**Masks and Stops:** What are the manufacturing tolerances?

**Large Optics:** Some designs require a single monolithic optical surface (face sheet) at each mirror, including the primary mirror. What is the largest single optic in the system, and is it feasible to produce such an optic?

**Aspheric Mirrors:** Is there a need for non-standard figured optics (aspherics) that need to be polished to a high accuracy?

**Beam Splitters:** Some designs rely on beam splitters, which must be manufactured precisely to achieve spatial uniformity and achromaticity, and to avoid multiple reflections.

### 6.1.3. Technological Maturity

**Simulation Maturity:** How well has the basic concept been simulated for the goals of a TEC? Have the major technical concerns been represented with adequate fidelity?

**Subsystems Lab Demonstration:** Have the critical subsystems of the DLSS been demonstrated in the lab?

**Integrated Lab Demonstration:** Has a fully working DLSS been tested in the lab? On the Sky Demonstration: Has the DLSS been used for ground-based observing?

## *6.2. Trade-Offs and Tolerances*

### 6.2.1. Two Fundamental Trade-Offs

A bird's-eye view of DLSS design reveals that two primary tradeoffs shape the design landscape.

The first fundamental tradeoff appears to be between coronographs that strictly use masks and stops, and those that require more specialized hardware, like shaped mirrors and beam combiners. In a sense, there is only one mask/stop coronograph, since masks and stops could be stored on filter wheels and swapped in and out as needed. This flexibility may be an advantage in a real space telescope; masks optimized for spectroscopy can be alternated with masks optimized for searching, and optics destroyed by cosmic rays can be exchanged for fresh ones.

However, alternative designs typically offer other advantages. Practical masks come in three varieties: binary amplitude masks, graded amplitude masks, and binary phase masks. Graded phase masks are conceivable, but they are not represented in the lists presented here. The challenge of making continuously variable phase masks limits the IWA of mask-stop coronographs; alternative designs can generally offer superior IWAs given the same size primary mirror. Since requiring large primary mirrors rapidly drives up the cost of a space telescope, achieving a small IWA through DLSS and WFSC is of critical importance.

A second fundamental tradeoff is between IWA, search area, throughput, and sensitivity to low-order aberrations. The tradeoff among these properties appears to be a consequence of Fourier math, and is loosely related to the Uncertainty Principle; localizing wave power comes at a cost. Some coronographs also sacrifice some image fidelity to achieve better IWA or otherwise score better in the above tradeoff.

The relationship between IWA and sensitivity to low-order aberrations is especially crucial. Pointing and focus errors are likely to be large since they are related to the position of the secondary mirror – e.g., the HST has roughly 1/20 of a wave of pointing error and 1/50 of a wave of focus, long-term averaged. The smaller the IWA, the more sensitive a coronograph is to these low-order aberrations.

In general a DLSS has a power-law sensitivity to each aberration. The sensitivity to pointing error could be described by a "null order" by analogy to nulling interferometers. For a very tiny angular offset $\theta$ from the point of greatest suppression, the transmitted flux typically varies as $\theta^n$ for some integer n. We call n the null order. Designs with prolate-spheroidal pupil plane apodization, because they can work with binary field stops, theoretically have infinite null order; but practically, the null order is finite because of red leak. Terrestrial planet finding requires a null order of at least 4. Equivalent orders can probably be described for other aberrations, though little work has yet been done in this area. Polarization effects can also be considered low-order aberrations. Since they are so important, we discuss low-order aberrations in more detail in the next paragraph.

### 6.2.2. Low Order Aberrations

As we have mentioned, low-order aberrations throughout the optical system have a special, and often unappreciated, role in coronograph design, because they are linked to rigid-body motion of the telescope mirrors, and because they directly impact the IWA. Although it is impossible to avoid sensitivity to mid frequency amplitude or phase errors (which give the same effect on contrast for all configurations), we can mitigate the effects of some low-order aberrations by our choice of DLSS, since these aberrations produce speckles that fall relatively near the image of the star in the final image plane.

These aberrations are:

1) Tip/tilt: the sensitivity of a coronograph to tip/tilt is directly related to the inverse of the IWA. A $\theta^4$ coronograph has a wider null depth than a $\theta^2$ but the planet intensity will be much reduced if the angular separation is about $\lambda/D$. In addition, the effect of tip/tilt or stellar radius is very localized in the center of the image (in contrast to nulling interferometry) and may not impact that much on the companion detectivity if the planet is angularly separated by more than 3 $\lambda/D$. The sensi-tivity of coronagraphs to tip-tilt should be studied in terms of null width but also regarding the planet peak transmission. Analytical calculations are not able to yield such comparison and numerical simulations are needed at this stage.

2) Focus (z), astigmatism (x,y,tx,ty), coma (x,y,tx,ty), and spherical (z) aberrations are all produced as a linear function of the error when the secondary mirror moves in despace (z), decenter (x,y), or tip/tilt (tx,ty). Other (higher order) aberrations are produced but generally with much lower amplitude and often quadratically or a higher power of the error.

For reference, the HST secondary mirror tolerances were 2 microns in z, 10 microns in x and y, and 2 arcsec tilt. HST failed to meet its z stability requirement and has a 5 micron variable focus error sometimes called breathing. Typical tolerances for TEC are 100 times tighter (if it is assumed to work at 3 $\lambda/D$). By relaxing the IWA to 4 $\lambda/D$, the tolerances become an order of magnitude easier but still very challenging.

Numerical simulations of low order aberrations have already been conducted for Lyot, AIC, and 4QPM coronagraphs, but the results have not been published yet.

For an on-axis telescope, the phase difference between the *s* and *p* polarizations add an unavoidable astigmatism ($R^2 \cos 2\theta$) term for at least one polarization direction at the focal plane arising equally from the fast primary and secondary mirrors. Off-axis systems generate a corresponding slice out of such a function in the pupil. An unavoidable amplitude apodization with a similar shape is also generated by the angle- and polarization-dependent reflectivities of real mirror coatings. These effects add to the low-order aberrations that will ultimately set the IWA of a planet-finding coronagraph.

### 6.2.3. DLSS Construction and Tolerances

The performance requirements of the DLSS are set by the limitations of the WFSC. This trade space has scarcely been explored. Most analyses treat DLSS and WFSC as decoupled problems. We expect this synergy to be a fruitful area of research that may shortcut the current struggle to construct precise masks and stops, for example. We are forced for now to discuss mask tolerance requirements without the benefit of a deep understanding of WFSC/DLSS synergy.

First of all, it is impossible to produce a pure amplitude mask or pure phase mask, or a curved mirror with no polarization effects. Changing the amplitude of a beam inevitably changes the phase, and changing the phase of a beam inevitably changes the amplitude, since the index of refraction of real materials always has some real part and some complex part. Even binary masks are impossible to construct perfectly. The opaque regions of the mask must be constructed of materials with finite resistivity and thickness, so they act as wave guides.

These "real-world effects" demand investigation using full vector scattering theory. All DLSSs described here have been designed using geometric and simple wave optics, usually Fraunhofer diffraction theory alone. *A key open question* is whether the numerical and analytic optimization techniques that led to the DLSS designs discussed here can also produce designs that work in the presence of vector propagation effects.

Though these vector effects remain largely unstudied, we can recite some general rules of thumb about the accuracies required of coronagraph optics – as derived using Fraunhofer theory. For example, image plane optics are generally smaller than pupil-plane optics, and they interact with focused light, so image plane optics must be accurate at smaller size scales than pupil plane optics. Errors in the locations of the edges of binary image masks at the level of roughly λf/3000 (where f is the focal ratio of the beam) in the illuminated region of the mask produce speckles in the final image plane comparable in brightness to the image of an extrasolar Earth (Kuchner and Spergel 2003). For λ = 0.5 micron and f = 60, this length translates to 10 nanometers. Pupil plane masks, on the other hand, would require cutting errors of D/3000, where D is the mask diameter, to produce the same level of speckles. For a beam of diameter 20 cm, this tolerance corresponds to 70 microns. A WFSC system can correct for speckles produced by masks, but we don't presently know the degree to which they can be corrected.

If the masks, stops, mirrors, and beam combiners in a DLSS were active, they could conceivably be adjusted to compensate for construction errors or other flaws in the optical system after the telescope is in space. For example, the slotted masks described by Vanderbei et al. (2004) could be built with active slots of controllable widths. Presumably, the optical elements of a DLSS will degrade to some degree during flight as they are hit by cosmic rays, for example. Such degradation might render active adjustment necessary.

Our final choice of DLSSs will be guided by factors we do not yet know. If we find low order aberrations impossible to control, we will aim for an eighth-order image mask or an apodized pupil. If we find bright exozodiacal clouds around all of our target stars, we will need a design without such apodization. If one system turns out to be especially robust to cosmic ray bombardment, or easy to make actively controllable, or naturally amenable to WFSC synergy, these factors will also influence our decision. Assessing all the relevant factors may require integrated observation scheduling models, like those being developed by Doug Lisman at JPL, which attempt to evaluate the total time to execute a planet search given a particular list of targets, a DLSS design and some assumptions about telescope performance.

It is very likely that any coronographs will not provide directly a $10^{-10}$ contrast at close angular distance; first, because coronographs are not perfect, and second, because the optical quality to achieve the TEC goal has not yet been demonstrated for large mirrors. However, the TEC goal could still be achievable even if the coronograph provides a lower contrast. Active wavefront compensation has been attempted at JPL and a reasonable gain of 1 or 2 orders of magnitudes could be expected. But above all, calibration is definitely the keystone of such a mission. There are a variety of calibration techniques that are expected to reduce the variance of the speckled halo. But the instrument should be conceived to allow such capabilities. For instance, speckles can be calibrated in real-time using simultaneous dual-imaging of 2 spectral bands or 2 polarization states. These kinds of technique could be applied on ground-based telescopes in a few years (NICI at Gemini South and Planet Finder at the VLT). Finally, the information needed to correct the wavefront is already contained in the coronographic image. A perfect coronograph removes most of the diffracted light from the on-axis object, and hence the residual light distribution is directly related to the phase aberrations which can therefore be reconstructed (with a smart algorithm) and applied to the deformable mirror. A comprehensive approach based on numerical simulations is now needed to study different instrumental concepts. The coronograph itself is not really the most critical component since the best that a coronograph can do is to remove the coherent part of the wavefront. The residual wavefront bumpiness is then the actual limitation for detecting terrestrial planets. Numerical simulations and laboratory experiments will allow us to identify the best instrumental concepts to achieve the TEC goal.

### 6.2.4. Open Problems

We reiterate here the key open problems in DLSS design and construction.

Addressing the following critical problems probably requires concerted activity by teams of scientists and engineers, and facilities available at major research laboratories.

- Understanding and optimizing DLSS/WFSC synergy: how accurate does a DLSS really need to be? What trade-off is there between DLSS and WFSC?

- How do we build masks and stops and shaped mirrors and beam combiners to the required accuracy?

- Can we optimize DLSS designs taking into account the vector (electromagnetic wave) character of light?

- How do masks, stops, mirrors, and beam combiners degrade in space?

- Can we build actively controllable DLSSs?

The following important problems can probably be tackled by individuals, perhaps even graduate students:

- Optimizing DLSSs for mirrors with many hexagonal segments. More and more telescopes use hexagonal segments – on the ground and in space.

- Analysis of DLSS sensitivity to low order aberrations beyond pointing: focus, astigmatism, and polarization effects.

- Comparing DLSS rejection of exozodiacal light.

Some gaps in our mathematical understanding of the Fraunhofer diffracted light problem remain – though filling in these gaps may be purely of academic use:

- Understanding and mapping the range of hybrid mask/stop designs using apodization in both image and pupil planes.

- Understanding and mapping the range of non-separable 2-D mask/stop designs.

## *6.3. Wavefront Sensing and Control for Coronographic Imaging*

Wavefront sensing and control (WFSC) is the central enabling technology for a coronographic telescope. First and foremost, WFSC detects and corrects the unavoidable amplitude and phase aberrations present in a telescope system. Second, WFSC has the potential to desensitize a coronograph design to errors in manufacturing, modeling, and implementation. In many cases WFSC may relieve the requirements upon a system model to demonstrate absolutely accuracy in its predictions. WFSC technologies present a unique opportunity to co-optimize a coronograph, control system, and telescope design to best enable high contrast imaging.

### 6.3.1. Sources of Error in a Coronographic Telescope

While WFSC technologies are central to achieving the goal of terrestrial planet detection and characterization, they cannot compensate for every error source that can be expected. In this section, we discuss some of the key error sources that can be anticipated in a coronograph space telescope.

### 6.3.1.1. Mirror Alignment, Figure and Surface Errors

Imperfection in the shaping and polishing of optics along with errors in the alignment of these telescope elements lead to optical path difference (OPD) errors evolving across the stellar wavefront as it propagates though the system. Because OPD errors can occur at any plane (not just at a pupil) along the optical train, they can induce a complex error. Left uncompensated, these errors translate into focal plane speckles that are many times brighter than any potential planet we hope to observe. By spatially adjusting the OPD across the pupil (by use of one or more DMs), such errors can largely be corrected. However, careful models of thermal and mechanical dynamics of any telescope design are necessary to understand the nature of the WFSC temporal bandwidth requirements.

### 6.3.1.2. Non-Uniformity of Optical Reflectivity / Transmissivity

Optical coatings on the optics have limitations. The variation in reflectivity across or transmission through an optical element induces wavefront amplitude variation that ultimately results in potentially bright speckles in the focal plane. Unlike OPD-induced speckles, amplitude errors create speckles that have magnitudes that do not scale with wavelength. While phase control can compensate for amplitude errors (see next section), the compensation will degrade as the optical band pass grows.

### 6.3.1.3. Mask Fabrication and Design Errors

Most coronograph concepts that have been developed rely on placing masks at pupil and focal planes within the telescope. The implicit physical assumptions and limited numerical accuracy of methods used to design these masks can yield sub-optimal designs. The masks under consideration generally employ binary phase variation, continuous amplitude apodization, or binary amplitude apodization. The extent to which any approach can successfully suppress or shape the diffraction process depends on the quality of its fabrications and design. The net wavefront errors that are induced by imperfections, however, are coherent to the optical system. As such, the employed WFSC technologies have the potential to compensate for such implementation limitations. Nevertheless, the locality of these errors (pupil plane versus focal plane) versus the locality of the WFSC influence may impose optical bandwidth constraints.

### 6.3.1.4. Mask-Induced Phase and Wave Guide Effects

Feature size and the properties of the material employed in coronographic masks cannot be ignored. For instance, there can be wave guide effects resulting from the interplay between the feature size in binary masks and the angle of incidence of the stellar light. These effects induce wavefront phase and amplitude variations. Furthermore, the variations will no doubt have a $\lambda$-dependence but not one that will scale the way that OPD errors usually do. Thus it may be very difficult for the WFSC system to produce a correction that can be useful for broadband imaging. Appropriate vector diffraction simulations must guide a careful selection of materials and mask design parameters. Otherwise, draconian optical bandwidth constraints during operations may have to be imposed.

### 6.3.1.5. Pointing Error

Pointing error refers the alignment of the optical axis of the coronograph to that of the star of interest. This error manifests itself as a phase-tilt across the coronograph entrance pupil. Because there will be dynamic pointing disturbances to the telescope through its reaction wheels and thermal variations, methodologies for maintaining this fine alignment are essential. The dynamic nature of pointing error generally calls for an ancillary WFSC sys-tem that traditionally uses a fast steering mirror (FSM). Because coronographs have very high aberration sensitivity, the optimal placement of a FSM in the optical train requires careful study.

### 6.3.1.6. Low-Order Dynamical Aberrations

Aside from line of sight jitter, system disturbances can lead to dynamic misalignments of the secondary mirror as well as to deformations of the primary mirror. These cause time evolving phase aberrations, which produce additional speckles near the IWA that increase quickly with time. Depending upon the WFSC strategies, the telescope may be designed to have sufficiently long time constants to avoid these issues. But it may be necessary to consider more complex ancillary WFSC systems to stabilize the key optical elements.

### 6.3.1.7. Finite Stellar Diameter

The finite size of the star presents a distribution of tilted incoherent wavefronts to the telescope. Depending upon the coronograph method and design, there may be a restrictive sensitivity to this simultaneous ensemble of wavefronts. Because of the incoherent nature of the wavefront ensemble, the WFSC cannot directly compensate for this effect.

### 6.3.1.8. Incoherent Scattering and Stray Light

Non-optical surfaces that are illuminated in the optical train will produce some level of incoherent scattering and stray-light. The lack of coherence of this light with the nominal system wavefront defies its being corrected by the WFSC system. Careful stray light analysis needs to guide the telescope baffling design. But it may be critical to employ tight contamination control procedures to guarantee that the background light does not excessively limit the achievable imaging contrast.

### 6.3.1.9. Induced and Cross-Polarization

The finite index of refraction at every optical surface induces amplitude and phase variations across the wavefronts from the *s* and *p* polarization states. Even if one were to use a perfect polarizer in the back-end of the telescope, there is a cross polarization term that represents an independent wavefront component that originated from the crossed state. After the polarizer, there are two incoherent wavefronts that may have significant phase and amplitude differences. A single WFSC system cannot compensate both simultaneously but merely find the best control state compromise. To best avoid these issues, the f-number of the primary may have to be made large, and on-axis designs may have to be considered. However, these solutions have their drawbacks due to stability and diffraction effects.

## 6.3.2. Wavefront Sensing and Control Implementations

A variety of methods are in use or have been proposed for controlling the wavefront in a coronographic telescope. By far the most experience with wavefront control has been for adaptive optics on ground-based telescope. All (or almost all) of these systems utilize a pupil plane sensor at the front end of the telescope (usually a Shack-Hartmann sensor) to reconstruct an estimate of the wavefront phase. This information is then used to adjust a deformable mirror, also at the front end, to correct the wavefront. While we can certainly benefit from the knowledge and experience of ground-based AO, it is generally agreed that such a front-end system is inadequate for planet finding. Foremost among the problems is the existence of non-common path errors in the sensing leg of the instrument.

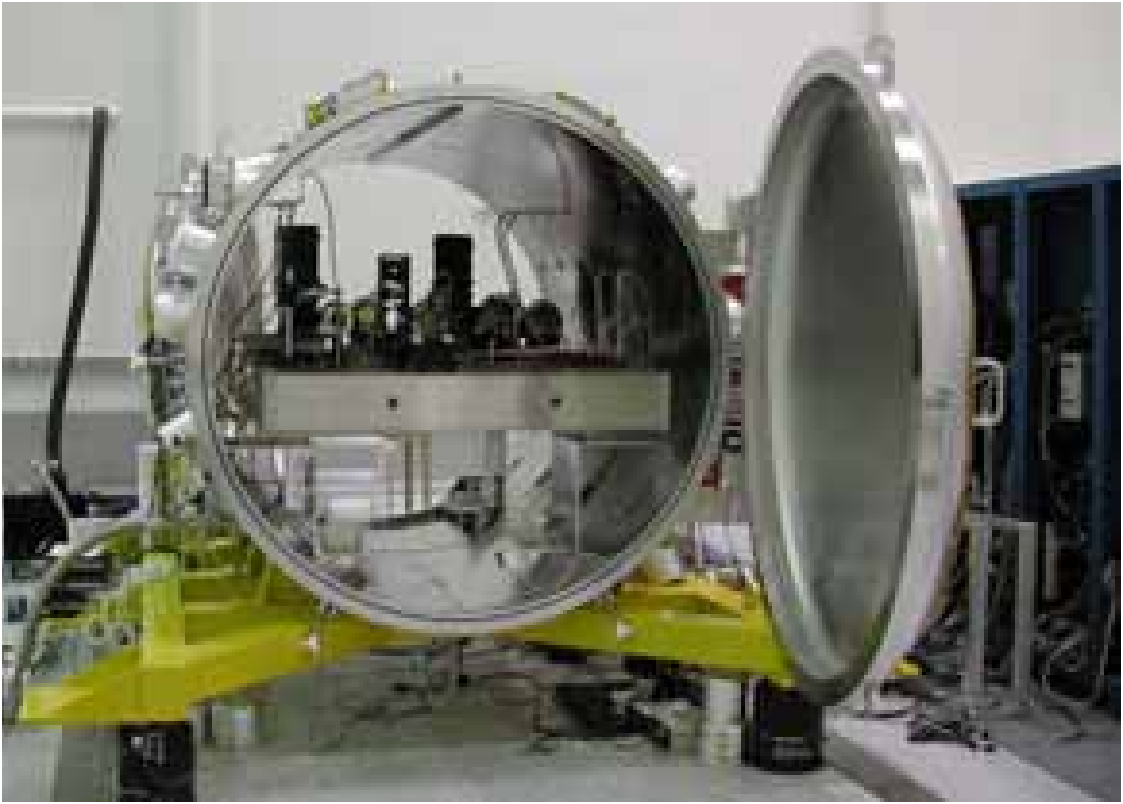

**Figure 5: The high contrast imaging testbed is shown within its vacuum chamber at the Jet Propulsion Laboratory in Pasadena, California.**

These errors can produce uncorrectable speckle far larger than the sought-after contrast of $10^{-10}$.

All WFSC approaches for a TEC must be common path; that is, sensing must occur in the same optical path as the science information (at least up until a light removal system). Nevertheless, there still remain quite a number of approaches to implementing wavefront sensing and control. In this section, we present several such systems, describe their salient characteristics, and highlight key limitations.

### 6.3.2.1. Direct Speckle Nulling

Speckle nulling refers to a closed loop system that removes speckle via a deformable mirror based only on measurements of the speckle in the image plane. By using a finite number of DM dithers, unambiguous information is acquired for speckle removal. No effort, however, is made to estimate the wavefront itself. If speckle is entirely due to phase errors in the wavefront, then a single DM can remove speckles in the focal plane to within the limitations of the DM. These limitations come from the actuator density (within the pupil), accuracy, dynamic range, and stability. For OPD errors that originate near pupil planes, the correction made by a DM is valid at any wavelength. However, if any speckle is produced by amplitude errors or by any error near a focal plane, then there are optical bandwidth limitations imposed by the correction. If there are amplitude errors in the system, a single DM can compensate by inducing a phase error that effectively cancels the amplitude speckle over half the controllable focal plane.

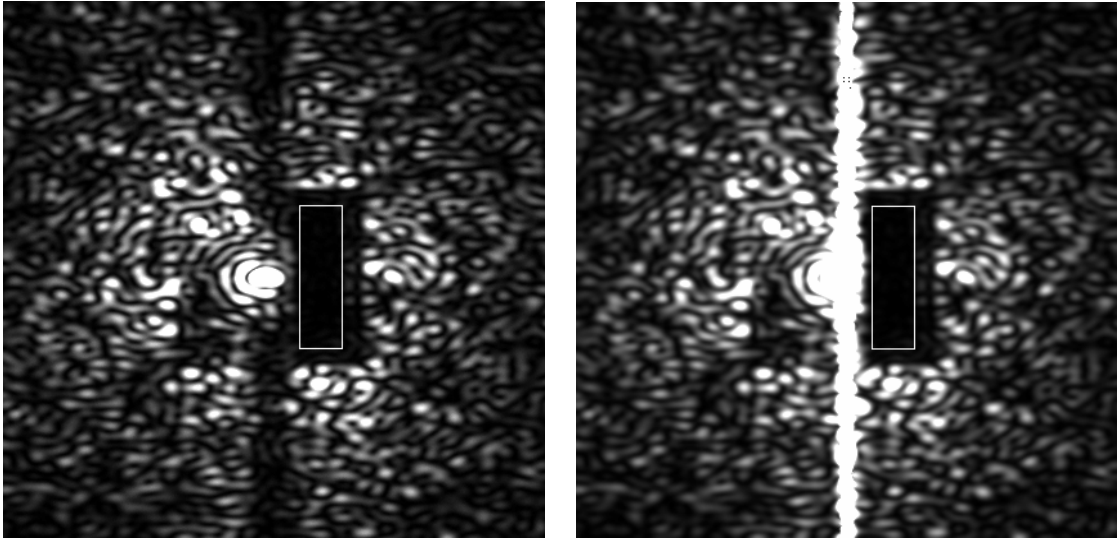

**Figure 6: Direct image of the scattered light at the focal plane of the HCIT coronograph, as obtained at JPL laboratory in January 2004 is shown on the left. The coronograph focal-plane occulting mask has a one-dimensional sinc-squared profile, the shadow of which is seen as the dark vertical feature passing through the center of the image. The contrast over the field is shown on the right. This is computed by dividing the coronographic PSF measurement by the occulter intensity transmission and then normalizing the result by the brightness of the central star (the flux that would be measured if the occulting mask were removed). In this experiment, the speckle nulling algorithm creates a "dark hole" to the right of the suppressed star image. Within the inscribed rectangle (leftmost side of which is at 4 $\lambda/D$), the average scattered speckle background level corresponds to an instrument contrast of $9 \times 10^{-9}$.**

A two DM system may be able to remove amplitude as well as phase everywhere in the search space, but will also impose an optical bandwidth limitation that scales with the magnitude of the amplitude errors. We must be prepared to accept the possibility that full control will require a separate DM for every optical surface whose defects can affect optical performance. There is a recognized need for an achromatic amplitude control system.

Using a single DM, speckle nulling has been successfully demonstrated on the TPF HCIT, shown in Figure 5. The high contrast achieved by this approach, depicted in Figure 6, shows the creation of a dark-hole over half the controllable focal plane. Currently, the depth of the dark hole (a contrast of $5 \times 10^{-8}$) is limited by the precision of the DM drive electronics.

### 6.3.2.2. Model-Based Correction

This category of systems is similar to ground-based AO in that corrections are made based on estimates of the wavefront phase and amplitude. There are two approaches to wavefront correction marked by where the sensing and estimation is accomplished. We note that in all of these approaches, single measurements are inadequate for unambiguously determining phase and amplitude as measurements are made of only intensity. Thus, multiple measurements with some form of diversity are required.

In the first type of system, which we call *Lyot Plane Nulling*, the wavefront estimate is built up from measurements in the Lyot plane of the coronograph (typically implemented by inserting a mirror or camera into the coronograph light path). By dithering the DM and taking multiple measurements, estimates can be made of the wavefront phase and amplitude. In the second type, which we refer to as *Image Plane Nulling*, estimates of the wavefront phase and amplitude are made using only measurements in the final image plane. In this case, diversity can be achieved through a variety of possible techniques, including focus, multiple pupils, multiple wavelengths, and DM dither.

In both of these approaches, a deformable mirror is used to make the final correction to the image plane. As before, a single DM can only correct phase achromatically in the entire search space, or both phase and amplitude in a smaller area of the search space (but at one wavelength). Two DMs can correct both phase and amplitude in the entire search space, but current methods only allow for narrow bandwidths of amplitude control. Again, technology development of achromatic amplitude control is desirable.

We also point out that this category of correction methods is model dependent as it relies on computational propagation of the wavefront to the image plane in order to compute the proper DM setting to achieve a dark hole. They also require an adequate model of the deformable mirror. This modeling requirement introduces a potential source of error not present in speckle nulling.

### 6.3.2.3. Subaperture Nulling

Subaperture nulling is a technique for wavefront correction using a single DM and a fiber bundle fed by lenslets. This technique only works in concert with a shearing/nulling coronograph. It is an example of coherent design of a coronograph and control system.

The technique splits the pupil into N ~ 1000 subapertures. Each subaperture's light is fed into a single mode fiber. A DM in one arm of the nulling interferometer is used to adjust amplitude/phase to totally null out the starlight. The output of the array of optical fibers is imaged to form a reconstituted pupil that the back-end optics can use to form an image of the planet.

Amplitude mismatch in the two arms is measured by using a shutter to block one arm of the interferometer. Intensity matching to 0.05% is needed for a leakage of $3\times10^{-11}$.

Phase matching is performed by dithering the DM in piston. Optical path must be matched to 0.1 nm for a null of $2.7\times10^{-10}$. The demonstrated lab results of nulling at visible wavelengths is $5\times10^{-6}$ steady state, which in a 1000 fiber system theoretically converts to $5\times10^{-9}$ starlight suppression (recall that Earth detection requires $1\times10^{-10}$).

### 6.3.2.4. Cascaded Coronographs

This technique is similar to speckle nulling in that it looks at the speckles in the output of the coronograph and attempts to remove them. However, instead of using a DM at the front to remove the speckles, in this scheme two DMs behind the first coronograph are used to shape the wavefront of the speckles so that at a subsequent pupil, the light hits a stop. The chromaticity issues associated with speckle nulling also apply to this technique.

### 6.3.2.5. Fixed Correctors

Systems in this family of WFSCs do not rely on deformable mirror and closed loop control, but rather remove speckle using a fixed optical arrangement designed for static or quasi-static speckle removal. Examples include holographic projection, and interferometric removal. The corrector fitted to HST in orbit is a good example of such a system.

Fixed correctors fall into two categories. In its simplest form, one mirror is used as a phase corrector for all the other optics in the system. The major challenge is the precise measurement of the wavefront of all the other mirrors in the system (and their alignment). In addition, optical wavefronts may change at the 0.1 nm level between lab measurements and on orbit deployment. Additionally, the compensation of amplitude errors must still be addressed.

A second type of phase corrector is a hologram. Methods have been invented to achieve quasi-achromatic operation of holographic correction (by using a $2^{nd}$ hologram to un-disperse the effects of the $1^{st}$ hologram). In theory both amplitude and phase can be corrected by a hologram. However current holographic materials used to make holograms have significant non-linearities that are a 2 to 3 orders of magnitude from the degree of amplitude control needed for $1\times10^{-10}$ suppression of starlight. (And even if amplitude control is not needed, the hologram may introduce significant amplitude errors.)

### 6.3.3. Performance Limitations

Each of these control approaches has certain categories of limitations. Foremost among the issues is *chromaticity*. Ideally, we like to see the WFSC system be able to correct both amplitude and phase at all wavelengths in the desired science band and in the entire coronograph search space. Unfortunately, that has not yet been achieved. For instance, the holographic approach is inherently monochromatic. In most cases, the severest chromaticity limitation occurs in the amplitude control technique. Research is essential on achromatic amplitude control devices that can be integrated with the above systems.

As currently conceived, the ultimate performance of many of the systems is determined by the capabilities of the deformable mirrors. These include dynamic range, accuracy, stability, and actuator density.

The temporal bandwidth of the systems also presents unique challenges. By temporal bandwidth we mean both the rate of response of the closed loop system, and whether the controller can be used in real time during an observation. All of the approaches above, save the subaperture nuller, are quasi-static. Corrections are made over a longer time and stability is relied upon during an observation. The implicit requirement of the quasi-static assumptions is the need for a very stable telescope. Mechanical vibrations and thermal variations must be controlled to very strict tolerances. This is an area where accurate and validated predictive models must be developed.

As alluded to above, for systems that rely on modeling of the DM or propagation, certain types of errors will be introduced. Only experiments will fully characterize the extent of these errors.

Finally, only certain categories of the errors described above may be addressable by any given wavefront control system. For instance, while sensitivity to finite stellar size is a critical characteristic of a coronograph, the resulting error is uncontrollable by any of the wavefront control systems due to the incoherence of the arriving wavefronts. In Table 4 we present the salient features, performance characteristics, and limitations of each of the above control systems. We also indicate where there may be an incompatibility between a given WFSC system and a particular coronograph approach.

**Table 4: Overview of wavefront sensing and control techniques**

| WFSC Techniques | Speckle Nulling | | | Lyot-Plane WFC | | | FP Phase Diversity | | | FP pupil Diversity | | | Sub-Aperture Nulled Pupil | | | Fixed Correctors | | |
|---|---|---|---|---|---|---|---|---|---|---|---|---|---|---|---|---|---|---|
| | only phase | croma amp | acrom amp | only phase | croma amp | acrom amp | only phase | croma amp | acrom amp | only phase | croma amp | acrom amp | only phase | croma amp | acrom amp | only phase | croma amp | acrom amp |
| **WFSC Constraints and Capabilities** | | | | | | | | | | | | | | | | | | |
| Required Spatial Bandwidth | | | | | | | TBD | TBD | TBD | TBD | TBD | TBD | | | | | | |
| Temporal Bandwidth | | | | | | | TBD | TBD | TBD | TBD | TBD | TBD | | | | | | |
| Null Depth Capability | | | | | | | TBD | TBD | TBD | TBD | TBD | TBD | | | | | | |
| Dynamic Range | | | | | | | TBD | TBD | TBD | TBD | TBD | TBD | | | | | | |
| All Common Path | | | | | | | TBD | TBD | | | TBD | TBD | | | | | | |
| Operational Complexity | | | | | | | | | | | | | | | | | | |
| Imposed Optical Bandwidth Limitations | 5~20% | | | 5~20% | | | | | | | | | 5~20% | | | | | |
| **WFSC Ability to correct:** | | | | | | | | | | | | | | | | | | |
| *Wavefront Errors* | | | | | | | | | | | | | | | | | | |
| Low Order Phase Aberrations | | | | | | | | | | | | | | | | | | |
| High Order Phase Aberrations | | | | | | | | | | | | | | | | | | |
| Mid-spatial Frequency Phase Aberrations | | | | | | | | | | | | | | | | | | |
| Amplitude Errors | | | | | | | TBD | TBD | | TBD | TBD | | | | | | | |
| *Coronagraph Mask & Stop Errors* | | | | | | | | | | | | | | | | | | |
| Mask Transmission Errors | | | | | | | TBD | TBD | | TBD | TBD | | N/A | N/A | N/A | N/A | N/A | N/A |
| Mask Shape Errors | | | | | | | TBD | TBD | | TBD | TBD | | N/A | N/A | N/A | N/A | N/A | N/A |
| Mask Waveguide Effect | | | | | | | TBD | TBD | | TBD | TBD | | N/A | N/A | N/A | N/A | N/A | N/A |
| Mask Phase Errors | | | | | | | TBD | TBD | | TBD | TBD | | N/A | N/A | N/A | N/A | N/A | N/A |
| Coronagraph Modeling Errors | | | | | | | TBD | TBD | | TBD | TBD | | N/A | N/A | N/A | N/A | N/A | N/A |
| *System Errors* | | | | | | | | | | | | | | | | | | |
| Polarization Effects | | | | | | | | | | | | | | | | | | |
| Pointing Errors | | | | | | | | | | | | | | | | | | |
| **Coronagraph Architecture Applicability:** | | | | | | | | | | | | | | | | | | |
| Interferometric Nulling | | | | | | | | | | | | | | | | | | |
| Diffraction Suppression | | | | | | | | | | | | | | | | | | |
| Diffraction Shaping | | | | | | | | | | | | | | | | | | |

## 6.3.4. Key Technology Issues

### 6.3.4.1. Wavefront Control

- Low power, high accuracy, high actuator density, highly stable deformable mirrors
- Achromatic ($\Delta\lambda/\lambda > 20\%$) amplitude apodization mechanism and schemes
- Optimal arrangement of controllable surfaces for high contrast imaging

### 6.3.4.2. Wavefront Sensing

- Photon efficient algorithms and analysis methods
- Real-time implementations
- Ancillary WFSC concepts for dynamic error control

### 6.3.4.3. Telescope Architecture

- Minimal telescope configurations that optimally incorporate controllable surfaces
- Designs cross-optimized with coronagraphs and their related aberration sensitivities
- Designs that minimize stability and polarization issues simultaneously

### 6.3.4.4. Coronograph Architecture

- Designs that are co-optimized with the presence of a control system
- Designs that have high efficiency with respect to the planet light
- Designs that have low sensitivity to telescope dynamical (temporally uncontrollable) aberrations

## *6.4. Detectors and Spectrographs*

### 6.4.1. Status

Optical spectra of terrestrial planets – both observations and computations – are described in *Biosignatures and Planetary Properties to be Investigated by the TPF Mission* (the Des Marais report). This report is available on the web at (http://planetquest.jpl.nasa.gov/TPF/TPF_Biomrkr_REV_3_02.pdf).

The requirements for spectroscopic requirements are being refined. See the Version 5.1 of the *Science Requirements Document* available from Wesley Traub (wtraub@cfa.harvard.edu).

The performance specifications implied by the Science Requirements Document are described in the *TPF-C Handbook* available from Karl Stapelfeldt (karl.r.stapelfeldt@jpl.nasa.gov).

Observing strategies are starting to be considered.

### 6.4.2. Open Issues

The pixel count-rates for spectroscopy are so low that all sources of background must be understood and, where possible, suppressed. The main areas of concern are astronomical background (zodiacal and exo-zodiacal light), speckle noise, detector noise (read noise and dark noise), and jitter-induced background. Extensive modeling and simulations are needed to understand the major sources of noise and to devise strategies for their suppression in the instrument and minimization in post-observation data analysis. See Section 8 for further details.

### 6.4.3. Potential Areas for Collaboration

The Darwin/TPF community would benefit from a close collaboration of scientists and engineers in Europe and the U.S. in two areas:

#### 6.4.3.1. Integral Field Unit (IFU)

It is expected that the optical spectrograph will be some form of an IFU because of its potential use for wavefront sensing and its advantages over conventional slit spectrographs as listed in Table 5.

Table 5: Comparison of integral-field units with slit spectrographs

| Property | Slit | IFU |
| --- | --- | --- |
| Transmission (using prism) | ~90% | ~80% |
| Telescope roll alignment for point source | needed | not needed |
| Alignment (slit to star) | difficult | easy |
| Multiple planets imaged in single exposure | no | yes |
| Fraction of FOV sampled | ~1% | 100% |

There has been a good deal of work on IFUs in the U.S. and Europe, e.g. in Lyon (Bacon et al.), Edinburgh (Todd et al.), Garching (Genzel et al.), and Padova (Gratton et al.), but most applications to date have been to ground-based telescopes. A detailed trade-off study of IFU approaches for TEC is needed. Not only should various hardware implementations be considered, but also methods of extracting the planet spectrum from the residual speckle background.

We recommend that a Darwin/TPF working group for spectroscopy be formed with membership including scientists who have extensive experience in the development and use of IFUs including data processing. One objective of the collaboration would be the design, fabrication, and testing of a "proof-of-concept" IFU.

### 6.4.3.2. Advanced Detectors

The detector requirements for TEC imagery and spectroscopy are listed in the table below. See Appendix 11.7 for an assessment of current optical imaging detectors.

**Table 6: Detector requirements**

| Property | Imagery | Spectroscopy |
| --- | --- | --- |
| Detector format | 1k × 1k | 2k × 2k |
| Wavelength range | 0.4 μm – 1.0 μm | 0.4 μm – 1.0 μm |
| Readout noise | ~2 $e^-$ rms | <0.01 $e^-$ rms |
| Quantum eff. | >80% | >80% |
| Dark current | <$10^{-3}$ $e^-$ $sec^{-1}$ $pixel^{-1}$ | <$10^{-4}$ $e^-$ $sec^{-1}$ $pixel^{-1}$ |
| Environment § | radiation tolerant | radiation tolerant |
| Read-out §§ | CCD-type, or non-destructive | non-destructive |

§ Barth et al. (JWST report 2002)

§§ Non-destructive readout desired to avoid contamination of a long exposure (e.g. 20 minutes) by stellar leakage during a short (e.g. 1 sec) pointing instability ("jitter").

Such detectors do not exist today. However, there are several European laboratories such as MPI Halbleiterlabor (Munich) or E2V Technologies (Chelmsford, UK) that are working to develop advanced detectors. We propose that a Darwin/TPF working group on detectors be formed to refine the detector requirements and to work with European laboratories in developing suitable detectors for exoplanet spectroscopy.

# 7. Telescopes for Coronographs

## *7.1. Introduction*

### 7.1.1. Telescope – Mask System Approach

Coronographic telescopes, and their primary mirrors, have characteristics that differ from typical astronomical telescopes. The estimate of performance, and in some cases even viability, of a form of coronographic telescope requires consideration of specific matching focal plane and/or image plane masks. As such, the primary mirror, telescope metering structure, and mask set constitute a linked system.

The Strehl ratio does not sufficiently define wavefront performance for coronographic functions. The mirror optical wavefront requirement applies to all surface spatial wavelengths ($g$), and is described by a two-dimensional power spectral density (PSD). The planetary search inner and outer working angles define a critical domain of surface error spatial frequencies. These surface frequencies lie between approximately $g = 10$ m$^{-1}$ to $g = 0.3$ m$^{-1}$, and include the spatial domain often associated with the optical finishing artifacts of

(1)   Mirror substructure print through,
(2)   Aspheric polishing zonal error.

The deformable mirror (DM) may correct phase errors over much of this frequency domain. Nevertheless, the magnitude of phase errors, resulting from any shear of the DM with respect to the primary mirror, closely relate to the primary mirror surface error in this frequency band. The use of the word "wavefront" in this section refers to not just the terms captured by the first 36 Zernike coefficients, but rather the complex wavefront error, addressing phase and amplitude over all spatial frequencies.

While industrial teams conducted an extensive survey of telescope forms during the recent TPF Architecture Study (completed June 2002), the coronographic community's understanding of the telescope-mask system has evolved since this study. We note the progress in developing novel and powerful mask forms, in defining the science and operations, and the recognition of issues associated with polarization as examples. Therefore, it is appropriate to reevaluate the conclusions of the architecture studies in the context of current perspectives. For example, we should evaluate if mirrors, segmented into two or more sections and deployed, are more viable today than at the time of the TPF Architecture Study, since deployed mirrors may offer advantages in accomplishing the mission science requirements.

### 7.1.2. Figures of Merit

We consider observing efficiency, i.e., the time required to observe a planet at the required spatial and spectral resolution, with a signal-to-noise ratio of five, to be the primary figure of merit for the detection and characterization of extrasolar planets.

As discussed below, many system characteristics contribute to observing efficiency, including the telescope's aperture, collecting area, thermal stability, point spread function, and mirror coating reflectivity, as well as the coronograph's mask and filter transmission; the detector's quantum efficiency, dark current and read noise; the observatory's slew rate, settling time and pointing stability, and the coronograph's calibration requirements and setup time.

Other parameters of interest to the system designer include the launch vehicle's fairing size, lift capability and launch environment plus the observatory's size, mass and power requirements, and packaging and deployment approach.

### 7.1.3. European Base

There are many European companies that could fabricate optics for TEC and large ground based telescopes. For example Sagem Reosc (F) has had experience in the production of large monolithic mirrors including the 8 m primary mirrors for the VLT and Gemini, and 1.9 m hexagonal segments for the GranTeCan 10 m telescope in La Palma. Carl Zeiss (D) and Advance Mechanical Optical Systems Ltd (AMOS; B) have also produced 2-4 m diameter mirrors for ground-based telescopes.

For mirror materials, Schott has had a long involvement in the production of large low thermal expansion glass blanks, whilst Boostec (F) has developed a silicon carbide (SiC) mirror capability. The largest SiC mirror that has been produced to date is a 3.5 m diameter mirror with 6 micron surface accuracy for the Herschel space telescope. Boostec has also produced a 0.6 m mirror with 40 nm form accuracy for the Rosat II satellite. Another material that might be of interest is carbon fiber composite, and there are several European companies such as QinetiQ (GB) and Cobham Composites (GB) with experience in the production of this material for space systems and research into using the material for lightweight optics.

Several groups in Europe are developing computer controlled polishing techniques such as the UK-based Zeeko Ltd's "Precessions" polishing technology and Dutch research by Nederlandse Organisatie voor Toegepast Onderzoek (TNO) / Delft University into a fluid jet polishing technique.

There is ongoing European research into optical metrology techniques such as that being carried out by Heriott-Watt University, Edinburgh, in collaboration with the UK National Physical Laboratory and University College London. TNO is also involved in a picometer metrology test bed for the GAIA mission and the University of Neuchatel (CH) is researching metrology systems for the VLTI.

The development of MEMS deformable mirrors is being undertaken by several groups including the Laboratoire d'Electronique de Technologie de L'Information (LETI) in collaboration with the University of Grenoble; Marseille University; and in the Netherlands by TNO. There are also companies involved in other forms of deformable mirrors such as Oko Tech Ltd (NL) and BAe Systems (GB).

## *7.2. Telescope / Mirror Shapes and Factors*

### 7.2.1. Requirements

Ideally, a telescope designed for use with a coronograph will have a symmetric point spread function; be thermally and mechanically stable for a period much greater than the time of an observation (24 to 30 hours); have no obstructions to diffract the incident light; and have a primary mirror with a very smooth surface (< 5 Å rms), a well corrected figure, no edge roll-off, and the smallest possible edge length to surface area ratio (i.e., a monolithic circular aperture). It should also have as large an f-number as possible to minimize polarization effects, yet be compact and light enough to be launched with existing launch vehicles. The primary should also be large enough to place an Inner Working Angle of ~ 40 mas at ~ 4$\lambda$/D, and have enough collecting area to detect an Earth-like planet with a reasonable integration time (2 to 4 hours). Since all of these requirements cannot be met due to the limitations of the currently available launch vehicles, a number of design compromises must be made.

The following sections describe several of the telescope forms that are being considered for use with coronographs, the factors that must be considered while designing the telescope, and the key trades to be performed.

### 7.2.2. Forms

As shown in Figure 7, many forms have been suggested for the primary mirrors of telescopes designed for use with coronographs. Off-axis designs with monolithic mirrors appear to be ideal since there is no secondary mirror support structure to interfere with the incident light; one has to contend only with diffraction from the mirror's outer edge. However, monolithic mirror sizes are limited by the largest current launch vehicle fairings (5 meters in diameter) to ∼ 4 meters for a circular aperture, or ∼ 4 × 10 m for elliptical or rectangular apertures.

Unfortunately, the low density of stars in the solar neighborhood means the telescope for the TEC may nine-segment circular mirrors. Rectangular mirrors have the advantage that their edges require an aperture of 12 to 14 meters to meet the requirements for the full science mission. Thus the TPF project team is considering a number of segmented telescopes that can be folded up for launch and deployed on orbit, including two-segment rectangular mirrors, four-segment square mirrors, six-hexagonal segment mirrors, and diffract light preferentially in two orthogonal directions, and the nine "pie-shaped" segment mirror provides a circular aperture, while the single-ring six-hexagonal segment mirror can be expanded to a two-ring 18-segment mirror or three-ring 36-segment mirror that are launched with a single launch vehicle and assembled on orbit. Even larger hexagonal segment mirrors have been proposed for ground based telescopes, including the 90 segment mirror (Figure 8) and the OWL mirror with ~ 3000 segments! Hexagonal- and pie-segment mirror forms also have symmetric point-spread functions that eliminate the requirement to obtain images at multiple roll angles to search the image plane at maximum spatial resolution. In addition to these geometric forms, several less conventional forms have also been proposed, including the Phase Induced Amplitude Apodization (PIAA) telescope (Figure 9) that uses highly aspheric mirrors for "pupil mapping" and multiple telescope architectures (Figure 10) that utilize delay lines and beam combiners to coherently overlay the beams and null out the light from the star while preserving the light from the planet(s).

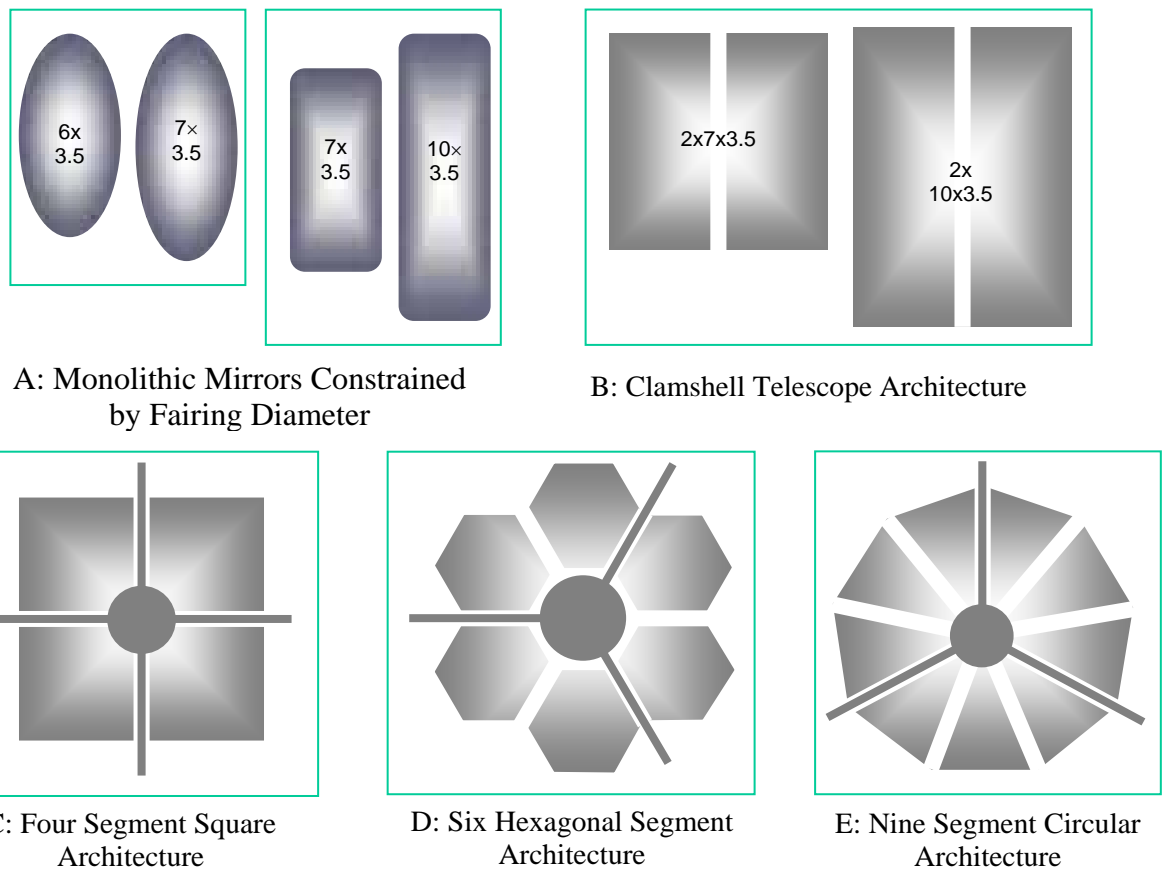

**Figure 7: Primary mirror forms that have been suggested for the use with coronographs.**

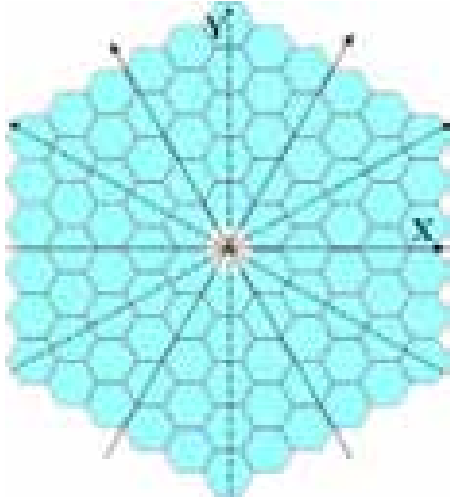

**Figure 8: Ninety-Segment Mirror**

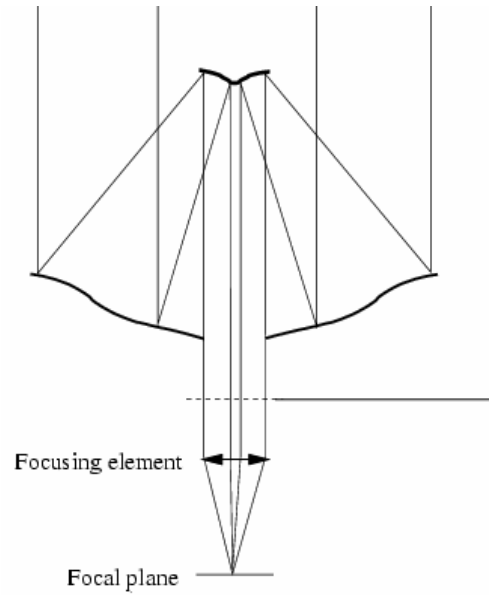

**Figure 9: Phase-Induced Amplitude Apodization**

### 7.2.3. Mirror Shapes and PSF

For the design of coronographic masks one must take into account the specific properties of the PSF produced by the different shapes of the primary mirror. The PSF characterization includes the following factors:

1) Central peaks of the Airy disk: May be round or elongated in one direction.

2) Degree of symmetry: PSF dos not change if rotated by definite angle.

3) Main direction of diffraction: Axis of strongest light scattering (X, or X and Y).

4) High order diffraction by staircase mirror border (for segmented mirrors).

5) Characteristic properties due to imperfect segment alignment.

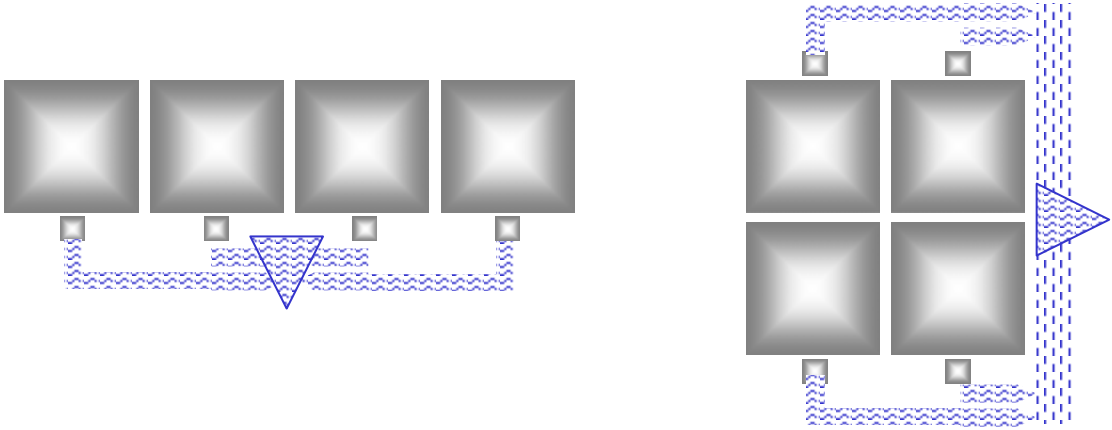

**Figure 10: Multiple telescopes with delay lines and beam combiners**

**Table 7: Diffraction properties of different apertures**

| Aperture Type | Airy Disk Shape | Diffraction | | | |
| --- | --- | --- | --- | --- | --- |
| | | Degree of Symmetry | Main Directions | By a Staircase Border | Due to Imperfect Segment Alignment |
| Round | round | $2\pi$ | no | no | No |
| Elliptical | elongated | $\pi$ | no | no | No |
| Square | round | $\pi/2$ | X, Y | no | No |
| Rectangular | elongated | $\pi$ | X, Y | no | No |
| Clamshell Square (Rectangular) | round (elongated) | $\pi/2$ ($\pi$) | X (Y) | no | 2 peaks |
| Quad Pupil | round | $\pi/2$ | X | no | ~ 7 peaks |
| N "Pie-Segments" | round | $2\pi/N$ | X | no | ~ $N^2/2$ peaks |
| Hexagonal N-segmented | round | $\pi/3$ | X, Y | high order peaks | ~ $N^2/2$ peaks plus higher order |

### 7.2.4. Segmentation Issues

Large deployable, segmented telescopes in space and giant segmented telescopes on the ground are currently being considered for extrasolar planet detection and characterization.

The larger telescope diameters of these telescopes provide both increased collecting areas and larger plate scales for a given Inner Working Angle (IWA). Thus for a 10m telescope operated at 0.5 micron, an IWA of 40 milli-arcseconds would correspond to the 4$^{th}$ Airy ring (~ 4 $\lambda$/D), where the diffraction halo has a contrast of $10^{-2}$ to $10^{-3}$ compared to the central peak. For 30m to 100m telescopes the IWA will correspond to 10 to 30 $\lambda$/D where the contrast is $10^{-5}$ to $10^{-6}$, which makes suppression of the light diffracted by the telescope aperture much easier.

On the other hand, segmented telescopes have several special features, which must be taken into account.

Table 8: Location of an IWA = 0.040 arcseconds vs. Telescope Diameter

| Telescope Dia. | λ = 0.5 microns | λ = 10 microns |
|---|---|---|
| 10 m | 3.8 λ/D | 0.19 λ/D |
| 30 m | 11.5 λ/D | 0.58 λ/D |
| 100 m | 38.4 λ/D | 1.9 λ/D |

### 7.2.4.1. Phasing

A segmented telescope, without accurate segment alignment, would have the resolution of a single segment. The position of segments with respect to each other can be measured with an accuracy of a *few nanometers* by edge sensors located between segments, or by phase and wavelength diversity techniques. These measurements are translated into position corrections applied by position actuators supporting the segments through suitable interfaces. Periodic calibration of the sensors is normally required to ensure sufficient stability and accuracy of the system. Several concepts for segment displacement measurement are being studied now in Europe: Mach-Zehnder interferometry (ESO, LAM), curvature sensor (GranTeCan) and pyramid sensor (Arcetri), and phase diversity combined with redundant spacings calibration (Heriot-Watt). The comparison analysis of these techniques is planned in a framework of the ELT design studies. Random tip-tilt-piston errors cause the appearance of speckles in the image plane, which make any coronograph masking inefficient.

### 7.2.4.2. Segment Fabrication

Errors in the radius of curvature and conic constant of segments for highly segmented mirrors may cause the appearance of speckles and diffraction patterns with periodic peaks. The intensity of these peaks is ∼ $10^{-5}$ to $10^{-6}$ of the central peak, which is not negligible for planet detection requirements. The roll-off of the segment edges resulting from segment polishing will have similar diffraction effects. For OWL, the preliminary specifications for edge roll-off of SiC segments are: edge width ∼ 5 mm and depth ∼ 1λ. The diffraction effects from the segmentation have been analyzed (ESO), but the top-level requirements for segment phasing and edge roll-off (Figure 11) for high contrast imaging still need to be established.

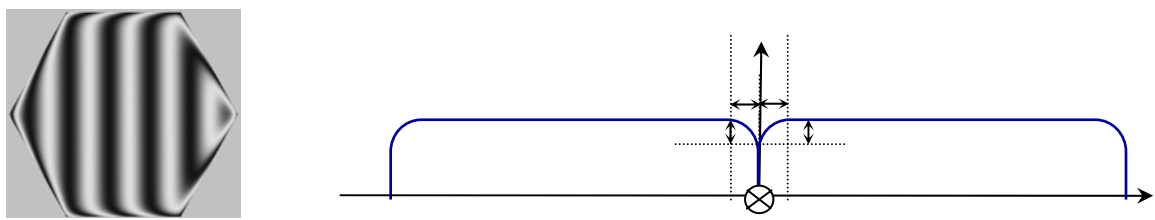

Figure 11: The Diffraction Effects of Edge Roll-Off

### 7.2.4.3. Coronograph Parameters

The geometry of the mask must match the geometry of the mirror following the main direction of diffraction. Thus for hexagonal segmentation geometry the mask would have $\pi/3$ or $\pi/6$ symmetry. The light diffracted from gaps (~ 2 to 5 mm) and the roll-off edges cannot be suppressed by a focal mask and must be removed by the mask in the next pupil plane (Lyot stop).

Finding the optimal shape of the coronographic mask, Lyot stop, and primary mirror apodization for the segmented mirrors is a solvable problem. For a small number of segments ($\lesssim 6$), image masks work fine – rectangular segments are preferred. For a large number of segments, pupil masks are probably required for useful levels of throughput. Vanderbei et al. have calculated many one-dimensional (i.e. linear or radial) designs. Masking a two-dimensional segmented pupil using numerical optimization is a solvable problem that has never been done. Such optimization work represents a natural avenue for a European contribution.

### 7.2.5. Factors

#### 7.2.5.1. PSF, Symmetry, and Position Angle Coverage

The Point Spread Function (PSF) is circular only for a circular pupil and mask form. Either mask or the pupil form may create an inner working angle that varies with position angle. When the PSF is not symmetrical it is necessary to obtain observations at multiple roll angles in order to have access to the entire image plane. Figure 10 shows the set of observations with an elongated mirror required for full discovery position angle coverage.

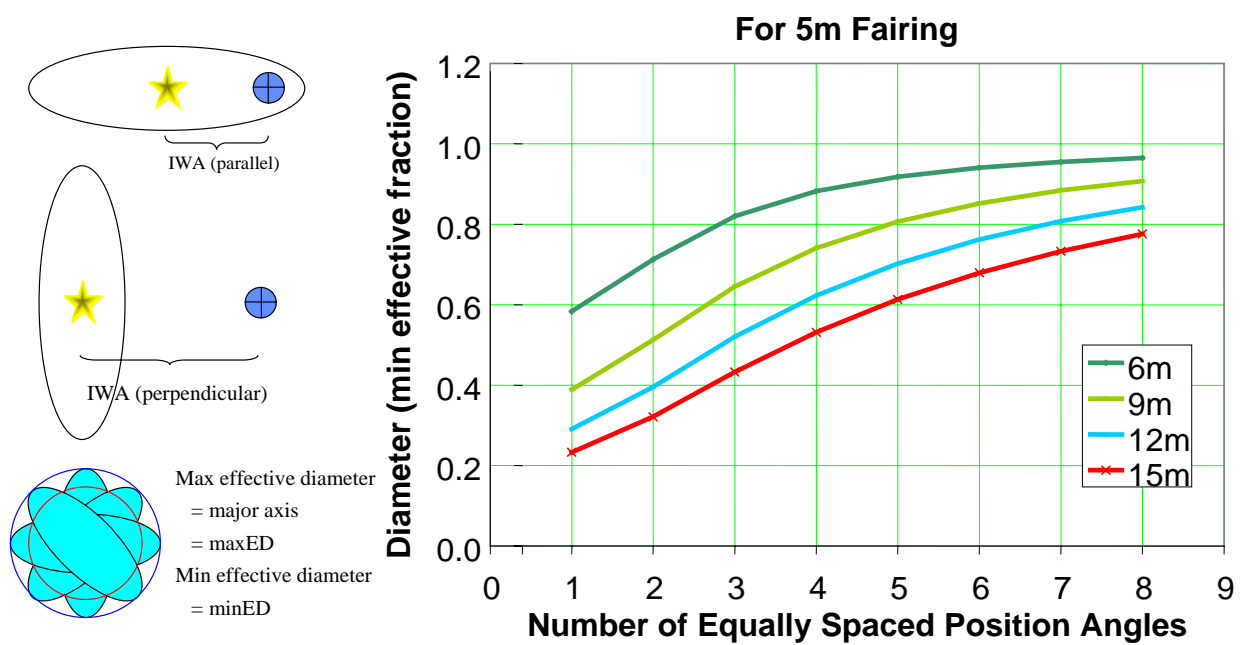

**Figure 12: Strategy for obtaining maximum spatial resolution by observing at different roll angles**

### 7.2.5.2. System Optical Performance and Performance Stability Factors

Several factors affect telescope wavefront performance:

1) Optical finish of the primary and all subsequent mirrors in the system;

2) Spatial uniformity of reflectance on each optical surface in the system;

3) Alignment of the optical path between the primary mirror and mask set;

4) Control of veiling glare due to contamination and other causes.

While mask vector phase and amplitude attributes also affect optical performance, and are considered in telescope systems analyses, mask errors are considered in Section 6.3.1.3.

Telescope control may occur over several timescales. The telescope, both mirrors and structure, will be subject to dimensional or reflective errors due to vibration, shock, thermal transients, aging and radiation effects, debris and micro-meteor damage, contamination, nanoyield and nanocreep, and ground validation imprecision. Both phase and amplitude errors will result from most types of degradation, or any change due to these effects. Compensations are feasible for components of these effects, while the timescale of an effect may range from microseconds to years. For example, at the shortest timescales, vibration isolation is crucial. At the onset of each observation, phase and amplitude errors can be mitigated via the DM set and complex wavefront sensing and control.

The phase PSD requirement is expected to be somewhat better than Hubble Space Telescope (HST), and better in some respects to even the requirements of the microlithography community (for example, AMSL and Alcatel). Lithographic mirrors, although polished to RMS surface values of the order of 1 nm, and satisfying the PSD objectives for a coronographic primary mirror, are typically less than 300 mm in diameter, and are not lightweighted.

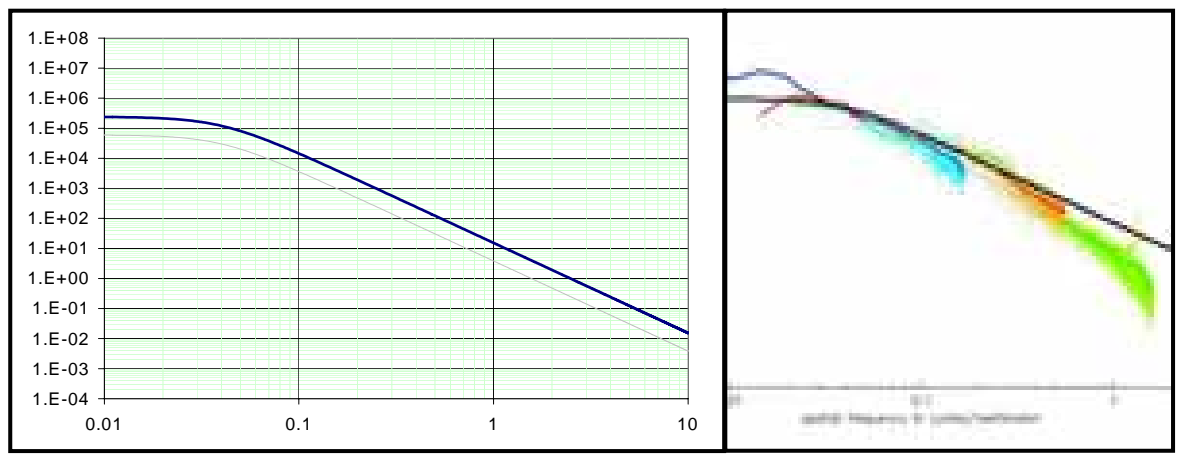

**Figure 13: TPF's 1.8m Technology Development Mirror (TDM) has mid-spatial-frequency (MSF) surface error specified by PSD. The TDM requirement is shown over the goal. The x-axis is in units of $cm^{-1}$, and the y-axis in $\Sigma^2 cm^2$. To the right, we show for context the measured PSD of 3 mirrors (the 6.5m diameter Magellan, the 2.4m HST, and a 1.5m research mirror).**

### 7.2.5.3. Control of Intensity, Polarization

Amplitude errors, as well as phase errors, produce speckles and need to be controlled to the same level in the critical spatial frequencies defining the inner and the outer working angles. Amplitude errors are derived from several sources, and can be controlled by minimizing the causes, by constraining the working area on the sky to a half-dark-hole, or by instituting a pair of DMs. The main causes are:

1) Reflectance variations across the entrance pupil, due to coating reflectance non-uniformity, or due to contamination, either molecular, or clumps of particles;

2) Mirrors not placed at pupils will mix phase errors into amplitude errors;

3) Reflectance variations on secondary, fold, DM, and other mirrors in the path to the masks;

4) Polarization variability, which can introduce both a phase and amplitude component.

Each of these is different in character. Items 1, 3, and 4 may be expected to vary on a long timescale when compared to an observation, and their effects, which must be considered in instrument design and specification, may be regarded to be constant during observations. Item 3 will be an element in the uncompensated error budget, since beam walk on the optics will induce amplitude effects at the focus.

Polarization has several manifestations, and understanding of these manifestations, and their mitigation, has commenced recently. Significant further contributions can be made in this area.

### 7.2.5.4. Dimensional stability, Controllable Surfaces, Operation and Test

Only through dimensional control is exo-Earth coronography viable. The pupil accuracy needs to be smooth to a few tens of picometers RMS, far better than feasible from optical finishing. The state-of-the-art in optical finishing is found in the photolithographic equipment industry, and while better than 1000 pm RMS has been achieved, this is typically on small stiff substrate optics, typically no more than 300 mm in diameter, 1½ to 2 orders of magnitude smaller than required for TEC. While it is viable to correct the pupil with a precision high authority deformable mirror (DM), dimensional stability of the pupil and the system alignment are of great concern. Metering members may be tens of meters in length, and these lengths are further magnified by optical power. Even small thermal drifts will affect the system performance. Creep and residual hydroscopic release of moisture may be significant.

Material and environmental factors determine the uncompensated dimensional stability of the system. Relevant material attributes are discussed in Table 9.

In some cases, considerable efforts will be needed to characterize material to the level that probability estimates can be made of system stability timescales. Only then can candidate architectures be designed and evaluated.

**Table 9: Material factors in dimensional control**

| Attribute | | Mechanical | Thermal | Optical | Thin Film |
|---|---|---|---|---|---|
| Nominal & σ | | E | CTE | Surface PSD | Coating stress |
| | | ρ | k | Radius of Curvature (ROC) | Undercoating |
| | | Poisson's Ratio | $c_p$ | Conic Constant | Coating thickness |
| | | Microyield | ρ | Continuity of discontinuous or segmented primary mirror optical surface | Overcoating |
| | | Microcreep | α | 6 DOF alignment of each element | Complex index of refraction |
| | | Discontinuous creep | ε | Position sensing | Scattering centers |
| | | Radiation compaction | Radiation effects on α,ε | Fiducial | Reflectance PSD |
| | | Damping | Contamination effects on α,ε | Position metrology | Contamination |
| | | Fracture Toughness | | Position actuation | With mirror geometry, polarization effects |
| | | Coefficient of Moisture Expansion (CME) | | | |
| inhomo-geniety & anisotropy | | E, ρ | CTE | Mounting and test condition | Thickness PSD |
| | | E, ρ | k | Surface response to stimulus at all spatial frequencies | Application variance or environmental change |
| Temporal | | E, ρ | CTE, k, $c_p$, Thermal Environment | Substrate or mount change | Gradual aging |
| BOL | | E, ρ | α, ε | | Reflectance & reflectance PSD |
| EOL | | E, ρ | α, ε | | Reflectance & reflectance PSD |

Furthermore, the telescope-mask system must be suitable for some level of ground qualification test. Ground performance of any TEC telescope will depend on both gravity release fixturing, and on extensive precise modeling. Just as for JWST and some other envisioned flight systems, full performance tests on the ground are not viable. It is appropriate to recognize the extensive exquisite wavefront sensing and correction ability in a TEC telescope, and this ability, coupled with high fidelity verified models (via test beds and test cases), may assure that the system can meet all requirements in space, even if it is not tested fully to these requirements on Earth.

Although TEC should start with an athermal design, vibration isolated, thermal isolation and using carefully selected materials for both optics and metering, sufficient stability will not be realized, especially considering the ground-to-space changes. Table 10 identifies likely sensing and actuations needed to address the stability requirements of a TEC.

### 7.2.5.5. Modeling, Integration & Test, and Functional Validation:

Special circumstances apply to integrating and testing a large lightweighted telescope system: (1) The system uses internal high precision wavefront sensing capable of detecting and isolating both pupil aberrations and any alignment errors. (2) The system has many degrees of freedom of actuation to correct these errors *in situ*. (3) The system is sensitive to gravity perturbations, test vibration and thermal effects that are inevitable on the ground, but will not be relevant to space-borne functionality. Only a small part of the space functionality will be able to be captured in ground tests. We propose that the primary solution to validating the system prior to launch involves (a) assuring that the models accurately cover both flight and validation requirements, and (b) that the system sensing and actuation range exceed the regime of model coverage by a considerable margin. Functionality under environmental stimulus needs to be investigated, but ground validation tests need not meet the full performance requirements. Figure 14 illustrates a top-level development and process flow.

**Table 10: Sensing and actuation functions needed to meet stability requirements**

| Mode | Type | Sensor | Actuator | Precision | Range | Flight Calibration | Real Time Control | Control Bandwidth | Constraint |
|---|---|---|---|---|---|---|---|---|---|
| Drift in centration on masks | pointing | Dedicated camera or quad cell | Fine stearing mirror | 0.1mas | 5 arcsec | Yes | Yes | ~10-100Hz | beamwalk, if correction too large => body pointing command |
| Despace, decenter or tilt of M2 | align optically | Wavefront Sensing, and thermal sensors | 6 DOF actuated mount at M2, and active thermal control of metering | nano-meters | microns | Yes | Probably | < 1Hz | |
| 6 DOF on primary | align optically | Wavefront Sensing | 6 DOF actuated mount at M1 | nano-meters | microns | Yes | Possibly | <<1Hz | |
| Phase up of primary mirror segments | align optically | Wavefront Sensing | Tip-tilt, piston and ROC of each element | nano-meters | microns | Yes | Unlikely | <<1Hz | |
| Correct low order primary aberrations | optical adjust | Wavefront Sensing | force or position actuators | nano-meters | nano-meters | Yes | Unlikely | <<1Hz | |
| Mid spatial frequency error on primary | optical adjust | Wavefront Sensing | DM, > $10^4$ DOF | ~10pm | <1um | Yes | Possibly | <1Hz | |
| Telescope metering structure | Thermal adjust | Thermal sensors | Heaters | 0.001C | 2C | Yes | Yes | ~1Hz | |
| Fold and relay mirror alignment | Thermal adjust | Wavefront Sensing, and thermal sensors | Heaters | 0.001C | 2C | Yes | Yes | ~1Hz | May need mechanical actuators in some instances |
| DM cavity | Thermal adjust | Thermal sensors | Heaters | <0.001C | 2C | Yes | Yes | ~1Hz | |

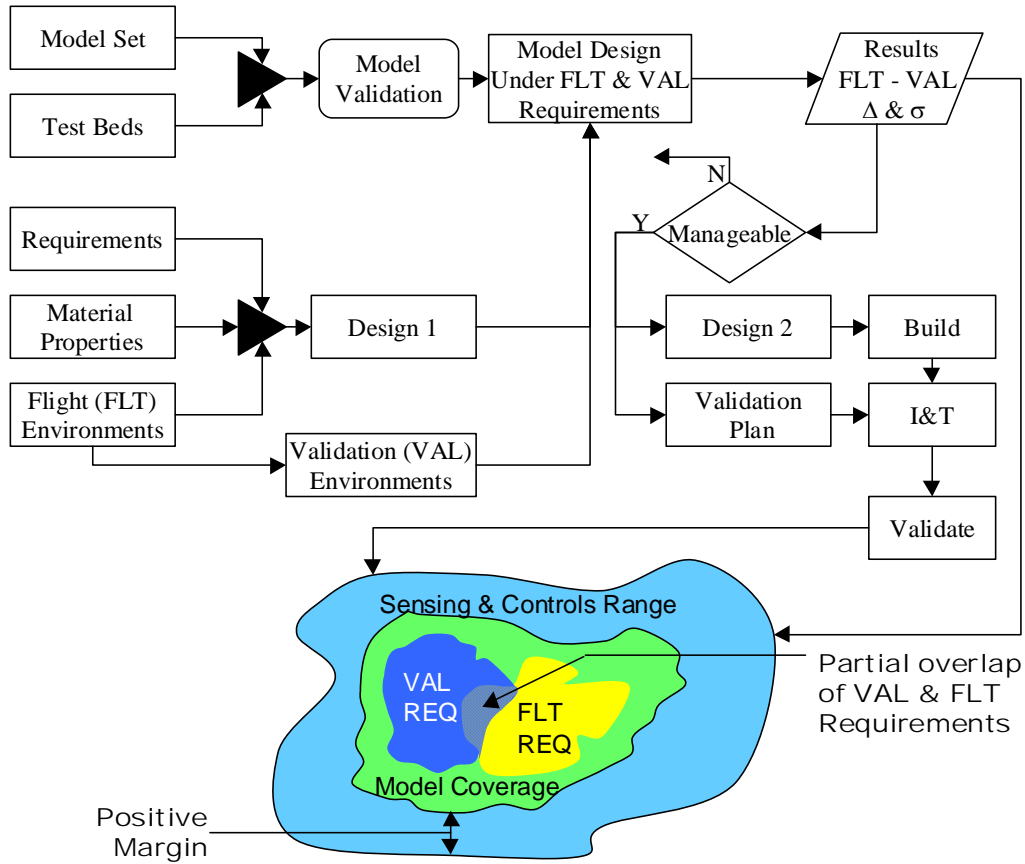

**Figure 14: Development and process flow for a large light-weighted telescope system**

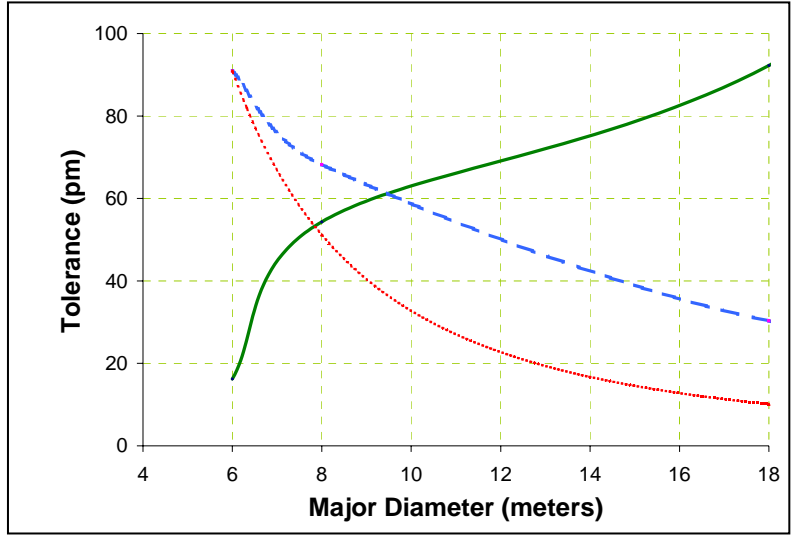

**Figure 15: Tolerance on focus error vs. telescope diameter (green curve). The two descending curves represent the relative time to collect the required number of photons (see text).**

### 7.2.6. Key Trades

#### 7.2.6.1. Inner Working Angle and Diameter

The inner working angle (IWA) available to a coronograph is determined by the closest Airy ring at which planet detection can be achieved with the telescope-mask system:

$$\text{IWA(arcsec)} = 206265 * n * \lambda / D \quad , \tag{1}$$

where $n$ is the Airy ring number, $\lambda$ the wavelength and $D$ the diameter. A smaller IWA and a smaller telescope diameter are both desirable, but these goals are mutually contradictory. If we regard the wavelength to be set by exo-Earth phenomenology and by thresholds in optical finishing, optical control and thin film coating performance, then $n$ is the free parameter. A small $n$ may be paired with a small $D$, or a large $n$ with a large $D$. While it is tempting to push for the former, we cannot state a priori that this is the optimum solution.

Figure 15 constrains the IWA and $\lambda$ to be constant, and for one case (representative only) describes how tolerance to focus error will depend on diameter. $n$ is then constrained to vary in proportion to $D$. As the diameter of the mirror becomes larger, the tolerance on focus error also becomes easier to meet. The two descending curves represent the relative time to collect the required number of photons. The dashed curve is for an elliptical primary mirror form, and the dotted red curve for a symmetric primary mirror form. A larger mirror, especially a symmetric mirror, collects more photons, or can collect photons faster. Faster collection both benefits the science productivity of the instrument, and shortens the time duration that the instrument must remain stable. Generally, Airy parameters $n$ in the range of 3 to 6 are being considered, but some instruments may transcend this range.

#### 7.2.6.2. Monolithic vs. Multi-Mirror Designs

A monolithic mirror of a given size will always be preferable to a segmented mirror of the same size, but the choice between a monolithic mirror of one size and a segmented mirror of a larger size is more complex. For space telescopes the dimensions of the mirror are strongly constrained by the size of the available rocket fairings, the largest currently available being 5 m in diameter. For this reason, the proposals for the minimum-mission TEC telescopes using monolithic mirrors require the use of elliptical or rectangular mirrors approximately 3.5 m × 6 m in size. The alternative to this is to use two or more mirror segments, which can be deployed in orbit. This has the advantage of providing a large collecting area and, possibly more significantly, providing a more symmetrical pupil and PSF. This makes it unnecessary to observe each star with multiple roll angles. The potential problems of using segmented mirrors include achieving the necessary phasing accuracy from one segment to another and the possible complexity of the pupil and image plane masks required to achieve the necessary contrast ratio.

### 7.2.6.3. On-Axis vs. Off-Axis Designs

Until recently, most coronograph designers had strongly favored an unobscured, off-axis telescope approach. Driven by wanting to minimize telescope aperture for a given inner working angle (IWA), most solutions operate at a few times $\lambda/D$, often at $3\lambda/D$ or less. In the Lyot coronograph case, small IWAs require intrusive Lyot stops, and low Lyot stop throughput. Throughput is made even smaller if Lyot masks must cover also a 4-blade spider and a secondary mirror shadow. Therefore, maximum Lyot stop throughput for such a telescope is achieved where there is no secondary mirror in the beam to be masked.

While this would seem attractive, there are counter indications in the complexity of the primary mirror itself. A large circular mirror for an unobscured telescope would have less than half the diameter of the hypothetical parent mirror, and the parent f/# would be less than half that of the used aperture. The result is more aspheric departure from the nearest sphere on the mirror blank. For a parabolic mirror, aspheric departure scales with the parent diameter over the cube of the f/#, suggesting the aspheric departure is sixteen fold greater for the unobscured circular blank than for the circular obscured blank.

Aspheric departure is both a traditional measure of optical fabrication difficulty and expense, and a measure of variation in slope over the surface. While modern optical fabrication methods have made reaching large aspheric departures feasible, the rapidly changing slope usually results in more zonal (bull's eye pattern) error. As a mirror becomes more elongated, and is rendered off-axis off the minor axis, the difference in optical complexity between unobscured and obscured forms is reduced. However, there are significant issues of position angle coverage using elongated mirrors, and multiple rotations are both time consuming, and put a higher premium on system dimensional stability.

Two new factors are being considered carefully now: (a) a rich variety of (mostly) pupil masks are being developed which can mitigate the deficiencies of complex pupil forms, including those created by an on-axis secondary mirror and its support, and (b) segmented mirrors, operating at a higher number of $\lambda/D$s, and at larger aperture are returning to trade space. Such mirrors may offer symmetry, and segments of such mirrors may be practically packaged in existing fairings.

### 7.2.6.4. Material Selection: Mirror and Metering

The selection of the material for a TEC will be dependent upon the mirror architecture adopted. For a coronographic telescope, the primary mirror figure and its stability will be the major drivers in the telescope's performance. The key elements that must be traded for each telescope architecture are shown in Table 11.

Table 11: Key trades for telescope architectures

| Trade | Description |
|---|---|
| Mirror Figure | - Figure requirements over applicable PSD regimes<br>- Primary mirror f/# drives size of the metering structure |
| Mirror Stability | - Stability requirements for the mirror figure(s) over applicable PSD regimes |
| Mirror Size | - Size of the telescope's primary mirror(s), and resulting requirements on mirror mass and mirror stiffness |
| Mirror Material | - Heritage from previous programs<br>- Ability to meet thermal and mechanical stability requirements<br>- Availability of mirror material in required quantities/quality<br>- Convergence time for mirror figuring/polishing<br>- Convergence limit for mirror figuring/polishing<br>- Complexity of figuring multi-component mirrors e.g. cladding<br>- Drives overall mass of the payload |
| Mirror Architecture | - Shape of primary mirror(s) and the resulting fabrication requirements, e.g. single mirror or multiple sections with a common faceplate.<br>- Facilities to process and test large aperture mirrors<br>- Light-weighting as a function of mirror stiffness, and print-through |
| Telescope Truss Material | - Choice of composite material for metering structure drives its mechanical and thermal stability over a range of timescales<br>- Material also drives overall mass of payload<br>- Long term performance of metering structure e.g. outgassing |
| Thermal/Mechanical Stability | - Thermal conductivity of mirror material<br>- Thermal stability of mirror design as a function of telescope truss and stray light shielding design<br>- Stability of the metering structure is function of its size |
| Launch Loads | - Requirements on μyield to maintain mirror figure following launch |
| Mirror Mount | - Impact on mirror figure from thermal/mechanical loads imparted into mirror from mounting concept<br>- Ability to correct mirror figure |
| Metrology | - The ability to test and verify the mirror performance is essential<br>- Metrology of primary mirror(s) over all PSD regimes<br>- Qualification of gravity models for aperture size(s)<br>- Ability to conduct full optical verification of mirror performance<br>- Ability to anchor analysis in verifiable mirror measurements |

### 7.2.6.5. Wavelength Coverage

The expected wavelength coverage for TEC is a driving issue for the primary mirror optical specification. As the coronograph is required to image terrestrial planets to shorter wavelengths, there is a corresponding impact upon the primary mirror's optical specification, especially the mid-frequency PSD. In the absence of well-defined science goals this can drive the cost of the mission. Clearly defined wavelength coverage is also required to facilitate a trade study of potential mirror coatings, since gold and silver have cutoffs that will factor into the planet finding and ancillary science programs. Key trades in the selection of a coating include uniformity, ability to coat the required size of optic, and polarization.

#### 7.2.6.6. Mirror Focal Ratio vs. Metering Length and Polarization

Packaging is a significant constraint for a TEC telescope system. By specifying the primary to have a faster f/#, the metering length between the primary and secondary is reduced, saving volume and deployment complexity. This shorter length may be easier to control thermally, and is stiffer to vibration. Also thermal shields could be made more compact, and be easier to deploy and less subject to radiation pressure torque.

However, while these advantages would seem attractive, there are also negative factors:

1) Faster f/#s are associated with greater aspheric departure, which usually is associated with smaller optical finishing tools and more zonal error in the critical mid-spatial-frequencies. Aspheric departure scales $\propto$ Diameter $\times$ (f/#)$^{-3}$.

2) Optical magnification will be greater, and the sensitivity to displacement errors in positioning of the secondary mirror will be unfavorable.

3) More curvature means more dispersion in *s* and *p* polarization due to angle of incidence range. This will result in a focus that is not compact.

While system analysis must weigh these advantages and disadvantages considering the pupil-mask system, we feel that it is likely that a primary mirror f/# of the order of f/2 will be selected. The implication is that a 10 m entrance pupil would require a metering structure of the order of 20 meters, and baffles and shields considerably longer.

### *7.3. Status, Issues and Future Work*

#### 7.3.1. Technology and Manufacturing

8-10 meter diameter class primary mirrors have been manufactured in both Europe and the United States, but these have been for ground use where the functional requirements are less exacting, and the areal densities are higher and more forgiving of optical processes. Manufacturing and validating deployment of the metering structure may be nearly as demanding as making the mirror, and the mirror and secondary optics are crucially dependent on this structure. In the case of a segmented mirror, the metering structure is integral to the mirror. We estimate that some facilitization and quality engineering can produce a mirror up to 8-10m size in existing facilities. To go larger than this in a monolithic or segmented mirror, a substantial amount of research and development and facilitization is required.

There are materials trades that should be made for both mirror and metering structure. Especially important is control of the homogeneity, isotropy, and creep of materials over this large scale. Both mirror and structure will be made of many pieces or even lots of material, and successful design and implementation is dependent on the characterization of material stability factors, and their dispersion in stability variables. Since the system will be deployed, development of stable high reliability joints and mechanisms will be needed. Material factors for dimensional control are tabulated in Table 9. Built-in metrology, both of linear distances, and precise temperature, is needed. Built-in actuation is expected, either for monolithic or segmented mirrors.

Every step of the manufacturing process is complicated by the scale and the accuracy needed by the coronograph. During and following each step, large scale tooling and metrology is needed. The steps are:

- Glass, SiC or composite selection and manufacture,
- Mirror substrate manufacture to required areal density; metering structure and hinge manufacture,
- Manufacturing mount of blank,
- Generate the surface,
- Grind and polish the surface,
- Metrology test set (radius of curvature or stitched auto-collimation test),
- Metrology mount,
- Final finish of the mirror's optical surface to $< \lambda/100$ rms ($\sim$ 5 nm rms),
- Thin film coat the mirror with uniform coating to $< 0.1\%$ reflectance variation, and proper polarization and stress and durability characteristics,
- Test of the completed mirror for phase and intensity uniformity and polarization properties,
- Mount and align mirror on metering structure and test,
- Assemble the rest of the telescope (with gravity offload fixtures) and test,
- Integrate instrument and test as payload,
- Environmental test of payload.

Technology development is needed to some level to

- Assess and control material dimensional stability properties for candidate materials, and conduct trades,
- Control and measure the optical surface PSD,
- Actuate the surface to correct for low order term errors,
- Deploy the mirror and metering structure,
- Implement thermal servo system and solar isolation system,
- Establish the coating that has optimum reflectance, reflectance uniformity, low scatter, thermal properties, and polarization properties, stress and survivability factors.

Since we know of no other application that imposes these coating requirements simultaneously and at this scale, metrology methods, especially for reflectance uniformity, should be developed.

### 7.3.2. Verification and Reliability

The major challenge for a TEC during Phase B and beyond will be the ongoing process of verifying the performance of its optical system in terms of the levied requirements. It is expected that the TEC observatory will be required to undergo an end to end test which verifies that it can meet its assigned Level 1 requirements either by demonstration, analy-sis, or more likely a combination of both. In the case of TEC, this requirement will be stated in terms of being able to detect a contrast ratio at a specific radial distance consistent with the detection requirements for a terrestrial planet. For TEC, therefore, even the process of performance validation will require new approaches, techniques, and technologies.

In Table 12 we outline a very high-level straw man summary of the major verification tasks for a telescope plus coronographic instrument implementation of TEC. It is too early to know whether the final stage of integration and test can be anything more than basic optical testing and functionality with the high level requirements left to verification by analysis. It is interesting to compare TEC with the verification proposed for JWST, where a full verification of imaging performance on the phased telescope is planned at a temperature of about 50 K. While the optical requirements do not approach those of TEC the test could be considered equally challenging in its own right since it will be performed at cryogenic temperatures. The problem of illuminating the telescope (at cryogenic temperatures) was solved by means of several specially developed concepts. The complexity of these tasks for TEC argues for consideration of performance validation and verification as a possible new technology development track for the program in its own right.

**Table 12: Major verification tasks for telescope plus coronographic instrument**

| TEC element | Item | To Be Verified |
|---|---|---|
| Telescope optics<br>- component level | Primary mirror | Mirror figure reqs.<br>Analysis verification<br>PSD characterization – all scales<br>Coating uniformity |
| | Secondary mirror | Mirror figure reqs.<br>PSD characterization – all scales<br>Coating uniformity |
| | Tertiary optical chain | Mirror figure reqs.<br>PSD characterization – all scales<br>Coating uniformity |
| Telescope optics<br>- integration level | Point source /<br>interferometry | - Image quality<br>- Optical alignment<br>- Limited PSD characterization<br>- Active optics capture ranges<br>- Mechanism capture ranges<br>- Alignment tolerances<br>- Focus tolerances<br>- Pupil tolerances |
| Telescope / SI interface | | - Optical alignment<br>- Active optics capture ranges<br>- Mechanism capture ranges<br>- Alignment tolerances<br>- Focus tolerances<br>- Pupil tolerances |
| Instrument<br>- fixed component level | Optical components | Wavefront / figure reqs.<br>PSD characterization – all scales<br>Coating uniformity<br>Transmission / reflectance |
| Instrument<br>- active components | Optical components<br>e.g. DM, tip/tilt etc. | Wavefront / figure reqs.<br>PSD characterization – all scales<br>Coating uniformity<br>Transmission / reflectance<br>Functionality<br>Capture ranges |

| TEC element | Item | To Be Verified |
|---|---|---|
| Wavefront sensor<br>- integrated | WFS requirements | - WFS can measure WF to required precision<br>- WFS can capture wavefront over required range with sufficient quantization over range |
| Instrument<br>- integration | | Wavefront reqs.<br>PSD characterization – all scales<br>Throughput<br>Scattered light at focal plane<br>DM / mechanism capture ranges<br>Meets contrast ratio reqs.<br>Operating mode functionality |
| Detector | | Reflectivity<br>Flatness<br>Basic detector functionality |
| Telescope + instrument<br>- integration | Interferometry | - Optical alignment<br>- Active optics capture ranges<br>- Mechanism capture ranges<br>- Capture range flow-down reqs.<br>- Alignment tolerances<br>- Focus tolerances<br>- Pupil tolerances |
| Telescope + instrument<br>- verification | Interferometry /<br>point source /<br>analysis | - Image quality<br>- Contrast ratio to TBD<br>- PSD characterization to TBD<br>- Wavefront measurement to TBD<br>- Stray light to TBD |

Reliability is a factor that must be considered during all phases of the design and development of a telescope, with robust margins for all design budgets and electrical and mechanical redundancy for all mechanisms. A Fault Tree and a Failure Modes Effects Analysis should be completed early in the design process and updated regularly as the design evolves.

### 7.3.3. Potential European Contributions

Besides the experience in fabrication and control of mirror optics mentioned above (Section 7.1.3), there are substantial areas where European institutes could contribute, including several key technologies that are being elaborated in Europe in the framework of the Extremely Large Telescope (ELT) design studies. Although these studies are being done for ground-based 30 – 100 m class telescopes, the lessons learned can have a broad application for design and performance analysis for any telescope. Potential European contributions include:

1) Studies (analytically and/or numerically based) of the influence of the different telescope errors (telescope aberrations, optics misalignment, polishing errors, segmentation, chromatic effect, etc.) on the image quality by PSF and PSD characterization.

2) Unique approaches for wave front analysis and correction for multi-conjugate adaptive optics, which might be adapted to other areas of coronography.

3) New phase control techniques that are being developed in parallel for telescope phasing and for adaptive optics. The general study of the optical basis of these methods could help to improve the efficiency of both, which are critical for the high contrast imaging areas.

4) The design and study of potential coronographs for extremely large ground-based segmented telescopes, a task where the European community could make a very substantial contribution.

## *7.4. Conclusions*

Telescopes intended for use with coronographs have special requirements that must be met to achieve the necessary performance. Monolithic off-axis architectures are preferred, but segmented telescopes are a viable option that can provide the larger collecting areas and higher spatial resolution required for many applications. These telescopes must be designed as part of a system that includes the coronograph optics and image and pupil plane masks that are optimized for the particular telescope architecture.

We recommend a rigorous investigation of masks co-optimized for segmented optics, and work on modeling the performance of coronographs during testing and in the space environment. A better understanding of the properties of mirror coatings and techniques for producing highly uniform coatings are also needed. We also recommend the formation of Working Groups for performance modeling, materials characterization, and coatings development.

The European community is well qualified to participate in many aspects of telescope development.

## *7.5. References for Section 7*

# 8. Simulation, Modeling, and Testbeds

## *8.1. Description and Purposes*

### 8.1.1. Design Process

Some early models can be constructed to answer simple questions about a broad category of designs. However, the more general and theoretical the analysis, the narrower the questions that can be answered. More detailed integrated models, needed to answer interdisciplinary questions like bandwidths and control system effectiveness, must have a starting point – they can't be fully general.

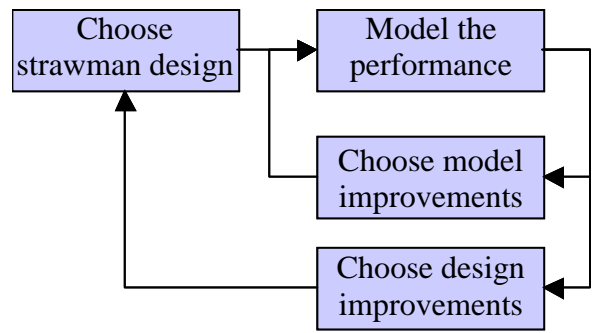

**Figure 16: Design iterations. Choose a strawman design, answer key questions about performance, then either improve model fidelity or change the coronograph design.**

An integrated model is a software tool-set and run-time environment which allow time- or frequency-domain simulations of system performance, incorporating discipline models: structures, optics, controls, and disturbance sources (Figure 17). Its purpose is to answer system questions which cross the boundaries of disciplines.

Models can be useful throughout the life of the project. During concept development, the models help understand requirements (Section 8.1.2) and compare designs (Section 8.1.3). Later, more advanced integrated models can predict science capability with consistent assumptions; assess system performance with measured hardware data; assemble results of several tests on flight hardware to predict in-flight performance; and test hypotheses to understand on-orbit behavior and performance. Thus modeling can help throughout the mission life cycle.

Early studies have covered simple questions for a wide variety of coronograph concepts. These computer models usually include the most basic physics, mainly to prove adequate suppression of diffraction or speckle. Higher-fidelity optical models have been used recently to answer more challenging questions, such as the effects of phase errors on each optic (not just the primary mirror) or polarization and phase effects on mirrors and masks. Integrated models are a more complete approach for understanding interactions within the observatory, such as the effects of structural vibrations, pointing control, and thermal control, and how each coronograph design can be vulnerable to such things.

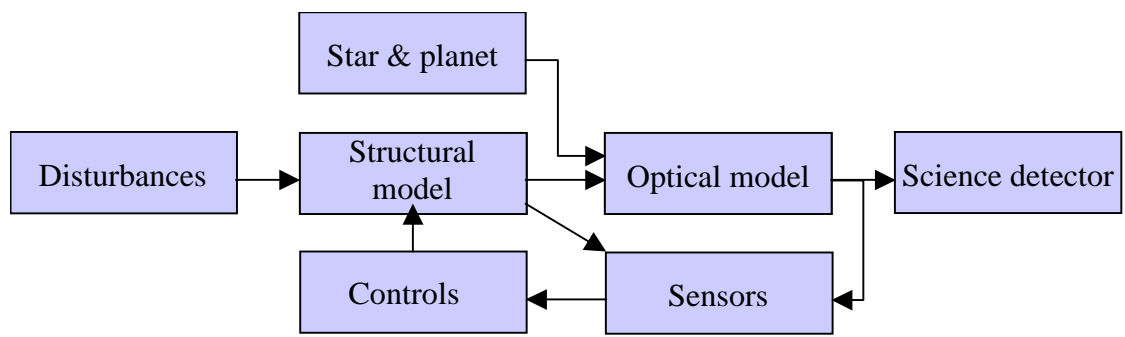

**Figure 17: Integrated model features. In a single run-time software environment, disturbances are applied to a structure model; structure elements move optical system elements, which affect signals on the science detector and perhaps other sensors. Control systems respond to the sensors to drive actuators represented in the structural model.**

### 8.1.2. Understanding Requirements, Including Data Analysis

Models have been used to determine or verify requirements on wavefront error (phase and amplitude), attenuation of red color filters, fabrication of shaped pupil masks, structural vibrations, thermal distortions, and pointing errors; and to check the spectral data quality. These have laid the groundwork for understanding the challenges of a TEC.

It turns out that many requirements are very tight, and so some recent ideas are aimed at easing requirements. Possible future modeling studies include

- What kinds of instrument errors can be compensated by DM control, or by data post-processing? How robust is the coronograph to partial failures like the loss of one or more pixels on the DM?

- What are the merits of different wavefront sensing approaches and calibration strategies? Are algorithms robust to stray light other than coherent starlight (sunlight, other stars, same star scattered from baffles, etc.)?

- What are the benefits of an integral field unit (IFU) in place of a simple CCD? Does this allow wavefront sensing and control, or post-processing to subtract speckles, with some relaxation of stability requirements? What IFU performance would this require?

- What are other impacts of these choices (e.g. optical bandwidth)?

- For segmented-pupil telescopes with various coronograph designs, we need to know the requirements for placement of the gaps and their width and variance of width.

### 8.1.3. Comparing Designs

Recent studies of sensitivity to low-order aberrations show interesting differences among the different candidate designs. By studying how these aberrations affect the detection of planets at the nominal working angle (IWA), we have begun to assess requirements on stability of the large telescope [1].

Many different coronograph designs have been proposed; how do they compare in throughput, inner working angle, contrast, and other measures of planet-detection ability?

If we had an integral field unit (IFU), would it help? It has been suggested one might analyze IFU data (spectra in each pixel of the focal plane) to characterize each speckle, sufficient to aid in extracting planets from speckles or even for correcting the wavefront. In the former case (post-processing only), does long-term drift in the speckle pattern degrade the accuracy of this correction? For wavefront corrections, how faint a star would be practical?

It has also been proposed that one might sense changes in low-order aberrations by analyzing the starlight that is stopped and reflected by the field occulter. Is this a practical way to measure and perhaps control those aberrations?

Some integrated modeling has produced estimates of the amount and effect of residual vibration assuming various types of vibration isolation. These results show that advanced but practical isolators that are within the state of the art today could be adequate for the TEC; newly developed designs could do even better.

### 8.1.4. Performance Prediction and Requirements Verification

At the time of the interferometer-coronograph architecture selection, integrated models can be used to produce a scientific throughput assessment for the final designs.

Near the start of the Formulation Phase, it is helpful to establish baseline estimates of system performance with a consistent set of design parameters and assumptions and reasonable fidelity. This also demonstrates the sufficiency of whatever requirements are adopted by the project at that time.

### 8.1.5. Support of System Testing

It seems reasonable to assert that a full end-to-end test of the TEC system with realistic light sources and thermal/vibrational environment is too expensive and impractical. More likely an ensemble of smaller tests on the flight system, assembled with the aid of integrated models, will verify that the flight hardware can meet the science requirements. This means there will be heavy reliance on credible high-fidelity models to support that proof.

## *8.2. Coronograph Modeling Activities (Including Bibliography)*

### 8.2.1. US Modeling Activities

In the US, coronograph modeling for TPF is comparatively well-funded. Major examples of different kinds of models are summarized below.

#### 8.2.1.1. Optical System Models for TEC

- JPL coronograph design team [1] (Joe Green):

  Fraunhofer modeling to examine sensitivity of Lyot coronograph designs to low-order aberrations. Pointing and other stability requirements are partially driven by these sensitivities, and these requirements are balanced against other figures of merit when choosing a design.

- Ball coronograph concept study [2]:

  Fraunhofer propagation (entrance pupil → focal plane → Lyot plane → image plane) with mirror static PSDs, low-order WFE, multiple wavelengths, 3-4 coronograph designs. Milliarcsec pointing on the field occulter; < 50 pm rms WFE in critical spatial frequencies, red leak, colored speckles. Later analysis (unpublished) indicates 15 mas body pointing requirement, and assumes challenging mirror PSDs.

- Goddard Space Flight Center [3] (Rick Lyon):

  Several optical modeling tools including rigorous vector wave propagation and OSCAR system modeling tools. Showed that thin masks used elsewhere are too idealized; real masks have phase and amplitude effects vs. polarization, color, and tip-tilt. Please refer to Appendix 11.4.3 and 11.4.5 for greater detail.

- Chris Burrows:

IDL models similar to Ball's simple Fraunhofer model.

- Olivier Guyon [4]:

  PIAA (pupil re-mapping) coronograph designs and performance.

- Princeton [5] (Kasdin, Spergel, Vanderbei):

  Optimization of binary mask designs, hybrid concepts.

**8.2.1.2. TEC Integrated Models Including Structures, Optics, and Controls**

- Ball coronograph concept study [2], [6]:

  Full integrated model including reaction wheel disturbances, structure dynamics, optical distortions and displacements, Fraunhofer model of coronograph performance, with attitude and pointing controls. Passive vibration isolation is very nearly adequate; active vibration isolation designs are available, probably succeed with some margin.

- JPL coronograph design team [7] (Joe Green):

  Integrated modeling that combines thermal transient analysis, finite element modeling, and a MACOS optical model. A recent study looked at how attitude changes induce a thermal gradient in the primary mirror which causes a small deformation. Also polarization effects, diffraction effects, the usual aberrations and the usual array of aberration sensitivities.

**8.2.1.3. Testbeds and their Models**

- JPL HCIT (John Trauger) [8]:

  MACOS based sensitivities derived from a model that employs Fresnel and near field diffraction calculations and gridded surfaces on every optic. Uses measured surface maps from Zygo measurements of the optics. The model has been validated to the $\lambda/200$ level in wavefront.

- Boeing study for Extra Solar Planets Advanced Concepts NRA [9] (Bob Woodruff) – a study of relevant issues in designing and operating a coronograph testbed.

## 8.2.2. European Modeling Activities

Most of the European studies of instruments aiming at detection of terrestrial planets have been focused on IR nulling interferometer designs. While ESA's long-term Darwin project is the main driver for these studies, a few laboratories have started to study coronographs as an alternative for exoplanet searches. The large number of coronographs that have been studied in Europe for the last 10 years were designed to work on ground-based telescopes with adaptive optic systems, but Europe is also involved in the study and qualification of the coronographs that will be placed on the instrument MIRI for the JWST.

The Observatoire de Paris/LESIA is responsible for the four coronographs to be placed on JWST. One Lyot coronograph and three Four-Quadrant Phase Mask (4QPM, [10]) coronographs will be placed in MIRI (Figure 18, Figure 19). An extensive study has been done on the capabilities of the 4QPM coronographs when used with the JWST defects. The same group has performed laboratory experiments in the visible, in the IR, at room temperature and at cryogenic temperature. The same team (LESIA) has placed a 4QPM coronograph behind the adaptive optic system of the VLT for ground-based testing [11].

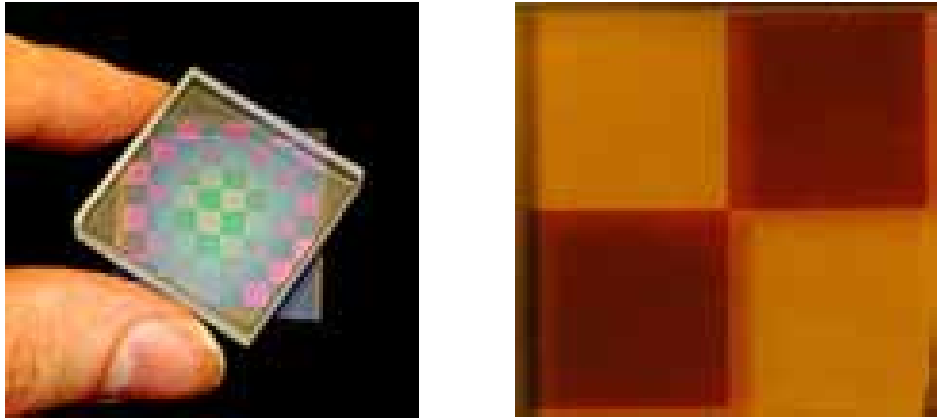

**Figure 18: Image of a multiple FQPM operating in the visible (left) for lab experiment and of a mid-IR 4QPM (right) for MIRI/JWST.**

The Achromatic Interfero Coronograph (AIC, [12]) has also been tested on ground-based telescope using adaptive optic system. This coronograph built by the Observatoire de la Côte d'Azur (OCA) has been tested in the laboratory at visible and IR wavelengths, and studied for ground-based observation. Observations with this coronograph took place at Observatoire de Haute Provence and at the Canada-France-Hawaii Telescope.

The Laboratoire Universitaire d'Astrophysique de Nice (LUAN) is studying coronographic concepts such as prolate apodized coronographs. LUAN also built a testbed for apodized coronographs.

The Observatoire de Haute-Provence also studied coronographic concepts for pupil densified interferometers.

The Laboratoire d'Astronomie de l'Observatoire de Grenoble (LAOG) has built several Lyot coronographs for ESO telescopes and possesses a strong background in simulating the effect of the adaptive optics on the coronograph performances.

European Southern Observatory recently launched a call for study of ground-based coronographic instrument for the detection of exoplanets. Two European groups are involved in answering this proposal. One is led by Markus Feldt (Germany, Switzerland, The Netherlands, Portugal, Italy) and the other by Jean-Luc Beuzit (France, UK, Switzerland, Canada).

**8.2.2.1. Simulations**

- LESIA/CEA (Anthony Boccaletti): MIRI/JWST coronograph study. Sensitivity of aberration, jitter, pointing errors, pupil shift, segment effects. Simulation for VLT Planet Finder (4QPM coronograph + Multi-spectral imaging), considering sensitivity to telescope aberrations, instrument aberrations, jitter, differential spectral pointing, differential pointing with respect to calibrator star, optimization of the Lyot stop.

- LUAN (Claude Aime): coronographic concept study

- OCA (Y. Rabbia): instrumental modeling (aberration, jitter, pupil-shape effects, polarization. Analytical and numerical studies).

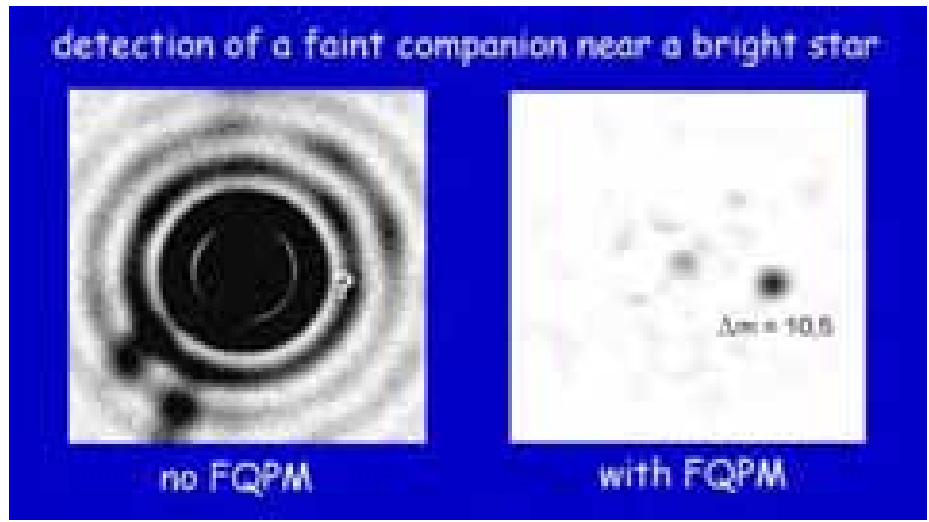

**Figure 19: 4QPM test in the lab at visible wavelength (monochromatic). Left: Image of the PSF. Right: Attenuated image of the on-axis target with a companion 10.5 mag fainter at 3λ/D.**

- LAOG (D. Mouillet): instrumental modeling and performance analysis for a ground-based coronograph.
- LISE (Antoine Labeyrie): conceptual studies on densified interferometer.
- MPIA (Markus Feldt): simulation for VLT Planet Finder.

#### 8.2.2.2. Testbeds
- LUAN (Lyu Abe): coronographic testbed.
- LESIA (P. Riaud): visible, IR coronographic testbed, testing with AO telescopes. MIRI/JWST coronograph qualification and test bench.
- OCA (Y. Rabbia): visible, IR coronographic testbed, testing with AO telescopes.

### *8.3. Modeling Issues*

### 8.3.1. Fidelity

Fraunhofer propagation handles only errors seen at pupils and focal planes; to propagate errors from other surfaces such as secondary and tertiary mirrors etc., we need surface-to-surface Fresnel propagation. Such codes are available, but there are risks of numerical errors against which the analyst must be vigilant.

Early treatment of pupil and phase masks is oversimplified, using scalar fields, and masks with zero thickness and no accompanying phase profile. Analysis for realistic masks (whose thickness and gap widths are comparable to the wavelength) needs rigorous vector field propagation. First results of that analysis [13] show phase profiles and polarization dependences much larger than the tolerances we think are needed. Both the models of these non-ideal behavior and the requirements we impose would benefit from further study.

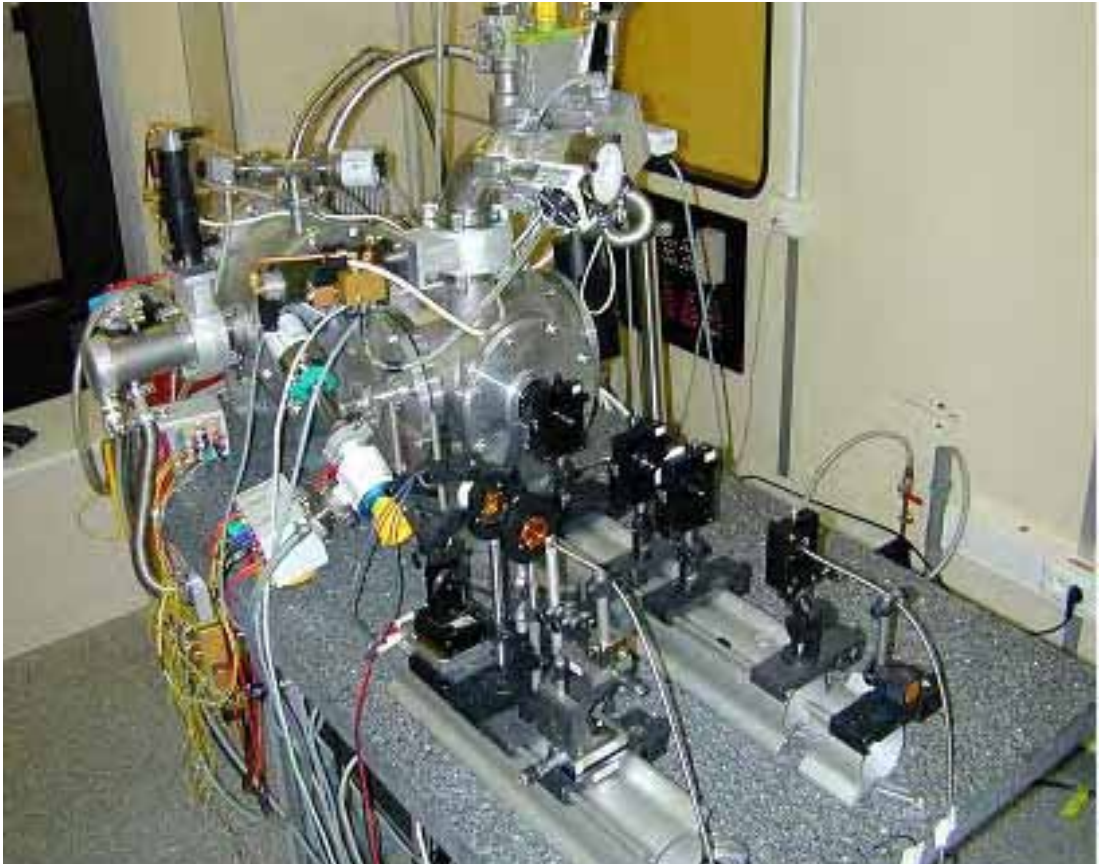

**Figure 20: Coronograph testbed for MIRI/JWST at Observatoire de Meudon/LESIA. The central cryostat (10K) contains the 4QPM. The fore-optics simulates the telescope beam and is fed with a fiber. The coronograph is now optimized for a wavelength of 4.75μm and is presently limited by the spectral resolution (R=10) and the angular size of the source.**

Polarization effects can arise in two main areas: from non-ideal behavior in masks, and from non-normal incidence on mirror coatings. For the latter, current thinking is to try augmenting both ray tracing and wave propagation techniques with physics-based polarization propagation using typical properties of coatings. Both approaches could be implemented easily enough with software written from scratch, but we are also investigating which commercial packages might provide this feature already.

During the workshop, the question was raised whether we need "full fidelity" models which can incorporate the effects of all relevant physics and all possible manufacturing errors. Instead, can we hope to achieve the required ~$10^{-10}$ contrast by empirically adjusting the DM, without a truly accurate understanding captured in models? We probably can, but it may come at a cost of reduced optical bandwidth.

But contrast is not the only issue; variations in speckle brightness are also important [14]. We expect to rotate the telescope around the line of sight to aid in distinguishing planets from speckles. We also expect to find the optimum setting of the DM based on observing changes in the speckles after applied changes in the DM. In both cases, we rely on very precise *knowledge* or *stability* of the pixel contrast during the long exposures and rotation. We will budget the contrast variations based on sensitivities calculated by the models. So how accurate must our contrast models be to allow accurate and reasonable budgeting of variations? This is a much more complicated question, and we don't know the answer yet.

## 8.3.2. Run-Time Speed

It is easy to construct a model with high enough fidelity that no ordinary computer can run it in a reasonable time. It is much more valuable to have a model which can produce results in a day or less. Of course one would include efficient run-time code as a key element in achieving that goal, but a more effective approach is to use appropriate simplification for answering each question.

Most modeling activities can be described as answering a particular question. For most questions, we don't need high fidelity in more than one or two parts of the model. The key is to answer each question with a model of appropriate fidelity.

For example, simple questions of low-order aberration sensitivity to telescope distortions could be answered by a simple ray trace optical model. To understand the basic effects on star leakage to the final science CCD, Fraunhofer propagation might well suffice. To add the effects of beam walk across surface errors on the secondary, tertiary, and other mirrors, Fresnel propagation is needed. To discuss the sufficiency of binary masks, vector field propagation is needed. To understand the interacting effects of multiple-mirror surface errors, attitude variations, and pointing control with a fine steering mirror, an integrated model with optics (Fresnel propagation) and controls is needed. To study pointing errors and vibration propagation through the structure, a full optics-structures-controls model is needed, but the optical model need not be rigorous vector field propagation through the entire system.

Another common shortcut is to use pre-computed sensitivities instead of a full optical propagation. This saves time by linearizing the treatment of small disturbances, and also vastly simplifies the run-time arithmetic.

For the earliest modeling efforts, it was obvious to try using standard commercial analysis tools (NASTRAN, Code-V, Zemax, etc.) as part of the run-time execution of an integrated model. For each instant in time, we re-compute the disturbed structure and optics, estimate sensor outputs, and repeat. This turned out to be impractical, because those commercial tools were never designed for efficiency in an iterative computation like that. Alternate tools have been developed in the US, at JPL and Goddard and elsewhere, to provide quick iteration times. Only recently have some commercial providers made specific efforts to provide tools for fast run-time exercise of their optical system models. Some of these solutions may now be commercially available.

## 8.3.3. Model Verification

The more we come to rely on model predictions to make decisions, the more we need verification that they are correctly representing reality. The issues to be addressed are both correct design ("wiring") of the model and correct parameter values within them.

Testbeds are the major tool for verifying models. We trust a model to the extent that we can do an experiment in the laboratory and duplicate its results with our model. The more the experiment matches the conditions we hope to achieve on orbit (such as the nominal $10^{-10}$ contrast), the better our confidence that no undiscovered physics has been omitted from our model. So in a sense the model is simply a tool for allowing a modest extrapolation of laboratory results to space using physics-based scaling.

In some cases we can compare model results to those from trusted commercial software, such as NASTRAN or Code-V. However, home-built codes from JPL and Goddard are in many cases more advanced than the "heritage" parts of those programs. Thus the com-parison is between new codes written by different groups, not between code to be verified and heritage code which has earned widespread trust.

## *8.4. References for Section 8*

# 9. Summary, Opportunities for US-European Collaboration, Recommendations

In this Section we summarize the status of coronograph development on both sides of the Atlantic, and the opportunities that we see for fruitful collaboration in the future.

## 9.1. Summary of the TPF-C Technology Program and Areas of European Interest

### 9.1.1. Status of TPF Coronograph effort

Most of the recent progress in coronography in the US has been made through the TPF Coronograph technology program, which funds the R&D necessary for the mission. The work being conducted at NASA centers, industry, and universities is proceeding on a broad front covering nearly all important areas:

1) High Contrast Imaging Testbed
   a) Contrast of $6\times10^{-9}$ (to date)
   b) Remote participation of scientists to control test bed
2) Masks and Stops
   a) HEBS glass fabrication success
   b) Alternates to HEBS glass
   c) Metrology testbeds
3) Primary Mirror Studies
   a) Materials studies
   b) Design study
   c) Metrology study
   d) Coating study
   e) Facilities limits
4) Visible Nulling Testbed
5) Integrated modeling
   a) Turn-key process to input performance parameters and output contrast
   b) Tool development performance
6) University Studies
   a) Vector wave front simulation – UC Berkeley
   b) Shaped Pupil Coronograph – Princeton University
   c) Pupil remapping coronograph – NOAO and University of Hawaii
   d) Vector optical modeling of coronograph system - GSFC
   e) Fabrication of MEMS Deformable for Visible Nulling –Boston University
   f) Fabrication of single mode fiber array for Visible Nulling – Penn State University

7) Observatory development and performance analysis

   a) Design concept

   b) Structural Analysis

   c) Thermal Analysis

   d) Attitude Control Systems

### 9.1.2. Areas of European Interest

European activities in coronography are not focused on a single instrument or mission, but rather cover a range of ground-based and space-borne instruments, as well as related applicable technologies:

1) High Contrast Imaging Testbed

2) Laser and other optical metrology

3) Mirror technology

   a) Manufacturing and polishing of large monolithic mirrors

   b) Alternative low-expansion materials

   c) Silicon carbide space-compatible mirrors

   d) Computer-driven mirror polishing, ion-beam figuring

   e) Active control of large monolithic mirrors; actuators for control loops

4) Dichroic beam splitters

5) Manufacture of masks, including achromatic phase-shifters, using semiconductor industry (ion and electron etching, lift-off, multi-layer deposition) techniques

6) Polarization preservation with ZOG or achromatic half-wave plates, and monomode fiber optics

7) Spatial filtering using monomode fiber optics on interferometric instruments

8) Stray light control, partly through ultra-smooth polishing

9) Spectroscopy of faint objects, especially on systems using Integral Field Unit (IFU) spectrometers (with slicer, fiber optics); also filters and wavelength-sensitive detectors

10) Optical Modeling

11) Temperature sensors

## *9.2. Opportunities for US-European Collaboration*

Whereas no centrally funded TEC technology program exists in Europe, a wide range of topics with direct relevance to TEC are currently being addressed by European research groups. In many cases, the expertise available in Europe is complementary to that in the US, and a number of European research institutes and companies are eager to contribute to the development of space coronography.

Cooperation between space agencies, universities, and industry is essential for an efficient and rapid development of TEC. Such collaborations have been established quite successfully during the past few years in the US; several TPF study teams funded by NASA in-clude European collaborators working at US institutions, European research institutions, or both. Results from research on coronography conducted at European institutions for ground-based applications and JWST can be integrated into the TPF-C technology roadmap if appropriate links are established. The present workshop is the first systematic step in this direction; similar follow-up meetings and bi-lateral collaborations can lead to a further cross-fertilization of the European and US research efforts.

The quest for studying habitable extrasolar planets is a common theme for the space agencies on both sides of the Atlantic. Even if they follow different strategies for the implementation of their first missions dedicated to this topic, there is a need for close scientific cooperation. Assuming that NASA will proceed first with a coronograph and ESA with an interferometer, it appears quite plausible that both missions could fly at nearly the same time in the middle of the next decade. Optimizing the return from both missions would then require some common planning, both scientifically and technically. Since combining visible and mid-IR data enables a much more detailed analysis of the planetary properties (see Section 3) than either wavelength range on its own, one would clearly want to maximize the overlap between the two missions in the targets than can be studied, and perhaps also in the time coverage for individual systems. This may require co-optimization of the inner and outer working angles, Sun exclusion angles, observing strategies, and other spacecraft and mission parameters.

## *9.3. Recommendations*

Suggestions for further work and recommendations to the space agencies have been interspersed throughout this document in the context of the discussions of individual topics. Here we summarize the most important top-level recommendations to NASA and ESA made by the workshop participants:

- Continue scientific studies of the prospects for coronographic characterization of terrestrial planets on both sides of the Atlantic. More detailed modeling of terrestrial planets will be needed to assess the reliability of biomarkers, the possible role of polarimetry and many other important issues for the formulation of the top-level requirements and for the interpretation of data from a TEC.

- Exchange technical information as freely as possible. The advisory groups to NASA and ESA (in particular, the TE-SAT and TPF SWG) have to be kept fully informed about the expected capabilities, schedules, difficulties, and risks of the relevant projects (currently TPF-C, TPF-I, and Darwin). The present practice of cross-appointing members between TE-SAT and TPF SWG should be continued; occasional joint meetings of the full TE-SAT and TPF SWG would also be desirable.

- Establishing a joint European-US technical working group could help identify specific opportunities for collaboration, facilitate the exchange of relevant technical information, and promote the use of European technologies for TEC.

- Continue the Darwin/TPF science conference series as an annual event promoting US-European collaboration in the area of terrestrial planet characterization.

# 10. Acknowledgements

The workshop organizers and participants would like to express their sincere thanks to the Lorentz Center at Leiden University and its staff, in particular Dr. Martje Kruk-de Bruin and Yolande van der Deijl, for their hospitality and the flawless organization of the meeting. The US participants were supported by NASA JPL, and ESTEC generously hosted a detailed tour of their facilities during the workshop.

# 11. Appendices

## 11.1. List of Workshop Participants

| | | |
|---|---|---|
| Claude Aime | (Nice, France) | Claude.Aime@unice.fr |
| Marc Barillot | (Cannes-La-Bocca, France) | marc.barillot@space.alcatel.fr |
| Pierre Baudoz | (Meudon, France) | Pierre.BAUDOZ@obspm.fr |
| Anthony Boccaletti | (Meudon, France) | Anthony.boccaletti@obspm.fr |
| Pascal Bordé | (Cambridge, MA, US) | pborde@cfa.harvard.edu |
| Christopher Burrows | (Edmonds, US) | chrisatmd@aol.com |
| Mark Clampin | (Greenbelt, MD, US) | clampin@stsci.edu |
| Daniel Coulter | (Pasadena, CA, US) | daniel.r.coulter@jpl.nasa.gov |
| Michael Devirian | (Pasadena, CA, US) | Michael.Devirian@jpl.nasa.gov |
| Peter Doel | (London, UK) | apd@star.ucl.ac.uk |
| Klaus Ergenzinger | (Friedrichshafen, Germany) | Klaus.Ergenzinger@astrium.eads.net |
| Virginia Ford | (Pasadena, CA, US) | Virginia.g.ford@jpl.nasa.gov |
| Malcolm Fridlund | (Noordwijk, Netherlands) | Malcolm.Fridlund@esa.int |
| Joseph Green | (Pasadena, CA, US) | joseph.j.green@jpl.nasa.gov |
| Alan Greenaway | (Edinburgh, UK) | a.h.greenaway@hw.ac.uk |
| Olivier Guyon | (Hilo, HI, US) | guyon@subaru.naoj.org |
| Sara Heap | (Greenbelt, MD, US) | sara.r.heap@nasa.gov |
| Tony Hull | (Pasadena, CA, US) | Tony.Hull@jpl.nasa.gov |
| Lisa Kaltenegger | (Noordwijk, Netherlands) | Lisa.Kaltenegger@esa.int |
| Anders Karlsson | (Noordwijk, Netherlands) | Anders.Karlsson@esa.int |
| N. Jeremy Kasdin | (Princeton, NJ, US) | jkasdin@princeton.edu |
| Steven Kilston | (Boulder, CO, US) | skilston@ball.com |
| John Krist | (Baltimore, MD, US) | krist@stsci.edu |
| Marc Kuchner | (Princeton, NJ, US) | mkuchner@cfa.harvard.edu |
| Charles Lillie | (Redondo Beach, CA, US) | chuck.lillie@ngc.com |
| James Lloyd | (Pasadena, CA, US) | jpl@astro.caltech.edu |
| Richard Lyon | (Greenbelt, MD, US) | Richard.g.lyon@nasa.gov |
| Frantz Martinache | (Saint Michel, France) | martinache@obs-hp.fr |
| Charley Noecker | (Boulder, CO, US) | mcnoecke@ball.com |
| Alan Penny | (Didcot, UK) | ajp@astro1.bnsc.rl.ac.uk |
| Andreas Quirrenbach | (Leiden, Netherlands) | quirrenb@strw.leidenuniv.nl |
| Yves Rabbia | (Grasse, France) | rabbia@obs-azur.fr |
| Pierre Riaud | (Liège, Belgium) | riaud@astro.ulg.ac.be |
| Huub Rottgering | (Leiden, Netherlands) | rottgeri@strw.leidenuniv.nl |
| Daniel Rouan | (Meudon, France) | daniel.rouan@obspm.fr |
| Jean Schneider | (Meudon, France) | Jean.Schneider@obspm.fr |
| Michael Shao | (Pasadena, CA, US) | mshao@huey.jpl.nasa.gov |
| Anand Sivaramakrishnan | (Baltimore, MD, US) | anand@stsci.edu |
| Rémi Soummer | (Baltimore, MD, US) | soummer@stsci.edu |
| Karl Stapelfeldt | (Pasadena, CA, US) | karl.r.stapelfeldt@jpl.nasa.gov |
| Stephen Todd | (Edinburgh, UK) | spt@roe.ac.uk |
| Volker Tolls | (Cambridge, MA, US) | vtolls@cfa.harvard.edu |
| Wesley Traub | (Cambridge, MA, US) | wtraub@cfa.harvard.edu |
| John Trauger | (Pasadena, CA, US) | john.t.trauger@jpl.nasa.gov |
| Zlatan Tsvetanov | (Washington, DC, US) | Zlatan.Tsvetanov@nasa.gov |
| Robert Woodruff | (Boulder, CO, US) | robert.a.woodruff@lmco.com |
| Natalia Yaitskova | (Garching, Germany) | nyaitsko@eso.org |

## *11.2. Specific Coronograph Designs*

Table 13 summarizes the characteristics of some leading coronograph designs. Most of these designs are strictly DLSSs, in which the presence of a WFCS is implicit. Here we define these characteristics.

*Diffracted Light Suppression System (DLSS):* An optical design that dramatically reduces the diffracted light of a star at the position of a nearby planet, compared to the diffraction that would be present from the shape of the pupil alone.

*Fundamental Properties:*

*Theoretical Maximum Extinction:* The extinction of a monochromatic on-axis point source provided by the DLSS given a perfect wavefront.

*Inner Working Angle (IWA):* The closest a planet can be to a star and still be conveniently detectable using a particular Diffracted Light Suppression System. Rather than trying to define convenience, we quote the throughput of each device at the quoted IWA.

*Throughput at Inner Working Angle:* The detectable fraction of the flux from a planet located at the inner working angle. An ordinary telescope with no coronograph is considered to have unity throughput on this scale.

*Throughput Outside Inner Working Angle:* The detectable fraction of the flux from a planet located in the search area far from the inner working angle.

*Outer Working Angle:* The farthest a planet can be from a star and still be conveniently detectable. The outer working angle of most DLSSs is limited by classical optical design constraints. In this case, the outer working angle is "large". In a few cases, the DLSS concept itself imposes a smaller limit to the outer working angle.

*Achromaticity:* All DLSSs need to be able to work across a broad band pass. Some designs aim to achieve broadband diffracted light control using special materials with intrinsic chromatic properties (e.g., 4QPM coronograph). This row flags these devices, which potentially face extra manufacturing problems. For example, phase masks rely on chromatic optical materials, while the AIC coronograph is achromatic by principle.

*Planet PSF FWHM:* The Full Width at Half Maximum of the Point Spread Function (PSF) of the image of the planet. Except in coronographs with band-limited or notch filter masks (and the visible light nuller), the PSF shape varies with the location of the planet. But usually the core of the planet's PSF has a reasonably uniform width for planets outside the inner working angle.

Table 13: The most important characteristics of specific coronograph designs.

Although to first order the shape of the image is not crucial, any extension of the image decreases the contrast between planet images and residual host flux, making both detection and spectral analysis harder. The effective contrast ratio achieved is reduced by the number of times the image size exceeds that of the diffraction-limited image. This parameter also indicates the sensitivity of the coronograph to exozodiacal light; when the image quality is poor, the coronograph detects relatively more flux in each pixel from this extended source.

*Search Space / Useable FOV:* Most DLSSs block the light in some parts of the image plane beyond the vicinity of the star. The numbers in this row indicate roughly how much of the area in the image plane between the IWA and OWA is useable for planet hunting.

*Number of Telescope Roll Positions Needed:* When the Search Space is not 100%, rolling the telescope (or perhaps just the mask) can provide access to the blocked part of the image plane. This row suggests how many roll angles need to be sampled in order to search the whole annulus between the IWA and the OWA.

*Comments about Off-Axis Image Quality:* Most DLSSs in this table produce good off-axis images interior to their OWAs, but some require comment.

*Double / Field Star Compatibility:* Some DLSSs can easily be adapted to block the light from two stars at once – usually at the cost of some search space.

*Sensitivity to Pointing / Stellar Size:* The ability of the DLSS to suppress starlight given pointing errors and stars with large angular diameters. When possible, we have indicated the order of the null. Second order nulls are probably inadequate for terrestrial planet finding. Fourth order nulls present a pointing challenge. Devices that create nulls of $8^{th}$ order and higher are highly robust. Relatively little is usually known about sensitivity to other low order aberrations; see also the discussion in Section 6.2.2.

*Photometric Efficiency:* The product of the numbers in the "Throughput Outside Inner Working Angle" row and the "Search Space / Useable FOV" row. For the planet search mode, this quantity summarizes the overall search efficiency.

*Sensitivity to Telescope Reflectivity / Transmission:* How robust is the DLSS to moderate (~ 0.1%) amplitude errors across the pupil?

*Sensitivity to Red Leak:* Some DLSSs only work shortward of some fiducial wavelength. Starlight at longer wavelengths must be removed by a filter. This row flags DLSSs with this potential drawback.

*Telescope Pupil Shape:* Some DLSSs are only compatible with some telescope pupil shapes. This row flags those DLSSs.

*Compatible with Segmented / Diluted Pupil:* Can the DLSS, or a modified version of it, work with a segmented primary mirror?

*Compatible with On-Axis Telescope Design:* The secondary mirror and its support structure block some of the primary mirror in an on-axis telescope. Can the DLSS, or a modified version of it, work effectively with such a system?

*Compatible With Active Speckle Suppression:* Ability to use the final signal output (e.g. the final coronographic image) to actively control the wavefront in order to suppress residual speckles.

*Mask Fabrication Issues:*

*Feature Size:* If a coronograph needs a mask with "hard" edges (phase or amplitude), the manufacturing tolerance on the placement of the edge will always be approximately $f\lambda/3500$ for image masks (in the illumination region) and $D/3500$ for pupil masks (Kuchner & Spergel 2003). The "smoothness" of the edge is also a concern.

*Amplitude Control:* When an amplitude mask is needed, the control of transmission or reflectivity of the mask is critical.

*Phase Control:* When a phase mask is needed, the control of the phase shift of the mask is critical.

*Scattering Control:* When a mask is needed it should not be scattering light (very high order aberrations).

*Mask Thickness:* Sensitivity to edge effects in the mask.

*Aspheric Mirror:* Is there a need for non-standard shaped optics (aspherics) that need to be polished to a high accuracy?

*Beam Splitter:* Common technological challenges for beam splitters are spatial uniformity, multiple reflections, and achromaticity.

*Technological Maturity:*

*Simulation Maturity:* How well has the basic concept been simulated for the goals of terrestrial planet detection and characterization?

*Subsystems Lab Demonstration:* Have the critical subsystems of the DLSS been demonstrated in the lab?

*Integrated Lab Demonstration:* Has a fully working DLSS been tested in the lab?

*On the Sky Demonstration:* Has the DLSS been used for ground-based observing?

The remainder of this section gives an overview of the most important features, strengths, weaknesses, and areas in need of more technology development for individual coronograph concepts. The references refer to Section 11.5.

## 11.2.1. Visible Nuller

- Family: interferometric coronograph
- Visible version of angel cross but with fiber bundle to spatially filter high spatial frequency errors.
- Nulling is done in the pupil plane by 4 overlapped pupils
- IWA: 1.5 $\lambda/D$, minimum IWA
- Throughput: 100% of planet light in 25% of the sky; equivalent Lyot plane throughput 45% @3$\lambda/D$

**Strengths:**

- Experimental null demo to date $5\times10^{-6}$, which theoretically converts to $5\times10^{-9}$ with a 1000-fiber bundle in a following stage
- Explicit amplitude and phase control

*Weaknesses:*

- Amplitude control limited to ~25% BW (current scheme), a more achromatic amplitude scheme would improve the technique

*Technologies and study areas:*

- Fiber bundle / lenslet array
- DM (segmented, tip/tilt piston)

## 11.2.2. Achromatic Interferometric Coronograph

- Family: interferometric coronograph
- References: Gay & Rabbia 1996, [29]; Baudoz, Rabbia, & Gay 2000, [10]
- Entrance pupil: unapodized centro-symmetric aperture
- Splits light interferometrically, produces destructive interference by a through-focus achromatic π phase shift
- Throughput: 50%
- IWA: less than λ/D
- OWA: full FOV

*Strengths:*

- Achromatic
- Total rejection in theory

*Weaknesses:*

- Symmetrizes the images (i.e., each companion gives two images)

*Technologies and study areas:*

- Beam splitter properties (intensity matching, thickness, material properties)

## 11.2.3. Four-Quadrant Phase Mask, Phase-Knife Coronograph

- Family: Roddier phase mask coronograph
- References: Abe, Vakili, & Boccaletti 2001, [2]; Rouan et al. 2000, [72]; Riaud et al. 2001, [69]; Riaud et al. 2003, [67]
- Entrance pupil: unapodized preferably unobscured circular
- Image stop: π phase mask in two diametrically opposed quadrants; the phase mask can be either transmissive or reflective
- Lyot Stop: clear everywhere, undersizing may be needed to reduce effects of aberrations
- Throughput: 100% for unobscured pupil (penalty otherwise)
- IWA: λ/D
- OWA: full FOV

*Strengths:*

- Total rejection in theory, contrast of $10^{-6}$ at 3λ/D demonstrated experimentally to date (without wavefront control)

*Weaknesses:*

- Chromatic dependence for the phase shift only
- Blind areas in image field
- Sensitivity to classical circular central obstruction

*Technologies and study areas:*
- Phase mask fabrication in visible / near-IR / mid-IR (monochromatic)
- Achromatization issues in visible / near-IR

### 11.2.4. Apodized Phase Mask
- Family: Roddier phase mask coronograph
- References: Aime, Soummer, & Ferrari 2001, [5]; Soummer et al. 2002, [83]
- Entrance pupil: apodized aperture, transmission 80%
- Image stop: central $\pi$ phase mask
- Lyot stop: clear everywhere, slight undersizing may be needed to reduce effects of some aberrations
- IWA: $\lambda/D$
- OWA: full FOV

*Strengths:*
- Total rejection in theory
- Obscured or unobscured aperture

*Weaknesses:*
- High chromatic dependence (both in size and phase shift)

*Technologies and study areas:*
- Phase mask fabrication
- Entrance pupil apodization
- Polarization effects in mask

### 11.2.5. Apodized Two-Zone Phase Mask
- Family: Roddier phase mask coronograph
- Reference: Soummer, Dohlen, & Aime 2003, [84]
- Entrance pupil: apodized aperture, transmission 60%
- Image stop: central phase mask, surrounded by a second annular phase mask; the two phase shifts and diameters are selected to optimize spectral bandwidth; the phase mask can be either transmissive or reflective, and close to (but not at) focus
- Lyot Stop: clear everywhere, slight undersizing maybe needed to reduce effects of some aberrations
- IWA: $\lambda/D$
- OWA: full FOV

*Strengths:*
- Low chromatic dependence
- Obscured or unobscured aperture
- Can be optimized for two separate passbands

*Weaknesses:*
- Rejection depends on spectral bandwidth
- Rejection limited for wide band pass ($R = 2$)

*Technologies and study areas:*
- Phase mask fabrication
- Entrance pupil apodization
- Polarization effects in mask

## 11.2.6. Band-Limited Coronograph
- Family: Lyot coronograph
- Reference: Kuchner & Traub, 2002, [44]
- Entrance pupil: unapodized aperture
- Image stop: field strength multiplier $(1 - m(x))$, using a band-limited function $m(x)$, with a bandpass = $b \sim D/n$ ($n \sim 5\ldots10$), and with $m(0) = 1$ (opaque at center).
- Lyot stop: clear everywhere except within $b$ of any pupil boundary; undersizing can reduce effect of some aberrations
- IWA: set by $b$
- OWA: full FOV
- Possible use with centrally obscured geometry, but with poor throughput

*Strengths:*
- Quartic dependence on tilt error
- Slightly achromatic behavior

*Weaknesses:*
- Trades IWA for throughput

*Technologies and study areas:*
- Apodized mask fabrication
- Polarization, phase effects in mask

## 11.2.7. MAPLC / Hybrid
- Family: Lyot coronograph
- References: Aime, Soummer, & Ferrari 2002, [6]; Soummer, Aime, & Fallon 2003, [82], Soummer (2005)
- Entrance pupil: apodized, preferably unobscured aperture
- Image stop: hard edge opaque spot
- Lyot stop: identical to the entrance aperture, undersizing can reduce effect of some aberrations
- Throughput: 30% to 60% depending on mask size and central obstruction
- IWA: 2 to 4 $\lambda/D$
- OWA: full FOV

*Strengths:*
- Arbitrary high rejection in theory
- Multistage application

*Weaknesses:*
- Medium chromatic dependence (size of the PSF)

## 11.2.8. Continuous Apodization

- Family: continuous apodized pupils
- Entrance pupil: smoothly apodized, either radially or in a square (rectangle)
- Modifies the point spread function to concentrate the light in the central lobe (or strip) and lower the halo to that of the planet irradiance
- Image stop: hard edged to prevent scatter of starlight
- Throughput: 10 to 30%
- IWA: 3 to 4 $\lambda/D$
- OWA: from 20 $\lambda/D$ to full FOV

*Strengths:*
- Single optic
- Insensitive to pointing or stellar size
- No chromaticity

*Weaknesses:*
- Very difficult to manufacture
- Poor for obscured, on-axis designs
- Potentially low throughput
- Large planet PSF
- Phase loss in mask limits performance

*Technologies and Study Areas:*
- Mask fabrication
- Polarization effects

## 11.2.9. Binary Apodization

- Family: shaped pupils
- Entrance pupil: binary hard edged mask with multiple openings, either circular or square
- Modifies the point spread function to concentrate the light in the central lobe (or strip) and lower the halo to that of the planet in the search space
- Image stop: hard edged to prevent scatter of starlight
- Throughput: 10 to 40%
- IWA: 2 to 4 $\lambda/D$
- OWA: 20 $\lambda/D$ to full FOV

*Strengths:*
- Extremely inexpensive and easy to manufacture
- No chromaticity
- Insensitive to pointing or stellar size
- Less sensitive to aberrations
- Can be designed for on-axis (obscured) apertures (though at throughput and / or IWA penalty)

*Weaknesses:*
- Potentially low throughput
- Large planet PSF

- Unknown interactions of field with mask edges (polarization)
- Unknown scatter from mask edges
- Sensitivity to speckle noise outside OWA

*Technologies and Study Areas:*
- Mask fabrication
- Edge effects
- Polarization and vector field effects

### 11.2.10. Phase-induced Amplitude Apodization, Pupil Mapping
- Family: pupil apodized coronograph
- Reference: Guyon 2003, [36]; Traub & Vanderbei 2003, [86]
- Entrance pupil: unapodized unobstructed aperture
- First mirror to redistribute light
- Second mirror to correct phase
- IWA: 2.5 $\lambda/D$
- OWA: full FOV

*Strengths:*
- Slightly achromatic behavior

*Weaknesses:*
- Off-axis image quality

*Technologies and study areas:*
- State-of-the-art optics
- Polarization

## *11.3. A Few Mathematical Notes*

Mask/Stop DLSSs can be described using three functions:

      $L(u)$, the Lyot stop amplitude transmission function;

      $A(u)$, the entrance pupil amplitude transmission function;

      $M(u)$, the Fourier transform of image mask amplitude transmission function.

Suppressing on-axis light requires

      $L(M*A) \sim 0$ in search area (between IWA and OWA) .     (2)

This formalism can also describe some schemes that don't use masks and stops, like the visible light nuller and the infrared nulling interferometer, and it is relevant to all DLSS designs.

Most well studied mask/stop DLSSs fall into two categories: those with tophat entrance apertures, $A(u) = tophat(u)$, and those with tophat image masks ($M(x) = 1 - C\,tophat(x)$), where C is usually 1 or 2. The tophat function $f(x) = 1$ for $|x| < 1/2$, $f(x) = 0$ elsewhere. The general problem of designing useful *hybrid* coronographs, those without at least one tophat function, has not been studied. One hybrid solution is the Dual Zone phase mask (Soummer et al. 2002).

For tophat entrance pupils, the exact solutions (those with $L(M*A)=0$) are notch-filter functions, functions with no power over some range of mid-spatial frequencies. For top-hat image masks, the exact solutions are notch-filter entrance pupils. Only the low-wave number part of the notch filter function (the band-limited part) affects the light in the final image plane – the high wave number components direct light onto a Lyot stop or onto an image stop.

While a variety of band-limited image masks are buildable in principle, band-limited pupil masks are not, because the telescope primary has finite size. But one band-limited function, the prolate-spheroidal wave function (Slepian 1965), turns out to be quite useful for designing pupil masks, in part because it minimizes the uncertainty product (core width times core width of its Fourier transform) for band-limited functions (Aime, Soummer, & Ferrari 2002; Kasdin et al. 2003) and pupil mappings (Guyon 2003; Traub and Vanderbei 2003). The Kaiser function is a very good approximation to the truncated prolate spheroidal wave function. The exact calculation of a prolate spher-oidal wave function was recently implemented by P.E. Falloon in Mathematica (see Aime et al. 2002 for linear prolates and Soummer et al. 2003 for circular prolates).

The 1-dimensional solutions to Equation (1) can be used to generate separable mask/stop coronographs, i.e., those using masks and stops of the form A(x)B(y), where x and y are Cartesian coordinates (Vanderbei, in prep.). An example of a non-separable 2-D DLSS is the 4-quadrant phase mask. The non-separable 2-D problem is unsolved in general, and may contain additional useful solutions.

## *11.4. Optical Modeling*

### 11.4.1. Foundations

The optical field, in its most general form, is a vector field that consists of both electric and magnetic field components transverse to the direction of propagation. The vector nature of light is usually represented by its polarization state which can change upon reflection and transmission through components such as mirrors, lens, and occulting stops. Optical field propagation through proposed TPF/Darwin coronographic and interferometric architectures have to date been treated by ignoring this vector nature, i.e., treating the optical fields as scalar entities. The full implications of this for TPF/Darwin are as yet not understood.

TEC requires contrasts of $10^{-10}$ in intensity, or ~ $10^{-5}$ in the electric field. Thus any theory which is used for the design, optimization and modeling of the optical structures for TEC ideally should be accurate to better than $10^{-5}$ in electric field. The question remains as to the accuracy of the scalar diffraction theory and under what conditions, if any, do we need invoke the full vector theory. The issue has been raised that if the field's amplitude and phase can be accurately controlled, in closed-loop, then are there enough degrees of freedom to bring the contrast to $10^{-10}$ without necessarily understanding why the contrast is at this level? For example, the HCIT testbed is already approaching $10^{-9}$ in monochromatic light. This is really a question of sensitivity, i.e., are the locations of the minimum in the curve of contrast versus an error source the same for the scalar and vector theories? And if not, does it matter? While not understanding the details of the theory is certainly a viable option for a laboratory testbed it is not an option for the full flight architecture. It would be remiss not to fully understand the details of any theory applied to the design, development, testing, and deployment of TEC. It may indeed be true that we need not invoke the full theory but a justification for this decision is needed.

A TEC would require detection of a terrestrial planet which is ~$10^{-10}$ times the central stellar source. Ideally the in-pixel contrast ratio, i.e. the ratio of planetary flux to stellar diffracted/scattered flux, collected within a pixel, should exceed unity or approach a level where background subtraction is viable. A number of coronographic methods have been analyzed [1, 2], and a multitude of other approaches are in the theoretical and/or development stage. Many of these appear to be viable candidate approaches for TEC provided that wavefront errors can be lowered to less than l/10,000 rms in the critical spatial frequency range of 3 – 30 cycles per aperture. However, the modeling and simulation based results are founded upon scalar diffraction theory and hence non-rigorous, thus, the theory which has been used to design and to analyze these approaches may or may not be accurate enough. This may yield a design which would, in the framework of scalar diffraction theory, yield a $10^{10}$ contrast but when fabricated may be found to be lacking. We are currently developing full vector propagation methods to evaluate the errors induced in ignoring the vector nature of the light and ultimately will determine the relevance to the design of ground testbed work and ultimately TEC. These models will be validated against the results from ground testbed work.

## 11.4.2. Scalar Diffraction Theory

Scalar diffraction theory refers to a set of approximations which are used either explicitly or implicitly in the evaluation of diffraction for an optical system. These approximations are:

<u>Fields are vector, not scalar fields</u> – This tends to hide the effects due to polarization since at each reflection the polarization state of the field can change, and propagation through occulting masks and pupil plane masks will introduce amplitude and phase shifts which will change the polarization state and ultimately the contrast.

<u>Kirchhoff approximation</u> – This assumes that all masks and apertures are infinitely thin perfect conductors. In reality masks have a finite thickness with a shaped edge. The net effect of this approximation is to ignore effects induced by the beveling of the mask edge, small-scale guiding effects, and evanescent modes and leakage induced by finite conductivity (imaginary part of the permittivity) and dielectric constant (real part of permittivity). These effects are typically negligible for most space imaging systems but may become important for occulting masks and shaped pupils.

<u>$2^{nd}$ order in phase</u> – The Green's function propagator's phase is expanded to only $2^{nd}$ order in phase (Fresnel approximation), thus, all spherical waves are treated as parabolic waves. This is a reasonable assumption for slow systems (large F/#) used nearly on-axis (paraxial) which is typical of TEC, however some errors are introduced due to this paraxial approximation.

<u>Aberrated pupil analysis</u> – In much of the analysis to date only single plane diffraction is used. All phase and amplitude errors due to residual design aberrations, misalignments, deformations, and mid- and high-spatial frequency errors are assumed to come from a single plane (pupil) within the optical system.

<u>Numerical effects</u> – In order to develop models and simulations the above approximate theory is relegated to the discrete domain (computers), thus errors are introduced due to discrete sampling, floating point representation, Fast Fourier Transforms, aliasing, programming errors etc.

### 11.4.3. Diffraction Theory

In free space, away from sources and matter, the spatial components of the monochromatic electric field, $\vec{E}(\vec{r})$, are completely described by the vector spatial Helmholtz equation:

$$\nabla^2 \vec{E}(\vec{r}) + k^2 \varepsilon_0 \vec{E}(\vec{r}) = 0 \quad . \tag{3}$$

$k = 2\pi/\lambda$, $\lambda$ is the wavelength and $\varepsilon_0$ is the free space permittivity. The magnetic permeability is taken here as unity for simplicity. Equation (3) essentially states that electric field is proportional to its local curvature and is mathematically an eigenvalue equation for the Laplacian operator; solutions are representable as a linear superposition of eigenmodes. The optimal choice of modes is generally dictated by the boundary conditions and/or coordinate systems chosen. In free space the spatial Helmholtz equation can be separated into three scalar equations, one equation for each of the vector components $\vec{E}(\vec{r}) = (E_x, E_y, E_z)$ and the free space eigenmodes are vector plane waves. This is the essence of scalar diffraction theory; one specifies the boundary conditions, solves for a single component of the electric field, then calculates the intensity as proportional to the modulus squared $I(\vec{r}) = |E_x(\vec{r})|^2$ of that component. Generally, this scalar theory is further approximated by assuming that all boundaries, such as masks, apertures, mirrors, lenses etc., are infinitely thin and that the edges of a boundary are perfect conductors (Kirchoff approximation). A Green's function, involving the fields and their normal derivatives on the boundaries, can be found which converts Equation (3) to integral form. This Green's function is subsequently approximated to 2$^{nd}$ order in phase (Fresnel approximation) and yields the integral form:

$$E(x, y, z) = \frac{-ie^{ikz}}{\lambda z} \iint E(x', y', 0) T(x', y', 0) e^{-i\frac{\pi}{\lambda z}\left[(x-x')^2 + (y-y')^2\right]} dx' dy' \quad , \tag{4}$$

where the integral is taken over the area of the mask or pupil located at $z = 0$ in the $(x', y')$ plane. $E(x', y', 0)$ is the field over the boundary (incident field), and $T(x', y', 0)$ is the complex transmitting function of the object (e.g. lens, mirror, grating, aperture etc.) within the boundary. Many optical systems can be reduced to evaluation of this Fresnel integral by a clever choice of the transmitting function or can be reduced by successive applications for multiple plane diffraction. The Optical Systems Characterization and Analysis Research (OSCAR) software package essentially reduces an optical system to this form and evaluates it on single or multiple user-specified diffraction planes. With a re-arrangement of the terms in the integrand this integral can be cast into the form of a 2D discrete spatial Fourier transform which can be evaluated numerically by fast Fourier transform techniques. This form can actually be used to evaluate systems with lens, mirrors, gratings, aberrations, misalignments, deformations and to a limited extent stray and scattered light. It generally agrees well with measurement and has become the dominant method of diffraction analysis for TEC and most other systems.

OSCAR has and is currently being used in our TEC studies (Lyon et al. 2002, Woodruff et al. 2002). It is capable of modeling filled aperture, segmented aperture, sparse and interferometric aperture systems, as well as spectrometers and coronographs. It also includes wavefront sensing capability in terms of phase retrieval [4, 7] and phase diversity and some modeling of controls [8].

The vector component level transfer functions will be developed for the masks and occulters using the rigorous model and be subsequently integrated into OSCAR for propagation throughout the system. This will allow us to calculate optical point spread functions (PSF) for an entire optical system with a rigorous model for the components and a Fresnel based propagator for the free space propagations. The results will be compared to propagation through the system with only the Fresnel model.

### 11.4.4. Vector Diffraction Model

The vector electric field in an inhomogeneous medium is governed by the inhomogeneous Helmholtz equation:

$$\nabla^2 \vec{E}(\vec{r}) + k^2 \varepsilon(\vec{r}) \vec{E}(\vec{r}) + \nabla \left( \nabla \ln \varepsilon(\vec{r}) \cdot \vec{E}(\vec{r}) \right) = 0 \quad . \tag{5}$$

Equation (5) represents a homogenous differential equation, but is physically inhomogeneous since the permittivity, $\varepsilon(\vec{r})$, can be a general function of coordinates. Thus, for example, a lens with spherical surfaces would be represented by a change in the permittivity along the curved surfaces, with the free space permittivity being used outside the lens and the square of the index of refraction being used inside the lens. Equation (5) is rigorously derivable from Maxwell's equations without any approximations and thus theoretically represents a starting point for the full vector solutions for shaped apertures, apodizers, occulting stops and fiber bundle wavefront correctors for TEC.

We have developed both a 2D and a full 3D vector solver for the inhomogeneous Helmholtz equation. The 2D solver is fully parallelized and can handle problems on up to ~8000 × 8000 grid sizes on a 1024 processor Beowulf cluster. It employs sparse matrix methods and an iterative bi-conjugate solver with pre-conditioning for rapid convergence. Our 3D solver rigorously enforces the condition that $\vec{\nabla} \cdot \vec{D} = 0$ by employing a tetrahedral mesh structure. This has been used to date to model photo-refractive polymers (Shiri et al. 2000) for volume data storage and micro-electro-mechanical (MEMS) devices. It has also been validated using a number of textbook cases. The 3D solver has been developed in "C" on a single processor architecture and is currently being parallelized using "C" with message passing interface. Six more months of development effort will be required to fully parallelize the 3D version.

The only level of approximation is the numerical issues such as floating-point representation (a small contributor), grid density and sampling, and convergence to the solution.

Vector Optical Model (VOM)

There are currently two separate programs which constitute the VOM, a two dimensional optical modeling, and a three dimensional optical model.

The two-dimensional optical modeling has been developed R. Shiri [9, 10] and R. Lyon, at NASA/Goddard Space Flight Center. This model is developed in C with Message Passing Interface (MPI) on GSFC's Beowulf cluster. It has been used to model two-dimensional electromagnetic wave propagation through homogeneous and inhomogeneous media. For instance, this tool has been used to model photo-refractive polymers for dense data storage, mask edges for higher order diffraction calculations, and to investigate the wave propagation through micro-electro-mechanical (MEMS) devices. The core algorithm is based on scalar nodal finite element model of wave propagation in two dimensions. Given an input geometry, boundary conditions, optical properties of material and beam properties, the program solves iteratively, in parallel, a sparse linear system representing the two-dimensional inhomogeneous Helmholtz equation.

In order to solve a three-dimensional inhomogeneous Helmholtz equation representing the electromagnetic field propagation we have developed a vector finite element method. This method utilizing the edge-base elements instead of node-base elements overcomes a number of issues such as modal propagations. The baseline algorithm has initially been developed in "C" and is currently under expansion to "C" with MPI for execution on the Beowulf cluster.

VOM will then be used to locally model focal plane occulter masks, and pupil plane masks and the outputs will be subsequently coupled with the OSCAR physical optics model.

### 11.4.5. OSCAR: Systems Level Modeling

The Optical Systems Characterization and Analysis Research (OSCAR) software package has been incrementally developed since 1987 and has seen use on the Hubble Space Telescope (HST) [3], James Webb Space Telescope (JWST) [4], the Wavefront Control Testbed (WCT) [5], the Earth Atmospheric Solar Occultation Imager (EASI), the Solar Viewing Interferometry Prototype (SVIP), a number of military missions, and recently on the Terrestrial Planet Finder (TPF) [6]. OSCAR has been developed by R. Lyon to analyze optical systems and includes the capabilities of:

- filled, segmented, sparse and interferometric aperture systems
- low-, mid- and high-spatial frequency aberrations
- misalignments and deformations
- single plane and multiple plane Fresnel diffraction
- near field angular spectrum propagation
- scattering approximations
- coronographic masks and pupils
- deformable mirrors
- point jitter
- some limited active/adaptive controls
- wavefront sensing (phase retrieval / phase diversity).

OSCAR will be used in conjunction with the 2D and 3D vector optical models and will be coupled together to perform the systems level modeling to generate focal plane images (PSFs) as "seen" through the entire system. The interface between OSCAR and VOM will be in the form of *component transfer functions*.

The component transfer function is built by simulating a plane wave, with a fixed polarization at a fixed angle and propagating it through the component via the full vector diffraction model, e.g. a metallic focal plane occulting mask on a dielectric substrate, and then solving the rigorous inhomogeneous vector Helmholtz equation via finite element methods on 2D (rectilinear) and/or 3D (tetrahedral) meshes. This is then repeated as a function of plane wave input angle and for both polarizations. The ratio of the output reflected field to the incident field is the complex reflection function for each component of the angular plane wave spectrum and the ratio of output transmitted field to the incident field is the complex transmission function for each component of the angular plane wave spectrum. These reflection and transmission functions are a function of location on the input plane of the device. Thus the procedure is to decompose any input field, e.g. a focusing beam, into an angular spectrum of plane waves, multiply the angular spectrum by the transmission and reflection component transfer functions, then recompose the resultant forward and backward propagating angular spectrum for each polarization. This result can then be propagated by the more conventional methods in the OSCAR model. Solving for the component transfer function is computationally intensive and generally requires the Beowulf cluster; however, once the component transfer function is constructed the full propagation can be performed by inserting the transfer function into the OSCAR model and thus the entire model can be run on a single processor machine. Note that this approach can be used to model rigid body misalignments of the component without recalculating the component transfer function. Thus full system level sensitivity studies can be performed with this approach.

### 11.4.6. References to Appendices 11.3 and 11.4

## *11.5. A Brief Bibliography on Coronography*

In this section we compile a list of useful references on the subject of coronography. Emphasis has been given to publications on technical issues; only very few papers on predominantly astronomical topics have been included. Our selection is very subjective, of course, but it should provide useful entry points into the literature.

*trophysics Missions and Instrumentation*. Edited by J. Chris Blades, Oswald H. W. Siegmund. Proceedings of the SPIE, Volume 4854, pp. 1–8, February 2003.

[89] R. J. Vanderbei, D. N. Spergel, and N. J. Kasdin. Circularly Symmetric Apodization via Star-shaped Masks. *ApJ*, 599:686–694, December 2003.

[90] R. J. Vanderbei, D. N. Spergel, and N. J. Kasdin. Spiderweb Masks for High-Contrast Imaging. *ApJ*, 590:593–603, June 2003.

[91] S. M. Watson, J. P. Mills, S. L. Gaiser, and D. J. Diner. Direct imaging of nonsolar planets with infrared telescopes using apodized coronagraphs. *Appl.Opt.*, 30:3253–3262, August 1991.

[92] D. W. Wilson, P. D. Maker, J. T. Trauger, and T. B. Hull. Eclipse apodization: realization of occulting spots and Lyot masks. In *High-Contrast Imaging for Exo-Planet Detection*. Edited by Alfred B. Schultz. Proceedings of the SPIE, Volume 4860, pp. 361–370, February 2003.

[93] S. C. Woods and A. H. Greenaway. Wave-front sensing by use of a Green's function solution to the intensity transport equation. *Optical Society of America Journal*, 20:508–512, March 2003.

## *11.6. European Institutions with Relevant Expertise*

Here we list a number of European institutions and industries with relevant expertise. The list is non-exclusive and heavily biased towards entities known well to the workshop participants.

### 11.6.1. Mirror / Telescope Technology
- Carl Zeiss (Jena, D)
- Sagem Reosc (F)
- TNO-TPD (Delft, NL)
- Seiko (GB)
- European Southern Observatory (Garching, D)
- Alcatel (Cannes, F)
- LAM (Marseille, F)

### 11.6.2. Materials (SiC, Carbon Fiber)
- Boostec (F)
- Quentin (GB)
- Cob ham Composites (GB)

### 11.6.3. Deformable Mirrors (Manufacturing)
- Laboratoire d' Electronique de Technologie de L'Information (F)
- TNO-TPD / TU Eindhoven (NL)
- Oko Technologies (Delft, NL)
- BAe Systems (GB)
- Cilas (Paris, F)
- LAOG (Grenoble, F) in collaboration with LPMO (Besançon, F), IEMN (Lille, F)
- LAM (Marseille, F) in collaboration with LAAS (Toulouse, F)

### 11.6.4. Deformable Mirrors (Control)
- European Southern Observatory (Garching, D)
- Leiden University (Leiden, NL)
- MPI für Astronomie (Heidelberg, D)
- Observatoire de Paris (Meudon, F)
- ONERA (Chatillon, F)
- TNO-TPD (Delft, NL)

### 11.6.5. Wavefront Sensors
- Isaac Newton Group (La Palma, GB / NL / E)
- Observatory of Arcetri (Arcetri , I)
- Onera (Paris, F)
- Observatoire de Paris (Paris, F)
- Imagine Optics (Paris, F)

### 11.6.6. Optical Metrology
- TNO-TPD (Delft, NL)
- Heriott-Watt University (GB)
- University of Neuchatel (CH)

### 11.6.7. Phase Mask Fabrication and Control
- Institut Fresnel (Marseille, F)
- CSL (Liège, B)
- CEA (Saclay, F)
- REOSC/Sagem (Paris, F)
- Uppsala University  (Uppsala, S)
- Carl Zeiss (Jena, D)
- Observatoire de Paris (Meudon, F)
- Observatoire de la Cote d'Azur (Nice, F)

### 11.6.8. Amplitude Mask Fabrication and Control
- Optimask (F)

### 11.6.9. Polarimetry
- Amsterdam University (NL)
- ASTRON (Dwingeloo, NL)
- ETH (Zürich, CH)
- University of Hertfordshire (GB)
- Univeristy of Nice (F)

### 11.6.10.   Detectors
- MPI Halbleiterlabor (Munich, D)
- E2V Technologies (Chelmsford, GB)
- European Southern Observatory (Garching, D)

### 11.6.11.   Integral-Field Spectrographs
- Observatoire de Lyon (Lyon, F)

- Max-Planck-Institut für Extraterrestrische Physik (Garching, D)
- UK Astronomy Technology Center (Edinburgh, GB)
- Padova University (Padova, I)

## *11.7. Current Imaging Detectors*

The standard visible light array detector for space imaging is the Charge Coupled Device (CCD). CCDs available from SITe, E2V, Fairchild Imaging, and other foundries typically provide similar levels of performance. Dark currents are typically ~$10^{-3}$ e$^-$pix$^{-1}$sec$^{-1}$ with Multi-Phase Pinning (MPP). Read noise levels have reached performance floors at ~2-3e$^-$ rms. The primary problem with CCDs is the cumulative effect of damage from high energy protons and secondary neutrons which give rise to traps, and pixels with anomalously high dark rates (hot pixels). The net effect of traps is to degrade the charge transfer efficiency (CTE) of the detectors, which rapidly starts to impact performance at low light levels. The problem of radiation damage can be mitigated by using p-type rather than the standard n-type CCD architectures. A good example of p-type CCDs are the LBL devices, which have demonstrated excellent resistance to radiation (Bebek et al. 2003). These LBL devices are not necessarily the best CCD design for TEC however, since they are thick, fully depleted devices (~200-300 μm thick). They generate considerable bulk dark current, and are susceptible to cosmic rays, which limit exposure times. For TEC the optimum design would be a CCD based on the p-type architecture, but fully depleted over a smaller thickness of ~40-50 μm, which would require backside thinned, custom devices. It should be noted that MPP is a mandatory capability for space borne CCDs since dark current increase due to radiation damage is somewhat mitigated and hot pixels are also relatively quieter.

Alternatives for visible array detectors are CMOS hybrid technologies such as the Rockwell HYVISI detectors (Raytheon make a similar device). These detectors are arrays of PIN diodes built on CMOS multiplexers, which are typically used for HgCdTe/InSb detectors. These devices offer single pixel addressing, which is useful for very high contrast imaging since it permits very high dynamic ranges. Hybrid devices are claimed to have reached performance capabilities similar to CCDs, with slightly higher read noises of ~5e$^-$ rms with multi-sampled read outs. Detailed performance charts for these detectors are available at the Rockwell and Raytheon web pages. Quantum efficiency is comparable to CCDs, since similar techniques are being used to backside-thin these devices. While these hybrids do share some of the problems of IR arrays such as cross-talk in the multiplexers, they offer improved radiation hardness. These devices have yet to be characterized for flight operation, so their dark current performance after radiation damage is not well understood. As yet little work has been done with these detectors for astronomical imaging in ground-based instruments.

CMOS imagers are also catching up with CCDs, but there remain a number of issues to be addressed, primarily the read noise which is still too high for practical astronomical imaging. Many of these devices also have partially filed pixel structures and so are best operated in a backside-thinned configuration. CMOS devices also offer single pixel addressing and are radiation hard. Once again, further work to characterize them for flight applications is required.

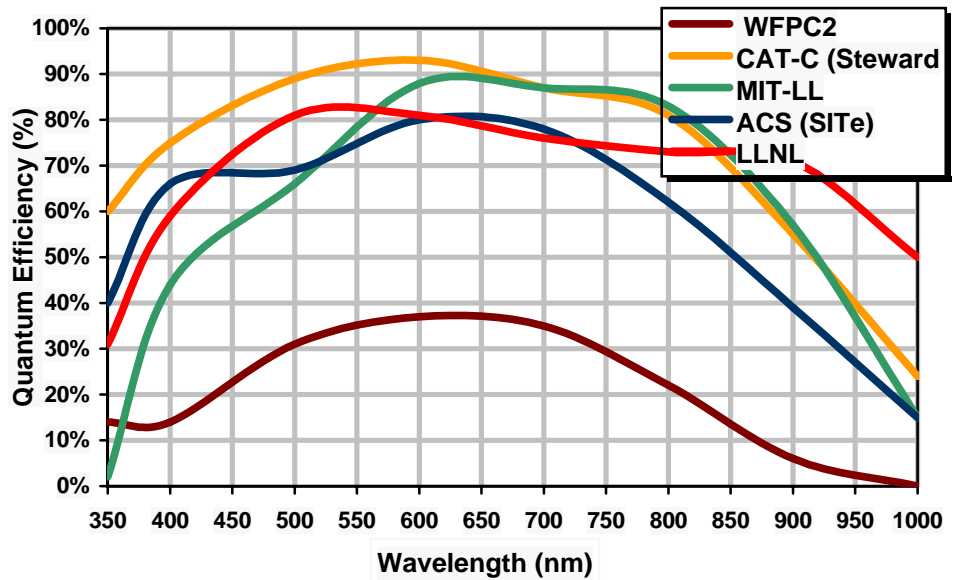

**Figure 21: Typical CCD quantum efficiency figures including the LBL CCD.**

A particularly desirable property in detectors for TEC would be the capability to image in three dimensions where the third dimension is a spectrum. STJs with a resolution of ~50 offer this capability, but their application for TEC is likely to be problematic since out of band blocking is a major challenge for coronographic imaging. In addition, array sizes are currently ~8×8, with a maximum of 18×50 proposed for development. These devices also exhibit large inter-pixel gaps. Finally, their operating temperatures of ~1K mean that they would severely impact the TEC telescope design since the cooling requires significant power, heat dissipation and possibility has associated dynamic disturbances from the cooler. This detector area is a field that offers considerable scope for future development. New concepts such as energy sensitive CCDs, or multi-layer CMOS detectors clearly need to be investigated.

Another area that needs further discussion is whether photon-counting performance is required. Currently, most photon counting detectors employ a microchannel-plate gain stage that cannot accommodate the large dynamic ranges found in coronography. Such limitations can impact the operational capabilities of the program, since a loss of pointing lock might, for instance, lead to loss of the detector. A photon counting detector is now available, the E2V $L^3$ CCD. It remains to be seen if this detector can be used in flight applications, since it employs multiple gain stages during readout. Hot pixels are produced by lattice damage in high field regions, so these devices might be more susceptible to high hot pixel production rates. Once again more flight characterization of this concept for TEC is required.

## *11.8. Acronyms and Abbreviations*

| | |
|---|---|
| AIC | Achromatic Interfero Coronograph |
| AO | Adaptive Optics |
| AU | Astronomical Unit |
| BW | BandWidth |
| CCD | Charge-Coupled Device |
| CHZ | Continuously Habitable Zone |
| CMOS | Complementary Metal Oxide Semiconductor |
| CTE | Charge Transfer Efficiency |
| CTE | Coefficient of Thermal Expansion |
| DLSS | Diffracted Light Suppression System |
| DM | Deformable Mirror |
| DOF | Degree Of Freedom |
| ELT | Extremely Large Telescope |
| ESA | European Space Agency |
| ESO | European Southern Observatory |
| ESTEC | European Space Research and Technology Center |
| FEEP | Field Emission Electric Propulsion |
| FOV | Field Of View |
| FSM | Fast Steering Mirror |
| FWHM | Full Width at Half Maximum |
| GSFC | Goddard Space Flight Center |
| HCIT | High Contrast Imaging Testbed |
| HEBS | High Energy Beam Sensitive |
| HST | Hubble Space Telescope |
| HZ | Habitable Zone |
| IFU | Integral-Field Unit |
| IR | InfraRed |
| ISO | Infrared Space Observatory |
| IWA | Inner Working Angle |
| JPL | Jet Propulsion Laboratory |
| JWST | James Webb Space Telescope |
| LBL | Lawrence Berkeley Laboratory |
| MACOS | Modeling and Analysis of Controlled Optical Systems |
| MAPLC | Multiple Apodized Pupil Lyot Coronograph |
| MEMS | Micro Electro-Mechanical System |
| MIRI | Mid-InfraRed Instrument |
| MPI | Max-Planck-Institut |
| MPI | Message Passing Interface |
| MPIA | Max-Planck-Institut für Astronomie |
| MPP | Multi-Phase Pinning |
| N/A | Not Applicable |
| NASA | National Aeronautics and Space Administration |
| NOAO | National Optical Astronomy Observatory |
| NRA | NASA Research Announcement |
| OPD | Optical Path Difference |
| OSCAR | Optical Systems Characterization and Analysis Research |
| OWA | Outer Working Angle |
| OWL | OverWhelmingly Large telescope |
| PIAA | Phase-Induced Amplitude Apodization |
| PSD | Power Spectral Density |
| PSF | Point Spread Function |
| RMS | Root Mean Square |
| ROC | Radius Of Curvature |
| RWA | Reaction Wheel Assembly |
| R&D | Research and Development |
| SI | Science Instrument |
| SIM | Space Interferometry Mission |
| SNR | Signal-to-Noise Ratio |
| STJ | Super-conducting Tunneling Junction |
| TBD | To Be Determined |
| TEC | Terrestrial Exoplanet Coronograph |
| TE-SAT | Terrestrial Exoplanets Science Advisory Team |
| TPF | Terrestrial Planet Finder |
| TPF-C | Terrestrial Planet Finder – Coronograph |
| TPF-I | Terrestrial Planet Finder – Interferometer |
| TRL | Technology Readiness Level |
| ULE | Ultra-Low Expansion |
| VOM | Vector Optical Model |
| VLT | Very Large Telescope |
| VLTI | Very Large Telescope Interferometer |
| VRE | Vegetation Red Edge |
| WF | Wave Front |
| WFE | Wave Front Error |
| WFS | Wave Front Sensor |
| WFSC | Wave Front Sensing and Control |
| ZOG | Zero Order Grating |
| 4QPM | Four-Quadrant Phase Mask |